\documentclass[12pt]{article}

\usepackage{graphicx}

\usepackage[intlimits]{amsmath}
\usepackage{amssymb}
\usepackage{bbm}
\usepackage{mathrsfs}
\usepackage{wrapfig}
\usepackage{caption}
\usepackage{subfigure}
\usepackage{enumerate}
\usepackage{epsfig}
\usepackage{cite} 
\usepackage{color}

\def\lsim{\mathrel{\raise.3ex\hbox{$<$\kern-.75em\lower1ex\hbox{$\sim$}}}}
\def\gsim{\mathrel{\raise.3ex\hbox{$>$\kern-.75em\lower1ex\hbox{$\sim$}}}}

\def\T{{\bf 10}}
\def\F{{\bf 5}}

\def\1{{\nu^c}}

\def\mt{\widetilde{m}_1}
\def\eq{\mathrm{eq}}

\def\bo{{\raise.15ex\hbox{\large$\Box$}}}               
\def\face{{\raise.2ex\hbox{$\displaystyle \bigodot$}\mskip-2.2mu \llap {$\ddot
        \smile$}}}                                      

\def\leftrightarrowfill{$\mathsurround=0pt \mathord\leftarrow \mkern-6mu
        \cleaders\hbox{$\mkern-2mu \mathord- \mkern-2mu$}\hfill
        \mkern-6mu \mathord\rightarrow$}       
\def\dvec#1{\vbox{\ialign{##\crcr
        \leftrightarrowfill\crcr\noalign{\kern-1pt\nointerlineskip}
        $\hfil\displaystyle{#1}\hfil$\crcr}}}           

\def\beq{\begin{equation}}
\def\eeq{\end{equation}}

\def\beqx{\begin{displaymath}}
\def\eeqx{\end{displaymath}}

\def\beqa{\begin{eqnarray}}
\def\eeqa{\end{eqnarray}}

\newcommand{\equi}{\textrm{eq}}

\catcode`@=11
\def\@citex[#1]#2{\if@filesw\immediate\write\@auxout{\string\citation{#2}}\fi
  \def\@citea{}\@cite{\@for\@citeb:=#2\do
    {\@citea\def\@citea{,\penalty\@m}\@ifundefined
      {b@\@citeb}{{\bf ?}\@warning
       {Citation `\@citeb' on page \thepage \space undefined}}%
\hbox{\csname b@\@citeb\endcsname}}}{#1}}

\def\citer{\@ifnextchar [{\@tempswatrue\@citexr}{\@tempswafalse\@citexr[]}}
\def\@citexr[#1]#2{\if@filesw\immediate\write\@auxout{\string\citation{#2}}\fi
  \def\@citea{}\@cite{\@for\@citeb:=#2\do
    {\@citea\def\@citea{--\penalty\@m}\@ifundefined
       {b@\@citeb}{{\bf ?}\@warning
       {Citation `\@citeb' on page \thepage \space undefined}}%
\hbox{\csname b@\@citeb\endcsname}}}{#1}}
\catcode`@=12

\begin{document}
\date{\mbox{ }}

\title{
\vspace{-2cm}
{\normalsize     
DESY 11-002\hfill\mbox{}\\
April 2011\hfill\mbox{}\\}
\vspace{1.5cm}
\bf Entropy, Baryon Asymmetry and\\ Dark Matter  
from Heavy Neutrino Decays \\[8mm]}
\author{W.~Buchm\"uller, K.~Schmitz, G. Vertongen\\[2mm]
{\normalsize\it Deutsches Elektronen-Synchrotron DESY, 22607 Hamburg, Germany}}
\maketitle

\thispagestyle{empty}

\vspace{1cm}
\begin{abstract}
\noindent
The origin of the hot phase of the early universe remains so far an unsolved 
puzzle. A viable option is entropy production through the
decays of heavy Majorana neutrinos whose lifetimes determine the initial 
temperature. We show that baryogenesis and the production of dark 
matter are natural by-products of this mechanism. As is well known, 
the cosmological baryon asymmetry can be accounted for by leptogenesis
for characteristic neutrino mass parameters. We find that thermal gravitino 
production then automatically yields the observed amount of dark matter, 
for the gravitino as the lightest superparticle and typical gluino masses. 
As an example, we consider the production of heavy Majorana neutrinos in the 
course of tachyonic preheating associated with spontaneous $B-L$ breaking. 
A quantitative analysis leads to constraints on the superparticle masses 
in terms of neutrino masses: For a light neutrino mass of 
$10^{-5}\,\mathrm{eV}$ the gravitino mass can be as small as $200\,\mathrm{MeV}$,
whereas a lower neutrino mass bound of $0.01\,\mathrm{eV}$ implies a lower 
bound of $9\,\mathrm{GeV}$ on the gravitino mass. The measurement of a light 
neutrino mass of $0.1\,\mathrm{eV}$ would rule out heavy neutrino decays as
the origin of entropy, visible and dark matter.
\end{abstract}

\newpage

\tableofcontents

\section{Introduction}

Detailed studies of the cosmic microwave background provide direct evidence
for the hot thermal universe close to its minimal temperature \cite{PDG10}. The
extrapolation to higher temperatures, beyond primordial nucleosynthesis, is
very uncertain, and we do not know how large the maximal temperature of the hot 
early universe has been. It is widely believed that the universe was `reheated'
by a transition from a preceding inflationary phase  where `vacuum energy' 
dominated the expansion \cite{lin07}. Knowing the resulting reheating 
temperature is of fundamental importance since it is closely related to the 
origin of the matter-antimatter asymmetry and the nature of dark matter. 

In a recent paper we have suggested that the entropy of the hot early universe
was produced in the decays of heavy Majorana neutrinos whose lifetimes 
determine the reheating temperature \cite{bsv10}. We have demonstrated 
that the baryon asymmetry and dark matter then naturally result as by-products:
The mechanism of baryogenesis is a mixture of thermal and nonthermal 
leptogenesis, and the dominant component of dark matter is the gravitino which 
is assumed to be the lightest superparticle.

Our work is closely related to previous studies of thermal leptogenesis 
\cite{fy86,bpy05} and nonthermal leptogenesis via inflaton decay 
\cite{ls91,ahx99}, where the inflaton lifetime determines the reheating 
temperature. 
In supersymmetric models with global $B-L$ symmetry the scalar superpartner
$\widetilde{N}_1$ of the lightest heavy neutrino $N_1$ can play the role
of the inflaton in chaotic \cite{msx92} or hybrid \cite{abx04} inflationary
models. Coherent $\widetilde{N}_1$ oscillations after inflation may even 
dominate the energy density of the universe \cite{hmy01}. Nonthermal
leptogenesis can be realized at significantly smaller temperatures than
thermal leptogenesis. In this way the `gravitino problem' for heavy unstable 
gravitinos \cite{we82,ens85,kkm05} can be easily avoided, which has been
one of the main motivations of nonthermal leptogenesis.

It is well known that the high temperatures characteristic for thermal
leptogenesis can become a virtue if the gravitino is the lightest superparticle
(LSP). For superparticle masses as they arise in gravity or gaugino
mediation, thermal production of gravitinos can then explain the observed 
amount of dark matter \cite{bbp98}. As pointed out in \cite{bsv10}, the
required high temperatures are indeed realized if the universe is reheated through 
the decays of the heavy Majorana neutrinos. The vacuum decay width of the
lightest heavy Majorana neutrino is given by
\begin{equation}\label{mtilde}
\Gamma_{N_1}^0 = 
\frac{\widetilde{m}_1}{8 \pi} \left(\frac{M_1}{v_\textrm{\tiny EW}}\right)^2
\sim 10^3\,\textrm{GeV} \ ,
\end{equation}
where we have used the typical values $M_1 \sim 10^{10}~\mathrm{GeV}$ for the 
$N_1$ neutrino mass, $\widetilde{m}_1 \sim 0.01~\mathrm{eV}$ for the 
effective light neutrino mass and $v_\textrm{\tiny EW} = 174~\mathrm{GeV}$ for 
the vacuum expectation value of electroweak symmetry breaking. The 
corresponding reheating temperature is given by 
\begin{equation}\label{Tr}
T_{RH} \approx \left(\frac{90}{8\pi^3 g_{\star,\rho}}\right)^{1/4} 
\sqrt{\Gamma_{N_1}^0 M_P} \sim 10^{10}~\textrm{GeV}\ ,
\end{equation}
where we have used $g_{\star,\rho} \sim 200$ for the effective number of 
relativistic 
degrees of freedom, and $M_P = 1.22 \times 10^{19}~\textrm{GeV}$ is the Planck 
mass. 

From Eqs.~(\ref{mtilde}) and (\ref{Tr}) one obtains the reheating temperature
$T_{RH}$ in terms of the neutrino masses $\widetilde{m}_1$ and $M_1$, 
the two key
parameters for thermal and nonthermal leptogenesis. Assuming the gluino to be
the heaviest gaugino, thermal gravitino production is dominated by QCD 
processes, which yields the gravitino abundance in terms of the gravitino 
mass $m_{\widetilde{G}}$ and the gluino mass $m_{\tilde{g}}$,
\begin{align}
\Omega_{\widetilde{G}} h^2 = C
\left(\frac{T_{RH}}{10^9~\textrm{GeV}}\right)
\left(\frac{10~\textrm{GeV}}{m_{\widetilde{G}}}\right)
\bigg(\frac{m_{\tilde{g}}}{1~\textrm{TeV}}\bigg)^2 \ ,
\label{eq:GDMestimate}
\end{align}
where the coefficient $C=0.26$ to leading order in the gauge coupling
\cite{bbb00,ps06}.\footnote{Note that $C$ has an $\mathcal{O}(1)$ uncertainty
due to unknown higher order contributions and nonperturbative effects 
\cite{bbb00}. Resummation of thermal masses increases $C$ by about a factor 
of two \cite{rs07}.} Since $T_{RH}$ depends on $\widetilde{m}_1$ and $M_1$,
the requirement\footnote{For the superparticle masses considered in this
paper the contribution to the gravitino abundance from the decay of the
next-to-lightest superparticle (NLSP) is negligible \cite{fst04,ste08}.}  
$\Omega_{\widetilde{G}} h^2 = \Omega_{\mathrm{DM}} h^2 \simeq 0.11$ \cite{wmap10}
yields a connection between neutrino and superparticle mass parameters.
The neutrino masses $\widetilde{m}_1$ and $M_1$ are in turn constrained
by the condition that the maximal baryon asymmetry is larger than the
observed one, $\eta_B \geq \eta_B^{\mathrm{obs}} = 6.2\times 10^{-10}$
\cite{wmap10}.

In the following sections we shall study in detail the connection between 
neutrino and superparticle masses, which is implied by successful leptogenesis 
and gravitino dark matter. As an example, we shall consider tachyonic
preheating \cite{fgb01}, associated with $B-L$ breaking, 
as a mechanism which can lead
to a phase where the energy density is dominated by heavy Majorana neutrinos.
As we shall see, the final baryon asymmetry and the dark matter abundance
can then be calculated in terms of several parameters of the Lagrangian, 
independent of initial conditions: the scale $v_{B-L}$ of $B-L$ breaking,
the heavy Majorana neutrino mass $M_1$, the effective light neutrino mass
$\widetilde{m}_1$, the gravitino mass $m_{\widetilde{G}}$ and the gluino mass 
$m_{\tilde{g}}$. Particularly interesting is the resulting connection between
the lightest neutrino mass $m_1$ and the gravitino mass.

Our analysis requires a flavour model which is flexible enough to allow for a 
large range of the neutrino masses $M_1$ and $\widetilde{m}_1$, the crucial 
parameters for leptogenesis. Such a model is described in Section~2. 
Subsequently, in Section~3, we discuss in detail the Boltzmann equations that 
describe the time evolution of the state produced in the tachyonic decay of 
the false vacuum with unbroken $B-L$ symmetry. A novel technical aspect of our
analysis is the separate treatment of thermal and nonthermal contributions
to the abundance of the heavy Majorana neutrinos. 

Using the set of Boltzmann equations derived in Section~3, we study an 
illustrative example of our mechanism in Section~4. The parameters are
chosen such that the nonthermal contribution to leptogenesis dominates.
Particular emphasis is
given to the time dependence of the various production and decay rates and
the emergence of a plateau where the temperature is approximately
constant. In Section~5 the analysis is extended to the entire parameter space.
First, the ranges of $M_1$ and $\widetilde{m}_1$ are determined for which 
leptogenesis is successful. The observed dark matter abundance then constrains
the superparticle masses in terms of the neutrino masses. 

Our results are summarized in Section~6. The appendices deal with various
technical aspects of our calculations: conventions for the Boltzmann
equations in Appendix~A, the distribution function of thermally
produced neutrinos in Appendix~B, analytical approximations for the reheating
temperature in Appendix~C and semi-analytical results for the
gravitino abundance in Appendix~D.
\section{Flavour model and leptogenesis}
\label{sec:flavor}

In the following we shall describe a flavour model which describes masses and
mixings of quarks, charged leptons and neutrinos, and which is flexible
enough to allow for a large range of the neutrino parameters $M_1$ and
$\widetilde{m}_1$ that are crucial for leptogenesis. The model is a variant
of \cite{by99} and satisfies all constraints from flavour changing processes 
\cite{bdh99}.

We consider the extension 
SU(3)$_c\times$SU(2)$_L\times$U(1)$_Y\times$U(1)$_{B-L}$
of the standard model gauge group. The Yukawa interactions of quarks and 
leptons with Higgs fields are described by the following superpotential 
for matter superfields,
\begin{equation}
\label{eq:potential}
W_{M} = h_{ij}^u \T_i\T_j H_u +  h_{ij}^d \F^*_i\T_j H_d 
+ h_{ij}^{\nu}
\F^*_i n^c_j H_u  +  \frac{1}{2} h_i^n n^c_i n^c_i S \ .
\end{equation}
Here the standard model fermions have been arranged in SU(5) multiplets, 
$\mathbf{10} = (q, u^c, e^c)$ and $\mathbf{5}^* = (d^c, \ell)$, and 
$i,j=1\ldots3$ are flavour indices. $n^c$ contain $\nu_R^c$, the charge 
conjugates of the
right-handed neutrinos, which are related to the heavy Majorana 
neutrinos $N$ through $N_i =\nu_{Ri} + \nu_{Ri}^c$. For simplicity, we have
used SU(5) notation assuming that the colour triplet partners of the Higgs 
doublets have been projected out.

Vacuum expectation values of the scalar 
Higgs fields, $\langle H_{u,d}\rangle = v_{u,d}$, break the electroweak 
symmetry. The superpotential
\begin{equation}
\label{eq:B-L}
W_{B-L} = \frac{\sqrt{\lambda}}{2} T \left(v_{B-L}^2 - 2 S_1 S_2\right) 
\end{equation}
enforces $B-L$ breaking, with 
$\langle S_1 \rangle = \langle S_2 \rangle = v_{B-L}/\sqrt{2}$,
via the supersymmetric Higgs mechanism. Using the field basis 
$S_1 = S'\exp{(i\Phi)}/\sqrt{2}$, $S_2 = S'\exp{(-i\Phi)}/\sqrt{2}$ and 
shifting around the vacuum expectation value, $S' = v_{B-L} + S$,  
$S$ and $T$ have a common Dirac mass term, whereas $\Phi$ and
the $B-L$ vector multiplet form together a massive vector
multiplet. Since $B-L$ is gauged the inflaton is
  identified with the scalar component of the singlet field $T$. The potential for the scalar
neutrinos is not sufficiently flat.

The pattern of Yukawa couplings $h_{ij}$ is determined by a Froggatt-Nielsen 
U(1) flavour symmetry, following Ref \cite{by99}. The 
matter fields $\psi_i$ and an extra singlet
$\Sigma$ carry charges $Q_i$ and $-1$ under the flavour symmetry
respectively, and are coupled together via effective
non-renormalisable interactions associated with a scale $\Lambda >
\Lambda_{\textrm{GUT}}$. The Yukawa couplings are generated once the
flavour symmetry is spontaneously broken by the expectation value of the 
$\Sigma$ field, and are given by
\begin{equation}
h_{ij} \propto \eta^{Q_i + Q_j}\,,
\end{equation}
where $\eta = \langle \Sigma \rangle /\Lambda$. The hierarchies
of quark and lepton masses are then naturally obtained
for $\eta^2 \simeq 1/300$ using the chiral charges
listed in Table \ref{tab:charges}. It is important to note that the
  Yukawa couplings are only specified up to factors of
  ${\cal O}(1)$. Cosmology further constrains the chiral charges. For
instance, successful thermal leptogenesis requires $a+d=2$ \cite{by99}.

\begin{table}[t]
\begin{center}
\begin{tabular}{c|cccccccccccccccc}\hline \hline
$\psi_i$ & $\T_3$ & $\T_2$ & $\T_1$ & $\F^*_3$ & $\F^*_2$ & $\F^*_1$ &
$n^c_3$ & $n^c_2$ & $n^c_1$ & $H_u$ & $H_d$ & $S_1$ & $S_2$ & $T$  \\ \hline
$Q_i$ & 0 & 1 & 2 & $a$ & $a$ & $a+1$ & $b$ & $c$ & $d$ & 0 & 0 & 0 & 0 & $e$  
\\ \hline\hline
\end{tabular}
\medskip
\caption{Chiral $U(1)$ charges.} 
\label{tab:charges}
\end{center}
\end{table}


We shall restrict our analysis to the case of a hierarchical heavy neutrino 
mass spectrum, $M_1 \ll M_{2,3}$, which is obtained for flavour charges of 
$N_1$ and $N_{2,3}$ separated by one unit, $b=c=d-1$. This is sufficient 
to illustrate our main point, the contraints imposed on the gravitino mass
by neutrino masses. The lepton asymmetry will mostly be generated by decays
of the lightest heavy neutrino $N_1$. 
The masses of the Majorana neutrinos are given by
\begin{subequations}
\label{eq:HNmasses}
\begin{align}
M_{1\phantom{,2}} &\simeq \eta^{2d}\ v_{B-L}\ , 
\label{eq:HN1mass}\\
M_{2,3}&\simeq \eta^{2(d-1)}\ v_{B-L}\ ,
\end{align}
\end{subequations}
where $M_2/M_3 = \mathcal{O}(1)$. The mass spectrum is now parametrised 
by the three remaining free charges $a$, $d$ and $e$, which can be related to 
the physical parameters $v_{B-L}$, $M_1$ and the Higgs mass $m_S$.

Using the Dirac mass matrix $m_D = h^\nu v_u$ and the Majorana
mass matrix $M = h^n v_{B-L}$ derived from
Eq.~\eqref{eq:potential}, the eigenvalues of the light neutrino mass matrix 
$m = - m_D M^{-1} m_D^T$ are given by
\begin{subequations}
\label{eq:LNmasses}
\begin{align}
m_1 &\simeq \eta^{2a+2} \frac{v_\textrm{EW}^2}{v_{B-L}}\ ,\\
m_2 \simeq m_3 &\simeq \eta^{2a\phantom{+2}}\frac{v_\textrm{EW}^2}{v_{B-L}}\ , 
\label{eq:LN23masses}
\end{align}
\end{subequations}
where we have used $\tan\beta = v_u/v_d > \mathcal{O}(1)$ and 
$v_u \simeq v_{\textrm{EW}}$.
Choosing $\overline m_\nu =\sqrt{m_2 m_3} \simeq 
|(m_1^2-m_2^2)(m_2^2-m_3^2)|^{1/4} \simeq 3\times 10^{-2}\ \mathrm{eV}$
(cf.~\cite{PDG10}), the $B-L$ breaking scale is determined by the flavour 
charge $a$,
\begin{align}
\label{eq:vBL}
  v_{B-L} \simeq \eta^{2a} \ \frac{v_{\textrm{EW}}^2}{\overline
    m_\nu}\,,
\end{align}
where $v_{\textrm{EW}}^2/\overline m_\nu \simeq 10^{15}~\mathrm{GeV}$ is the 
grand unification mass scale. Once $v_{B-L}$ is fixed, $M_1$ is directly 
related to the charge $d$ through Eq.~\eqref{eq:HN1mass}.

The ranges over which the chiral charges $a$ and $d$, and thus the
physical parameters $v_{B-L}$ and $M_1$, are allowed to vary is restricted.
First, the requirement that Yukawa couplings do not exceed the top-Yukawa
coupling imposes the lower bounds $a\geq 0$ and $d\geq 1$.
Furthermore, the upper bound $a\leq 1$ follows from  $\tan
\beta > \mathcal{O}(1)$. No corresponding upper bound on the charge $d$ 
exists but, as we shall see later, a CP asymmetry sufficiently large for
successful leptogenesis requires $d<3$. Using  Eq.~\eqref{eq:vBL}, the allowed 
range of $B-L$ breaking scales reads
\begin{align}
3\times 10^{12}\ \textrm{GeV}\leq v_{B-L} \leq 1\times 10^{15}\ \textrm{GeV}\ .
\label{eq:vBLrange}
\end{align}
For fixed $B-L$ breaking scale, the possible range of $M_1$ is given by
Eq.~\eqref{eq:HN1mass},
\begin{subequations}
\begin{align}
1\times 10^5\,\textrm{GeV}\leq M_1 \leq 1\times 10^{10}\,\textrm{GeV}
\qquad \textrm{for} \quad v_{B-L} &= 3\times 10^{12}\,\textrm{GeV}\,,\\
3\times 10^7\,\textrm{GeV}\leq M_1 \leq 3\times 10^{12}\,\textrm{GeV}
\qquad \textrm{for} \quad v_{B-L} &= 1\times 10^{15}\,\textrm{GeV}\,.
\end{align}
\end{subequations}
The ranges for $v_{B-L}$ and $M_1$ correspond to a continuous 
variation of the flavour charges, which can effectively be realised by
fractional charges. Note that the constraint from thermal leptogenesis,
$a+d = 2$, is now relaxed.

Given the Yukawa couplings, one easily obtains the decay widths of the
heavy Majorana neutrinos,
\begin{align}
\label{eq:Ndecayrate}
\Gamma_{N_i}^0 \simeq \frac{\widetilde m_i}{8\pi}
\frac{M_i^2}{v_{\textrm{EW}}^2}\ , 
\end{align}
with the effective light neutrino masses
\begin{align}\label{m1tilde}
\widetilde m_i = \frac{1}{M_i}\ (m_D^\dagger m_D)_{ii}
\simeq \eta^{2a}\ \frac{v_{\textrm{EW}}^2}{v_{B-L}}
\simeq \overline m_\nu \ .
\end{align}
The CP asymmetry in the heavy neutrino decays are given by \cite{crv96,bp98}
\begin{align}
\epsilon_i = \frac{1}{8 \pi (h^{\nu \dagger} h^\nu)_{ii}} \, \sum_{j\neq
  i} \mbox{Im}\left\{\left[ \left(h^{\nu \dagger} h^\nu  \right)_{ij} 
\right]^{2}\right\} \,F\left( \frac{M_j}{M_i} \right) \,,
\label{eq:epsiloni}
\end{align}
where we use the standard model expression for $F$.\footnote{The 
expression in the supersymmetric standard model would only
slightly increase the value of $\epsilon_i$.} 
Using our flavour model, one gets
\begin{align}
\epsilon_1 &\simeq 0.1\,\eta^{2(a+d)} = 0.1\,
\frac{\overline m_\nu \,M_1}{v^2_{\textrm{EW}}}\,, \qquad
\epsilon_{2,3} \simeq \epsilon_1\,\eta^{-2}\,.
\label{eq:epsilon123}
\end{align}
Note that this is the maximal CP asymmetry for fixed $M_1$ \cite{hmy01,di02}, 
which is obtained in the limit $\widetilde m_1 \rightarrow 0$.
For $M_1 \simeq 10^{10}~\mathrm{GeV}$, this yields $\epsilon_1 \sim
10^{-6}$. For other Majorana neutrino masses, the asymmetry scales like 
the mass ratio, $\epsilon_1 \sim 10^{-6}\,M_1/10^{10}~\mathrm{GeV}$.

Since the light neutrino mass matrix is not hierarchical, the ${\cal O}(1)$ 
uncertainties in the $h^\nu$ Yukawa couplings can lead to large deviations
from the relation (\ref{m1tilde}) between $\widetilde m_1$ and 
$\overline m_\nu$. The only rigorous inequality is  
$\widetilde m_1 \geq m_1$ \cite{fhy02}. We take these uncertainties
into account by varying the effective neutrino mass in the range
\begin{align}
10^{-5}\ \textrm{eV} \leq \widetilde m_1 \leq 0.1\ \textrm{eV}\ .
\end{align}
Since the heavier Majorana neutrinos $N_{2,3}$ only play a marginal role in 
our scenario (see below), we ignore possible deviations from the relation
(\ref{m1tilde}) and use $\widetilde m_{2,3} = \overline m_\nu$.

In the following we consider the `waterfall transition' from the false
vacuum $\langle S \rangle = 0$ to the true vacuum
$\langle S \rangle = v_{B-L}$, which may happen at
the end of hybrid inflation.
A tachyonic instability in the Higgs potential leads to spinodal growth of 
the long-wavelength Higgs modes. The true vacuum is reached after a rapid
transition at time $t_{\textrm{PH}}$ \cite{gbx01},
\begin{align}
\left.\langle S^{\dagger}S\rangle \right|_{t = t_\textrm{PH}} 
= v_{B-L}^2\ , \qquad
t_\textrm{PH} \simeq \frac{1}{2m_S}
\ln\left(\frac{32 \pi^2}{\lambda}\right)\ .
\label{eq:tPH}
\end{align}
The Higgs boson forms a massive supermultiplet together with a Dirac
fermion and three additional bosons. Its mass and the false vacuum
energy density are given by
\begin{align}
m_S^2 = \lambda v_{B-L}^2\ , \qquad 
\rho_0 = \frac{1}{4} \lambda v_{B-L}^4\ .
\end{align}
The size of the coupling $\lambda$ is determined by the flavour charge $e$
in Table~1. We shall restrict our analysis to the case $e=2(d-1)$, such
that $m_S \simeq M_{2,3}$. As a consequence, the Higgs boson only decays
to pairs of $N_1$ neutrinos and not to pairs of $N_2$ or $N_3$ neutrinos.
Like the heavy neutrino masses and the CP asymmetries, the Higgs mass only 
depends on the flavour charge $a+d$ whereas the false vacuum energy density 
is determined by a different combination of charges,
\begin{align}
m_S \simeq \eta^{2(a+d-1)}\ \frac{v_{\textrm{EW}}^2}{\overline m_\nu}\ ,
\quad
\rho_0^{1/4} \simeq \eta^{2a+d-1}\ \frac{v_{\textrm{EW}}^2}{\overline m_\nu}\ .
\end{align}
 
During the tachyonic preheating the energy of the false vacuum is converted 
mostly into a nonrelativistic gas of $S$ bosons ($|\vec p_S|/m_S\ll 1$), with 
an admixture of heavy neutrinos. Their contribution to energy density
and number densities is determined by their coupling to the Higgs
field \cite{gbx01},  
\begin{subequations}
\begin{align}
r_{N_i} &= \frac{\rho_{N_i}}{\rho_0} \simeq 1.5 \times 10^{-3}\, g_N\,
\lambda \,f(\alpha_i,0.8) \,,\label{eq:GarcBellRes0}\\
n_{N_i} &\simeq 3.6 \times 10^{-4} \,g_N \,m_S^3 \,f(\alpha, 0.8)/\alpha\,,
\label{eq:GarcBellRes}
\end{align}
\end{subequations}
where $g_N = 2$ and
$f(\alpha,\gamma) = \sqrt{\alpha^2 + \gamma^2} - \gamma$
with $\alpha_i = h^n_i / \sqrt{\lambda}$. For the heaviest neutrinos
$N_{2,3}$, one obtains
\begin{align}
r_{N_{2,3}} \simeq 1\times 10^{-3} (h_2^n)^2 \,,
\end{align}
while the relative contribution from the lightest right-handed
neutrinos, $r_{N_1}/r_{N_{2,3}} \simeq {\cal O}(\eta^2)$, is
negligible.

For simplicity, we neglect all superpartners as well
  as $B-L$ gauge bosons and inflaton modes ($T$) which are produced
  during tachyonic preheating. We expect their contributions to
  not significantly change our results, similarly as in supersymmetric
  leptogenesis \cite{Plumacher:1997ru}. A detailed discussion will be
  presented in \cite{Buchmuller:2011aa}. A further important aspect of
  tachyonic preheating is the production of cosmic strings
  \cite{fgb01}. Their effect on the baryon asymmetry is model
  dependent \cite{Jeannerot:2006dy}. It has to be analysed for the parameters
  of our model \cite{Buchmuller:2011aa} taking also into account
  non-minimal couplings of the inflaton \cite{Ferrara:2010yw}.

Our choice of flavour charges implies that the Higgs field $S$ decays 
exclusively into pairs of $N_1$ neutrinos. The resulting
decay rate is given by
\begin{align}
\Gamma_S^0 = \frac{(h^n_1)^2}{16 \pi} \ m_S\ \left[ 1-(2M_1/m_S)^2
\right]^{3/2}\ .
\label{eq:GammaS0}
\end{align}
Note that if the $S$ boson decayed into more than one heavy neutrino flavour, 
this would lead to an interplay between lepton asymmetries
which could have different signs and it
would also change the reheating temperature. In the present paper we
restrict our analysis to the simplest case.

\section{Entropy production through neutrino decays}
\label{sec:boltz}

In order to understand the reheating process subsequent
to false vacuum decay quantitatively, we have to track the evolution of the
following abundancies as functions of the scale factor $a$:
the $S$ Higgs bosons, the heavy $N_1$
Majorana neutrinos, the $B-L$ asymmetry, the standard model radiation $R$ and
the gravitinos $\widetilde{G}$.
The appropriate tool for this task are the Boltzmann equations
in an expanding Friedmann-Lema\^itre
universe, the formalism of which is summarized in
Appendix~\ref{appendix:BE}.
In what follows, we consider comoving number densities,
\begin{align}
N_X(t) = a(t)^3 n_X(t)\,,
\end{align}
where $a$ is normalized to $a(t_{\textrm{PH}}) = 1$. This
quantity exhibits the advantage of being well defined for times prior
to reheating of the universe.

\subsection{Initial conditions}

\label{subsec:IniCon}

After tachyonic preheating the universe is filled by a gas of
nonrelativistic $S$ bosons as well as heavy $N_2$ and $N_3$ neutrinos.
Given the flavour structure presented in Section~\ref{sec:flavor},
the latter decay into standard model particles on time scales much shorter
than the $S$ boson lifetime,
\begin{align}
\frac{\Gamma_{N_{2,3}}^0}{\Gamma_S^0} \simeq \eta^{2\left(a-d-1\right)}
\geq \eta^{-2}\,, \qquad 1\geq a \geq 0 \,, \qquad d \geq 1 \,.
\end{align}
Hence, we do not explicitly resolve the time dependence
of the $N_{2,3}$ number densities but approximate their evolutions
by step functions, \textit{i.e.} instantaneous drop-offs at times
$t_2 = t_{\textrm{PH}} + 1/\Gamma_{N_2}^0$ and
$t_3 = t_{\textrm{PH}} + 1/\Gamma_{N_3}^0 \simeq t_2$
(cf. Eq.~\eqref{eq:tPH}),
\begin{align}
n_{N_{2,3}}(t) \approx \frac{a(t_{\textrm{PH}})^3}
{a(t)^3} \,n_{N_{2,3}}(t_{\textrm{PH}})\,
\Theta(t_2 - t)\,, \qquad t\geq t_{\textrm{PH}}\,.
\label{eq:nN2t}
\end{align}
The sudden decay of the $N_{2,3}$ neutrinos sets the stage for the
reheating of the universe which is why we choose $t = t_2 \simeq t_3$
as initial time when solving the Boltzmann equations.
Let us now determine the initial conditions at this time.
When the $N_{2,3}$ neutrinos decay they transfer their energy
inherited from tachyonic preheating to radiation,
\begin{align}
\rho_R(t_2) = 2 r_{N_2} \frac{\rho_0}{a^3(t_2)}\,.
\label{eq:rhoRIni}
\end{align}
For a thermal bath of temperature $T$ the energy and number
densities $\rho_R$ and $n_R$ of radiation quanta are given by
\begin{subequations}
\label{eq:rnRT}
\begin{align}
\rho_R &= \frac{\pi^2}{30} \,g_{\star,\rho} T^4 \,,\label{eq:rRT}\\
n_R &= \frac{\zeta(3)}{\pi^2} \,g_{\star,n} T^3\,, \label{eq:nRT}
\end{align}
\end{subequations}
where $g_{\star,\rho}$ and $g_{\star,n}$ denote corresponding
effective sums of relativistic degrees of freedom
\begin{subequations}
\begin{align}
g_{\star,\rho} = \: \sum_{\textrm{bosons}}\left(T_i/T\right)^4 +
\frac{7}{8}\sum_{\textrm{fermions}}\left(T_i/T\right)^4 \,,\\
g_{\star,n} = \: \sum_{\textrm{bosons}}\left(T_i/T\right)^3 +
\frac{3}{4}\sum_{\textrm{fermions}}\left(T_i/T\right)^3\,.
\end{align}
\end{subequations}
In our numerical analysis we employ the values $g_{\star,\rho} =
915/4$ and $g_{\star,n} = 427/2$ $\left(T_i =T\right)$ for the minimal
supersymmetric standard model (MSSM).
Equating the energy densities in Eqs.~\eqref{eq:rhoRIni} and \eqref{eq:rRT}
yields the initial temperature $T(t_2)$ and thereby the
initial comoving number density $N_R(t_2)$,
\begin{subequations}
\begin{align}
T(t_2) &= \left(\frac{30}{\pi^2 g_{\star,\rho}}
\frac{2 r_{N_2}\rho_0}{a^3(t_2)}\right)^{1/4}\,, \\
N_R(t_2) &= a^3(t_2) \,\frac{\zeta(3)}{\pi^2} \,g_{\star,n}\,
T^3(t_2)\,.
\end{align}
\end{subequations}
Note that Eq.~\eqref{eq:rnRT} provides us with an expression
for $T$ as a function of $N_R$,
\begin{align}
T = \left(\frac{\pi^2 N_R}{\zeta(3)g_{\star,n}a^3}\right)^{1/3}\,.
\label{eq:TNR}
\end{align}
As we will argue in Section~\ref{subsubsec:RelCOs} this relation can be used
to determine the time evolution of the temperature.
The out-of-equilibrium decay of $N_{2,3}$ also produces an
initial $B-L$ asymmetry.\footnote{In Ref. \cite{gbx01}
only the initial $B-L$ asymmetry from tachyonic preheating is taken
into account.}
The corresponding comoving number density is given by
(cf. Eq.~\eqref{eq:epsilon123})
\begin{align}
N_{B-L}(t_2) &= \epsilon_2 \,N_{N_2}(t_2) +
\epsilon_3 \,N_{N_3}(t_2)\,, \nonumber\\&\simeq 0.2 \,\eta^{-2}
\frac{\overline m_\nu M_1}{v_{\textrm{EW}}^2} 
N_{N_2}(t_{\textrm{PH}})\,,
\end{align}
where the comoving number density $N_{N_2}$ at $t_{\textrm{PH}}$ follows from
Eq.~\eqref{eq:GarcBellRes}.

At $t=t_2$ the dominant contribution to the energy density
resides in the gas of non-relativistic $S$ bosons
(cf. Eq.~\eqref{eq:GarcBellRes0}),
\begin{align}
\rho_S (t_2) = \frac{1}{a^3(t_2)}
\left(1 - 2 r_{N_2}\right) \rho_0\,,
\end{align}
which corresponds to an initial comoving number density
\begin{align}
N_S(t_2) = a^3(t_2) n_S(t_2) =  a^3(t_2)\,\frac{\rho_S(t_2)}{m_S} \,.
\end{align}
According to Eq.~\eqref{eq:numdendef},
the initial phase space distribution function $f_S (t_2,p)$ can be
inferred from $n_S(t_2)$.
Guided by the results of Ref.~\cite{gbx01} we make the ansatz of a
delta-peaked momentum distribution, \textit{i.e.} $f_S(t_2,p) \propto \delta(p)$,
which leads to
\begin{align}
f_S(t_2,p) =
2\pi^2 N_S(t_2) \frac{\delta(k)}{k^2} \,,\qquad
k = a(t_2)\, p(t_2) = a(t) \,p(t)\,.
\label{eq:fSIni}
\end{align}
Because of the chosen mass hierarchy $M_1 \ll m_S \simeq M_{2,3}$, the
amount of $N_1$ neutrinos produced during tachyonic preheating 
as well as through $S$ decays up to $t_2$ is negligibly small.
Likewise, in absence of standard model radiation no gravitinos are produced
until $t_2$,
\begin{align}
N_{N_1}\left(t_2\right) = 0\,, \qquad N_{\widetilde{G}}\left(t_2\right) = 0\,.
\end{align}

The time dependence of the scale factor $a(t)$ is governed by the
Friedmann equation.
For a flat universe and constant equation of state
$\omega = \rho / p$ between some time $t_0$ and $t$, one has
\begin{align}
a(t) = a\left(t_0\right) \left[1+ \frac{3}{2}(1+\omega)
\left(\frac{8\pi}{3M_p^2}\, \rho_{\textrm{tot}}\left(t_0\right)\right)^{1/2}
\left(t - t_0\right)\right]^{\frac{2}{3(1 + \omega)}}\,.
\label{eq:scafac}
\end{align}
After preheating, until the decay of the $S$ bosons around
$t_S =t_2 + 1/\Gamma_S^0$, the system is mostly matter-dominated.
In view of the mass hierarchy $M_1/m_S \simeq \eta^2 \ll 1$
the subsequent $S$ decay into relativistic $N_1$ neutrinos, however,
entails a continuously changing equation of state.
We account for that behaviour by working with
two constant \textit{effective} equation of state coefficients $\omega_2$
and $\omega_S$ for times $t_2 < t \leq t_S$ and $ t > t_S$,
respectively.
For times $t_\textrm{PH} < t \leq t_2$, we take $\omega = 0$.
$\omega_2$ can be deduced from the decrease in the total
energy density until $t_S$.
Keeping only the leading contributions to $\rho_{\textrm{tot}}$
one has
\begin{align}
\frac{\rho_S(t_S,\omega_2) +
\rho_{N_1}(t_S,\omega_2)}{\rho_0 / a(t_2)^3}
= \left(\frac{a(t_2)}{a(t_S,\omega_2)}
\right)^{3\left(1+\omega_2\right)}\,.
\end{align}
With explicit expressions for $\rho_S$ and $\rho_{N_1}$ at hand
(cf. Sections~\ref{subsubsec:Higgs} and \ref{subsubsec:N1neutr})
this equation can be solved numerically for $\omega_2$.
Within the region in parameter space to which we restrict our study 
(cf. Section~\ref{sec:flavor}), $\omega_2$ typically turns out to be closer to $0$
than to $1/3$ indicating matter domination.
The $\omega_2$ mean value, standard deviation, minimum and maximum
values are found to be
\begin{align}
\omega_2 = \left(2.4 \pm 1.6\right) \times 10^{-2} \,,\quad
\omega_2^{\textrm{min}} \simeq 8.8 \times 10^{-3} \,,\quad
\omega_2^{\textrm{max}} \simeq 8.7 \times 10^{-2} \,.
\end{align}
At times $t > t_S$ most of the initial energy density has already been transferred
to $N_1$ neutrinos or subsequently to standard model radiation and we can safely use $\omega_S = 1/3$.

\subsection{Boltzmann equations}
Using the conventions introduced in Appendix~\ref{appendix:BE} we
now write down the Boltzmann equations relevant for our specific
scenario.
In Section~\ref{subsubsec:RelCOs} the Boltzmann equations for all
species but the gravitino ($S$, $N_1$, $B-L$, $R$) are given for
phase space distribution functions.
Subsequently, they are discussed one by one in Sections~\ref{subsubsec:Higgs}
through \ref{subsubsec:Rad}.
For the gravitino component we directly give the
integrated Boltzmann equation for the comoving number
density in Section \ref{subsubsec:Gravi}.

\subsubsection{Collision operators}
\label{subsubsec:RelCOs}
The dynamics of our system is dominated by three
types of particle interactions:
$S$ boson decays into pairs of $N_1$ neutrinos,
$N_1$ neutrino interactions with standard model lepton-Higgs pairs
$\ell H$ and $\bar{\ell} \bar{H}$ and supersymmetric QCD $2\rightarrow 2$
scatterings responsible for the production of gravitinos.\footnote{For
  notation convenience, we refer to $H_u$ as $H$ from now on.}

The Boltzmann equations for the $S$ bosons and the $N_1$ neutrinos
take the form
\begin{align}
\hat{\mathcal{L}} f_S = & \:  C_S(S \rightarrow N_1 N_1) \,,\\
\hat{\mathcal{L}} f_{N_1} = & \: 2\, C_{N_1} (S \rightarrow N_1 N_1)
+ C_{N_1}(N_1 \leftrightarrow \ell H, \bar{\ell}\bar{H}) \,,\label{eq:BEfN1}
\end{align}
where the factor of 2 in Eq.\,\eqref{eq:BEfN1} accounts for the fact
that two $N_1$ neutrinos are created per $S$ decay,
and $C_{N_1}(N_1 \leftrightarrow \ell H, \bar{\ell}\bar{H})$
encompasses $N_1$ decays into particles and antiparticles,
\begin{align}
C_{N_1}\left(N_1 \leftrightarrow \ell H, \bar{\ell}\bar{H}\right) =
C_{N_1}\left(N_1 \leftrightarrow \ell H\right) +
C_{N_1}\left(N_1 \leftrightarrow \bar{\ell}\bar{H}\right)\,.
\end{align}
Since we expect their effects to yield only minor corrections, we do
not include the rescatterings of $N_1$ neutrinos
into $S$ bosons as well as scatterings involving massive $Z^\prime$
bosons which arise in the course of $B-L$ breaking.
We have checked that the scatterings $N_1 \ell \leftrightarrow q \bar u$,
$N_1 \bar u \leftrightarrow \ell \bar q$ and $N_1 q\leftrightarrow
\ell u$ do not affect the final $B-L$ asymmetry.

The Boltzmann equation for the $B-L$ asymmetry is defined in terms of
the respective equations for lepton number $L$ and anti-lepton number $\bar{L}$
\begin{subequations}
\label{eq:BEelltot}
\begin{align}
\hat{\mathcal{L}} f_{B-L} &=  \hat{\mathcal{L}} \left(f_L - f_{\bar{L}}\right)\,,\\
\hat{\mathcal{L}} f_L &= C_L (\ell H \leftrightarrow N_1)
+ 2\, C_L^\textrm{red}(\ell H \leftrightarrow \bar{\ell}\bar{H}) \,,\label{eq:BEell}\\
\hat{\mathcal{L}} f_{\bar{L}} &=  C_{\bar{L}} (\bar{\ell}\bar{H} \leftrightarrow N_1)
+ 2\, C_{\bar{L}}^\textrm{red}(\bar{\ell}\bar{H} \leftrightarrow \ell H) \,.\label{eq:BEellbar}
\end{align}
\end{subequations}
The collision operators for decays and inverse decays are able to mimic
$\Delta L = 2$ scatterings of the type
$\ell H \rightleftarrows N_1 \rightleftarrows \bar{\ell}\bar{H}$
with on-shell $N_1$ neutrinos in the $s$-channel.
They, however, ignore off-shell scatterings even though
these will equally affect the final asymmetry.
This leads us to adding reduced collision operators
$C_L^\textrm{red}$ and $C_{\bar{L}}^\textrm{red}$ to Eqs.~\eqref{eq:BEell}
and \eqref{eq:BEellbar} that account for
the production and decay of off-shell neutrinos,
$\ell H \rightleftarrows N_1^* \rightleftarrows \bar{\ell}\bar{H}$.
In Ref.~\cite{bdp04} it has been shown that
the on- and off-shell contributions to the total CP asymmetry
in $\ell H\leftrightarrow \bar{\ell}\bar{H}$ scatterings cancel up to
$\mathcal{O}\left(\left(h_{i1}^\nu\right)^4\right)$.
Hence, one may equivalently say that the reduced collision operators
subtract scatterings with real intermediate states. %

\medskip
The temperature $T$ of the thermal bath can be determined as a function of
the scale factor from the covariant energy conservation
\begin{align}
\dot{\rho}_{\textrm{tot}} + 3H\left(\rho_{\textrm{tot}}
+ p_{\textrm{tot}}\right) = 0\,.
\label{eq:covengcon}
\end{align}
After inserting the explicit expressions for the total energy
and number densities this relation becomes
a non-linear first-order differential equation for $T$.
For simplicity, we assume that the energy transfer to the
thermal bath happens instantaneously, $\dot{\rho}_{\textrm{tot}} \approx 0$.
Consider a spatial volume $V$ in which $N_1$
neutrinos of average energy $\varepsilon_{N_1}$ decay
into lepton-Higgs pairs.
Per decay the energy density of the thermal bath
is then increased by $\varepsilon_{N_1}/V$ and a new
thermal equilibrium at a slightly higher
temperature is established right after the decay.
According to Eq.~\eqref{eq:rnRT} the latter entails an increase
in $n_R$,
\begin{align}
n_R \rightarrow n_R\left(1 + \frac{\varepsilon_{N_1}}{V\rho_R}\right)^{3/4}
\simeq n_R + \frac{3}{4}\frac{\varepsilon_{N_1}}{V\rho_R/n_R}
= n_R + \frac{r_R}{V}\,.
\end{align}
Therefore, producing two standard model particles adds
$r_R = 3\varepsilon_{N_1} / \left(4\rho_R / n_R \right)$
radiation quanta per unit volume $V$ to the thermal bath.
This leads us to an effective Boltzmann equation for the
number density of radiation
\begin{align}
\hat{\mathcal{L}} f_R = r_R \left(
C_\ell (\ell H \leftrightarrow N_1) +
C_{\bar{\ell}} (\bar{\ell}\bar{H} \leftrightarrow N_1)\right) \,,
\label{eq:BEfR}
\end{align}
where all CP violating contributions have been neglected.

Having formulated the Boltzmann equations for distribution functions we now calculate
the various collision operators explicitly and simplify the equations
as far as possible.

\subsubsection{Higgs bosons $S$}

\label{subsubsec:Higgs}
In terms of the $S$ decay rate
\begin{align}
\label{eq:gammaS0}
\Gamma_S^0 = \frac{1}{2 m_S} \int d\Pi\left(N_1,N_1\right)
\left(2\pi\right)^4 \delta^{(4)}\left(\textstyle \sum p_{\textrm{out}}
  - \textstyle \sum p_{\textrm{in}}\right)
\left|\mathcal{M}\left(S \rightarrow N_1 N_1\right)\right|^2\,,
\end{align}
the collision operator for $S$ decay is given by
\begin{align}
\hat{\mathcal{L}} f_S = & \:  C_S(S \rightarrow N_1 N_1)
= -\frac{m_S}{E_S} \,\Gamma_S^0 f_S\,.
\end{align}
This is a linear homogeneous ordinary differential equation
which has a unique solution for each initial value.
Given the initial distribution function $f_S(t_2,p)$ in
Eq.~\eqref{eq:fSIni} we find
\begin{align}
f_S(t,p) &=  f_S(t_2,p)
\exp\left[- m_S \Gamma_S^0 \int\limits_{t_2}^t dt' E_S^{-1}(t')\right] \,,\\
E_S(t') &= \sqrt{\left(a/a'\right)^2 p^2 + m_S^2}\,,\quad
a =  a\left(t\right)\,,\quad
a' = a(t')\,. \label{eq:ESt}
\end{align}
Thanks to the momentum delta function in $f_S(t_2,p)$ the time
integration becomes trivial,
\begin{align}
f_S(t,p) = f_S(t_2,p) \,e^{-\Gamma_S^0(t-t_2)} =
2\pi^2 N_S(t_2)\frac{\delta(k)}{k^2}\,
e^{-\Gamma_S^0(t-t_2)}  \,,\qquad
k = a(t) \,p\,.\label{eq:fSres}
\end{align}
Hence, the comoving number density $N_S$ 
simply falls off exponentially,
\begin{align}
N_S (t) = N_S (t_2) e^{-\Gamma_S^0(t-t_2)}\,.
\end{align}

\subsubsection{Heavy Majorana neutrinos $N_1$}

\label{subsubsec:N1neutr}

The Boltzmann equation for $N_1$ neutrinos also involves a collision
operator for $S$ decay. Using the tree-level amplitude squared
\begin{align}
\left|\mathcal{M}(S\rightarrow N_1 N_1)\right|^2 =
2 (h_1^n)^2 m_S^2 \left[1-\left(2M_1/m_S\right)^2\right]\,,
\end{align}
the general collision operator in Eq.~\eqref{eq:Cpsi} takes the form
\begin{align}
C_{N_1}(S\rightarrow N_1 N_1) =&\:
\frac{(h_1^n)^2 m_S^2}{2 E_{N_1}} \left[1-\left(2M_1/m_S\right)^2\right]
\int d\Pi(N_1|N_1;S)(2\pi)^4
\delta^{(4)}\, f_S\,.
\end{align}
With the explicit expression for $f_S$ in Eq.\,\eqref{eq:fSres} the phase
space integration becomes
\begin{align}
\int d\Pi\left(N_1|N_1;S\right)\left(2\pi\right)^4
\delta^{(4)} f_S =
\frac{1}{a^3} \frac{\pi N_S}{8 m_S E_{N_1}} \delta\left(E_{N_1}-m_S/2\right)\,,
\end{align}
where a symmetry factor $1/2$ results from the fact that two $N_1$
neutrinos are involved in the decay process.
Employing the result Eq.\,\eqref{eq:GammaS0} for the $S$ decay width we obtain
for the collision operator
\begin{align}
C_{N_1}\left(S\rightarrow N_1 N_1\right) =
\frac{1}{a^3}\frac{\pi^2 N_S \Gamma_S^0}{E_{N_1}^2}
\left[1-\left(2M_1/m_S\right)^2\right]^{-1/2}
\delta\left(E_{N_1}-m_S/2\right)\,.
\label{eq:CN1S}
\end{align}

The collision operator for $N_1$ decay into standard model particles
has the familiar form
\begin{align}
C_{N_1}\left(N_1 \leftrightarrow \ell H, \bar{\ell}\bar{H}\right) =
- \frac{M_1}{E_{N_1}}\Gamma_{N_1}^0 \left(f_{N_1} - f_{N_1}^\equi\right)\,.
\label{eq:CN1T}
\end{align}
In total, the Boltzmann equation for $N_1$ neutrinos encompasses two
production and one decay term. On the one hand, the collision 
operator in Eq.\,\eqref{eq:CN1S} and the term proportional to
$f_{N_1}^\equi$ in Eq.\,\eqref{eq:CN1T} represent $N_1$ production
from $S$ decay and from inverse decays in the thermal
bath, respectively. On the other hand, the term proportional to
$f_{N_1}$ in Eq.\,\eqref{eq:CN1T} accounts for $N_1$ decays.
It is convenient to decompose the $N_1$ population into two
independently evolving components:
Nonthermal neutrinos $N_1^S$ stemming from $S$ decays
and thermal neutrinos $N_1^T$ originating from the thermal bath. 
The respective Boltzmann equations are then given as
\begin{subequations}
\begin{align}
\hat{\mathcal{L}} f_{N_1}^S = & \:
- \frac{M_1}{E_{N_1}}\Gamma_{N_1}^0 f_{N_1}^S
+ 2\, \frac{1}{a^3}\frac{\pi^2 N_S \Gamma_S^0}{E_{N_1}^2}
\left[1-\left(2M_1/m_S\right)^2\right]^{-1/2}
\delta(E_{N_1}-m_S/2)\,, \\
\hat{\mathcal{L}} f_{N_1}^T = & \:
- \frac{M_1}{E_{N_1}}\Gamma_{N_1}^0 f_{N_1}^T
+ \frac{M_1}{E_{N_1}}\Gamma_{N_1}^0 f_{N_1}^\equi \,.\label{eq:BEfN1T}
\end{align}
\end{subequations}
The sum of the nonthermal and thermal distribution functions yields the total
$N_1$ distribution function, $f_{N_1} = f_{N_1}^S + f_{N_1}^T$.

The Boltzmann equation for the nonthermal neutrinos $N_1^S$ 
can be solved exactly. Starting from zero initial abundance we find
\begin{align}
f_{N_1}^S\left(t,p\right) = & \: 2\pi^2 \Gamma_S^0
\left[1-\left(2M_1/m_S\right)^2\right]^{-1/2}
\int\limits_{t_2}^t dt'
\Bigg[\delta\left(E_{N_1}\left(t'\right)-m_S/2\right) \nonumber \\
& \times   \frac{N_S(t')}{a'^3 E_{N_1}^2(t')}
\exp\left(-M_1 \Gamma_{N_1}^0 \int\limits_{t'}^t dt''
  E_{N_1}^{-1}(t'')\right) \Bigg] \,,
\label{eq:fN1S}
\end{align}
where $E_{N_1}(t')$ is defined analogously to $E_S(t')$
in Eq.~\eqref{eq:ESt}. The energies $E_{N_1}$ therefore redshift as 
\begin{align}
E_{N_1}(t) = E_{N_1}(t') \frac{a'}{a}
\left[1 + \left(\left(\frac{a}{a'}\right)^2 - 1\right)
\left(\frac{M_1}{E_{N_1}(t')}\right)^2\right]^{1/2} \,.
\end{align}
If we evaluate this relation with $E_{N_1}(t') = m_S/2$
we obtain the redshifted energy at time $t$ of a neutrino $N_1$
that has been produced in $S$ decay at time $t'$. Let us denote
this quantity by $\mathcal{E}_{N_1}\left(t',t\right)$,
\begin{align}
\mathcal{E}_{N_1}(t',t) = \frac{m_S}{2} \frac{a'}{a}
\left[1 + \left(\left(\frac{a}{a'}\right)^2 - 1\right)
\left(\frac{2 M_1}{m_S}\right)^2\right]^{1/2}\,.
\end{align}
The energy delta function in the integrand of Eq.~\eqref{eq:fN1S}
thus turns $E_{N_1}(t'')$ into $\mathcal{E}_{N_1}(t',t'')$.
Meanwhile, it can be rewritten as a function of $E_{N_1}(t)$,
the energy at time $t$,
\begin{align}
\delta(E_{N_1}(t')-m_S/2) =
\left(\frac{a'}{a}\right)^2 \frac{m_S/2}{\mathcal{E}_{N_1}(t',t)}
\delta\left(E_{N_1}\left(t\right)-\mathcal{E}_{N_1}(t',t)\right)\,. 
\end{align}
The final result for $f_{N_1}^S$ then reads
\begin{align}
f_{N_1}^S(t,p) = & \: \frac{1}{a^3}\,
2\pi^2 \,\Gamma_S^0
\left[\left(m_S/2\right)^2-M_1^2\right]^{-1/2} \int\limits_{t_2}^t dt' \Bigg[
\frac{a}{a'}\,
\delta(E_{N_1}\left(t\right)-\mathcal{E}_{N_1}(t',t))
\nonumber\\
& \times   \frac{N_S(t')}{\mathcal{E}_{N_1}(t',t)}
\exp\left(-M_1 \Gamma_{N_1}^0 \int\limits_{t'}^t dt''
  \mathcal{E}_{N_1}^{-1}(t',t'')\right)
\Bigg]\,.
\end{align}
By integrating over the $N_1^S$ phase space,  we obtain the
number density $n_{N_1}^S$ of the nonthermal neutrinos
\begin{subequations}
\begin{align}
n_{N_1}^S = & \: g_{N_1} \int \frac{d^3p}{\left(2\pi\right)^3} \,f_{N_1}^S
= \frac{2\Gamma_S^0}{a^3}
\int\limits_{t_2}^t dt' \Bigg[
N_S(t') \exp\left(-M_1 \Gamma_{N_1}^0 \int\limits_{t'}^t dt''
\mathcal{E}_{N_1}^{-1}(t',t'')\right)\Bigg]\\
= & \: \int\limits_{t_2}^t dt' \delta n_{N_1}^S(t',t)\,.
\label{eq:deltanN1S}
\end{align}
\label{eq:nN1S}
\end{subequations}
The corresponding results for the energy and interaction densities
$\rho_{N_1}^S$ and $\gamma_{N_1}^S = \gamma(N_1^S \rightarrow \ell H,
  \bar{\ell}\bar{H})$ can conveniently be expressed using $\delta
n_{N_1}^S(t,t')$ introduced in Eq.\,\eqref{eq:deltanN1S},
\begin{align}
\rho_{N_1}^S &= \int\limits_{t_2}^t dt' \mathcal{E}_{N_1}(t',t)
\delta n_{N_1}^S(t',t) \,,\label{eq:rnN1S}\\
\gamma_{N_1}^S &=  \int\limits_{t_2}^t dt'
\frac{M_1}{\mathcal{E}_{N_1}(t',t)}\,
\Gamma_{N_1}^0 \delta n_{N_1}^S(t',t)
= n_{N_1}^S \Gamma_{N_1}^S\,,
\end{align}
where $\Gamma_{N_1}^S$ denotes the $N_1$ decay width weighted with
the average inverse time dilatation factor for nonthermal neutrinos
\begin{align}
\Gamma_{N_1}^S = \left<\frac{M_1}{E_{N_1}}\right>_S \Gamma_{N_1}^0 =
\frac{1}{n_{N_1}^S} \int\limits_{t_2}^t dt' \frac{M_1}{\mathcal{E}_{N_1}(t',t)}\,
\delta n_{N_1}^S(t',t)  \,\Gamma_{N_1}^0 \,.
\label{eq:InvDila}
\end{align}

The exact phase space distribution function $f_{N_1}^T$ for thermal
neutrinos $N_1^T$ is given as the unique solution of
Eq.~\eqref{eq:BEfN1T} for the initial distribution
$f_{N_1}^T\left(t_2,p\right) = 0$,
\begin{align}
f_{N_1}^T(t,p) = \int\limits_{t_2}^t dt'
\exp\left(-M_1 \Gamma_{N_1}^0 \int\limits_{t'}^t dt'' E_{N_1}^{-1}(t'')\right)
\frac{M_1}{E_{N_1}(t')}\, \Gamma_{N_1}^0 f_{N_1}^\equi(t',p)\,.
\label{eq:fN1Tex}
\end{align}
As the thermal neutrinos are produced within a broad range of energies, it
cannot be integrated over phase space as simply as in the nonthermal case.
However, since the thermal neutrinos inherit their momentum distribution
from the thermal bath it is reasonable to assume that they are approximately
in kinetic equilibrium,
\begin{align}
f_{N_1}^T(t,p) \approx \frac{N_{N_1}^T}{N_{N_1}^\equi}\,
f_{N_1}^\equi(t,p)\,, \qquad
f_{N_1}^\equi(t,p) = e^{-E_{N_1}/T}\,.
\label{eq:fN1Tapprox}
\end{align}
This approximation holds if the quotient $f_{N_1}^T/f_{N_1}^\equi$,
with $f_{N_1}^T$ taken from Eq.\,\eqref{eq:fN1Tex}, is independent
of the neutrino momentum $p$.
In Appendix~\ref{subsec:PSDFN1T} we will demonstrate numerically in the
context of a specific parameter example that this can usually
be assumed to be the case in our scenario.
Under the assumption of kinetic equilibrium the comoving number
density $N_{N_1}^T$ is the unique solution of the integrated
Boltzmann equation
\begin{align}
a H \frac{d}{da} N_{N_1}^T = -\left(N_{N_1}^T - N_{N_1}^\equi\right)
\Gamma_{N_1}^T\,.
\label{eq:BEN1T}
\end{align}
Here, $\Gamma_{N_1}^T$ stands for the $N_1$ decay width weighted with
the average inverse time dilatation factor for thermal neutrinos
\begin{align}
\Gamma_{N_1}^T = \left<\frac{M_1}{E_{N_1}}\right>_T \Gamma_{N_1}^0 =
\frac{K_1(z)}{K_2(z)}\,\Gamma_{N_1}^0 \,,
\end{align}
where $z = M_1/T$, and $K_{1,2}(z)$ are modified Bessel functions
of the second kind.
Note that
\begin{subequations}
\begin{align}
\gamma_{N_1}^T &=  \gamma\left(N_1^T \rightarrow \ell
  H,\bar{\ell}\bar{H}\right) = n_{N_1}^T\Gamma_{N_1}^T\,,\\
\gamma_{N_1}^\equi & = \gamma^\equi\left(N_1 \rightarrow \ell
  H,\bar{\ell}\bar{H}\right) = n_{N_1}^\equi\Gamma_{N_1}^T\,.
\end{align}
\end{subequations}

\subsubsection{$B-L$ asymmetry}

The collision operators for decays and inverse decays present
themselves as
\begin{subequations}
\begin{align}
C_L \left(\ell H \leftrightarrow N_1 \right) \sim & \:
f_{N_1} \left|\mathcal{M}\left(N_1 \rightarrow \ell H\right)\right|^2 -
f_H f_\ell \left|\mathcal{M}\left(\ell H \rightarrow N_1\right)\right|^2 \,,\\
C_{\bar{L}} \left(\bar{\ell} \bar{H} \leftrightarrow N_1 \right) \sim & \:
f_{N_1} \left|\mathcal{M}\left(N_1 \rightarrow \bar{\ell} \bar{H}\right)\right|^2 -
f_{\bar{H}} f_{\bar{\ell}} \left|\mathcal{M}\left(\bar{\ell} \bar{H} \rightarrow N_1\right)\right|^2\,.
\end{align}
\end{subequations}
Using the definition of the CP parameter $\epsilon_1$ and
CPT invariance, the various partial amplitudes squared are
related to the total amplitude squared as follows
\begin{subequations}
\begin{align}
\left|\mathcal{M}\left(N_1 \rightarrow \ell H\right)\right|^2 = & \:
\left|\mathcal{M}\left(\bar{\ell} \bar{H} \rightarrow N_1\right)\right|^2 =
\frac{1}{2}\left(1 + \epsilon_1\right)
\left|\mathcal{M}_{N_1}\right|^2 \,,\\
\left|\mathcal{M}\left(N_1 \rightarrow \bar{\ell} \bar{H}\right)\right|^2 = & \:
\left|\mathcal{M}\left(\ell H \rightarrow N_1\right)\right|^2 =
\frac{1}{2}\left(1 - \epsilon_1\right)
\left|\mathcal{M}_{N_1}\right|^2\,,
\end{align}
\end{subequations}
where, at tree-level, $\left|\mathcal{M}_{N_1}\right|^2$ is given as
\begin{align}
\left|\mathcal{M}_{N_1}\right|^2 =
\left|\mathcal{M}\left(N_1 \rightarrow \ell H, \bar{\ell}\bar{H}\right)\right|^2 =
4 \left(h^{\nu \dagger} h^\nu\right)_{11} M_1^2\,.
\end{align}

The reduced collision operators in Eq.\,\eqref{eq:BEelltot} account
for the production of off-shell neutrinos $N_1^*$ which subsequently
decay into the CP conjugate of the lepton-Higgs pair from which they
were produced. 
Working up to leading order in $\epsilon_1$ we may take
the decays $N_1^* \rightarrow \ell H, \bar{\ell}\bar{H}$ to
equally branch into particles and antiparticles
\begin{align}
C_{L,\bar{L}}^\textrm{red}\left(\ell H \leftrightarrow \bar{\ell}\bar{H}\right) \sim
\pm\left[f_{\bar{\ell}} f_{\bar{H}} \cdot \frac{1}{2}
\left|\mathcal{M}\left(\bar{\ell}\bar{H} \rightarrow N_1^*\right)\right|^2 -
f_\ell f_H \cdot \frac{1}{2}
\left|\mathcal{M}\left(\ell H \rightarrow N_1^*\right)\right|^2
\right]\,.
\end{align}
For $M_1 \ll 10^{14}\,\textrm{GeV}$ the CP preserving parts
of the off-shell scatterings are negligibly small \cite{HahnWoernle:2009qn}.
We thus discard them keeping only the CP violating
contributions to the reduced collision operators.
Imposing that the total CP asymmetry of lepton-Higgs scatterings
be zero up to $\mathcal{O}\left(\left(h_{i1}^\nu\right)^4\right)$,
we deduce
\begin{align}
\left|\mathcal{M}\left(\ell H \rightarrow N_1^*\right)\right|^2 =
- \frac{1}{2} \left(-\epsilon_1\right) \left|\mathcal{M}_{N_1}\right|^2\,,
\qquad
\left|\mathcal{M}\left(\bar{\ell}\bar{H} \rightarrow N_1^*\right)\right|^2 =
- \frac{1}{2} \left(+\epsilon_1\right) \left|\mathcal{M}_{N_1}\right|^2\,.
\end{align}

The above results allow us to write all collision operators
as integrals over the total amplitude squared
$\left|\mathcal{M}_{N_1}\right|^2$.
Assuming kinetic equilibrium for leptons and antileptons
as well as thermal equilibrium for all other standard model particles, we obtain
\begin{align}
\hat{\mathcal{L}} f_{B-L} = \frac{1}{2 g_\ell p} \int d\Pi\left(\ell|H;N_1\right)
\left(2 \pi\right)^4 \delta^{(4)}
\frac{1}{2}\left|\mathcal{M}_{N_1}\right|^2
\left[2\epsilon_1\left(f_{N_1} - f_{N_1}^\equi\right) 
- \frac{N_{B-L}}{N_\ell^\equi} f_{N_1}^\equi\right]\,.
\end{align}
As for the $N_1$ Boltzmann equation, we split the $N_1$ distribution
function into its thermal and nonthermal parts.
After integrating over phase space we retrieve the interaction densities
$\gamma_{N_1}^S$, $\gamma_{N_1}^T$ and $\gamma_{N_1}^\equi$
(cf. Eq.\,\eqref{eq:gammadef}), 
\begin{align}
a H \frac{d}{da} N_{B-L} = a^3 \left[\epsilon_1 \left(
\gamma_{N_1}^S + \gamma_{N_1}^T - \gamma_{N_1}^\equi\right) -
\frac{N_{B-L}}{2 N_\ell^\equi} \,\gamma_{N_1}^\equi\right]\,.
\end{align}
In terms of comoving number densities and averaged decay rates
this Boltzmann equation then reads
\begin{align}
a H \frac{d}{da} N_{B-L} = \epsilon_1 N_{N_1}^S \Gamma_{N_1}^S
+ \epsilon_1 \left(N_{N_1}^T - N_{N_1}^\equi\right)\,\Gamma_{N_1}^T -
\frac{N_{N_1}^\equi}{2 N_\ell^\equi} \,\Gamma_{N_1}^T N_{B-L}\,.
\label{eq:BEBL}
\end{align}
Similarly to the $N_1$ abundance we also split the $B-L$ asymmetry
into two components:
A nonthermal asymmetry $N_{B-L}^S$ produced in $N_1^S$
decays and a thermal asymmetry $N_{B-L}^T$ generated from the
thermal bath,
\begin{subequations}
\begin{align}
a H \frac{d}{da} N_{B-L}^S = & \: \epsilon_1 N_{N_1}^S \Gamma_{N_1}^S
- \frac{N_{N_1}^\equi}{2 N_\ell^\equi} \,\Gamma_{N_1}^T N_{B-L}^S\,,\label{eq:BEBLS}\\
a H \frac{d}{da} N_{B-L}^T = & \:
\epsilon_1 \left(N_{N_1}^T - N_{N_1}^\equi\right)\Gamma_{N_1}^T -
\frac{N_{N_1}^\equi}{2 N_\ell^\equi} \,\Gamma_{N_1}^T N_{B-L}^T\,.\label{eq:BEBLT}
\end{align}
\end{subequations}
A comparison of the corresponding final baryon asymmetries
$\eta_B^S$ and $\eta_B^T$ will allow us to identify
the relative importance of nonthermal and thermal leptogenesis
in different regions of the parameter space (cf. Section \ref{sec:results}).

\subsubsection{Radiation $R$}

\label{subsubsec:Rad}

In order to obtain an effective Boltzmann equation for
the number density of radiation quanta we add up the
contributions coming from decays into standard model particles and antiparticles.
Neglecting any CP violating effects Eq.\,\eqref{eq:BEfR} can be
written as
\begin{align}
\hat{\mathcal{L}} f_R = r_R \,
\frac{1}{2 g_\ell p} \int d\Pi\left(\ell|H;N_1\right)
\left(2 \pi\right)^4 \delta^{(4)}
\frac{1}{2}\left|\mathcal{M}_{N_1}\right|^2
\left[2 f_{N_1} - 2 f_{N_1}^\equi\right]\,.
\end{align}
Splitting $f_{N_1}$ into its thermal and nonthermal
parts and integrating over phase space, one gets
\begin{align}
a H \frac{d}{da} N_R = r_R^S\, N_{N_1}^S \Gamma_{N_1}^S
+ r_R^T\,\left(N_{N_1}^T - N_{N_1}^\equi\right)\Gamma_{N_1}^T\,.
\label{eq:BENR}
\end{align}
Since the two different sorts of $N_1$ neutrinos possess different
average energies, two independent factors $r_R^S$ and $r_R^T$ have
been introduced in order to keep track of the radiation quanta
produced in the decays of nonthermal and thermal neutrinos
respectively,
\begin{align}
r_R^S = \frac{3 \varepsilon_{N_1}^S}{4 \varepsilon_R}\,, \qquad
r_R^T = \frac{3 \varepsilon_{N_1}^T}{4 \varepsilon_R}\,.
\label{eq:rRSTDef}
\end{align}
The average energies per (non-)thermal neutrino
as well as the average energy per radiation quantum are
obtained from the respective ratios of energy and number
densities (cf. Eqs.\,\eqref{eq:rnRT}, \eqref{eq:nN1S} and
\eqref{eq:rnN1S}),
\begin{align}
\varepsilon_R = \rho_R / n_R\,, \qquad
\varepsilon_{N_1}^S = \rho_{N_1}^S / n_{N_1}^S\,, \qquad
\varepsilon_{N_1}^T = \rho_{N_1}^T / n_{N_1}^T = 3T + \frac{K_1(z)}{K_2(z)} M_1\,.
\label{eq:eRN1ST}
\end{align}
$r_R^S$ and $r_R^T$ clearly depend on the temperature $T$
which, in turn, is deduced from the radiation number density $N_R$
according to Eq.\,\eqref{eq:TNR}.

\subsubsection{Gravitinos $\widetilde{G}$}

\label{subsubsec:Gravi}

Gravitinos are produced through scattering processes in the
thermal bath.
The evolution of their comoving number density is governed by
the Boltzmann equation
\begin{align}
a H \frac{d}{da} N_{\widetilde{G}} = a^3 \,\gamma_{\widetilde{G}}(T)\,.
\label{eq:BEGt}
\end{align}
The dominant contribution to $\gamma_{\widetilde{G}}$ comes from
QCD scatterings.
For supersymmetric QCD and up to leading order in the
strong gauge coupling $g_s$, one has \cite{bbb00}
\begin{align}
\gamma_{\widetilde{G}}(T) = \left(1 + \frac{m_{\tilde{g}}^2(T)}{3 m_{\widetilde{G}}^2}\right)
\frac{54 \zeta(3) g_s^2(T)}{\pi^2 M_p^2} \,T^6 \left[\ln\left(\frac{T^2}{m_g^2(T)}\right)
+ 0.8846\right]\,.
\label{eq:gammaGT}
\end{align}
Here, $m_{\tilde{g}}$ denotes the energy scale-dependent
gluino mass and $m_g$ is the gluon plasma mass,
\begin{align}
m_{\tilde{g}}(T) = \frac{g_s^2(T)}{g_s^2\left(\mu_0\right)} m_{\tilde{g}}\left(\mu_0\right)\,,
\qquad
m_g(T) = \sqrt{3/2}\,g_s(T) \,T\,,
\end{align}
where we choose the $Z$ boson mass $M_Z$ as reference scale $\mu_0$.
The scale dependence of $g_s$ is dictated by the corresponding
MSSM renormalization group equation
\begin{align}
g_s(\mu) = g_s\left(\mu_0\right) \left[1 + \frac{3}{8\pi^2} \,g_s\left(\mu_0\right)^2
\ln\left(\mu/\mu_0\right)\right]^{-1/2}\,,
\end{align}
with $\mu$ being the typical energy scale during reheating.
It can be estimated by the average energy per relativistic particle
in the bath:
$\mu \simeq \varepsilon_R \simeq 3 T$.
For instance, at temperatures $T = 10^8, 10^{10}, 10^{12}\,\textrm{GeV}$ the
strong gauge coupling takes on values $g_s = 0.90, 0.84, 0.80$.
The gravitino mass $m_{\widetilde{G}}$ and the low-scale gluino
mass $m_{\tilde{g}}\left(\mu_0\right)$ remain as free parameters.
\section{An illustrative example}
\label{sec:example}
\begin{table}
\begin{center}
\begin{tabular}{ccccccc}
Point & Label & $v_{B-L}$ [GeV] & $M_1$ [GeV] & $\widetilde{m}_1$ [eV] & $m_{\widetilde{G}}$ [GeV] & $m_{\tilde{g}}$ [GeV] \\
\hline\hline
Section~\ref{sec:example} & Red circle   & $5.8 \times 10^{13}$ & $1.4 \times 10^{10}$ & $3 \times 10^{-3}$ &  $100$ & $800$ \\
Ref.~\cite{bsv10}         & White circle & $3.0 \times 10^{12}$ & $1.0 \times 10^{10}$ & $1 \times 10^{-3}$ &  $100$ & $800$ \\
\hline\hline
\end{tabular}
\caption{Values of the input parameters chosen for the discussion
in Section~\ref{sec:example} and Ref.~\cite{bsv10}, respectively.
In Figs.~\ref{fig:treheat}, \ref{fig:BLasym}, \ref{fig:mGboundsM1},
\ref{fig:mGboundsT}, \ref{fig:gammant}, \ref{fig:mGrecon}
and \ref{fig:mGreconT} in Section~\ref{sec:results} and
Appendices~\ref{appendix:TR} and \ref{appendix:RC} the positions
of both parameter points in parameter space are marked with different
labels as indicated.}
\label{tab:parapoints}
\end{center}
\end{table}
In the previous section we have seen how the decay of the false
vacuum of unbroken $B-L$ symmetry generates -- via production and decay
of heavy neutrinos -- entropy, baryon asymmetry and dark matter.
We have numerically solved this inital-value problem by means of
Boltzmann equations, with the initial conditions
described in Section~\ref{subsec:IniCon}.
Before we turn to a detailed  discussion of the parameter space
we first describe, as an example, one solution
for a representative choice of parameters.
A similar study, albeit not as detailed, has already been
performed in Ref. \cite{bsv10}.
The values for the input parameters chosen in this section as well
as in Ref.~\cite{bsv10} are outlined in Tab.~\ref{tab:parapoints}.
In the present case, the selected values for $v_{B-L}$ and $M_1$
correspond to Froggatt-Nielsen charges of $a = 1/2$ and $d \simeq 1.5$.
Note that we have adjusted $M_1$ such that we obtain the right
gravitino abundance for dark matter.
The input parameters in Tab.~\ref{tab:parapoints}
directly determine a couple of further important parameters
\begin{align}
m_S,\, M_{2,3} \simeq 4.1 \times 10^{12}\,\textrm{GeV}\,, \:
\lambda \simeq 5.0 \times 10^{-3}\,, \:
\epsilon_1 \simeq 1.4 \times 10^{-6}\,, \:
\epsilon_{2,3} \simeq  -4.1 \times 10^{-4}\,.
\end{align}
We have chosen opposite signs for the CP asymmetries $\epsilon_1$ and
$\epsilon_{2,3}$, so that one can easily distinguish
their respective contributions to the final $B-L$ asymmetry.
\begin{figure}
\begin{center}
\includegraphics[width=12cm]{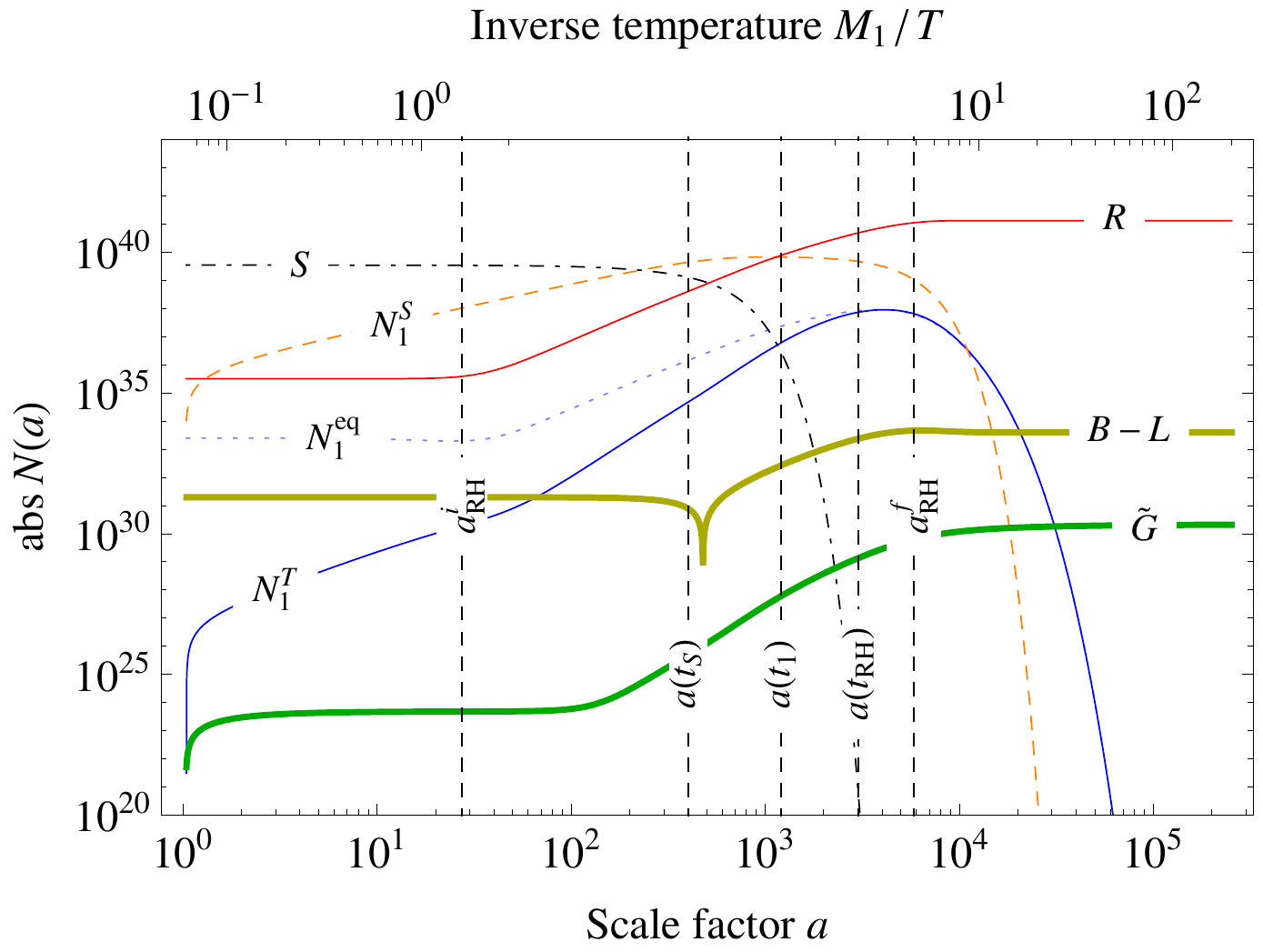}
\label{fig:numden}
\vspace{5mm}
\includegraphics[width=12cm]{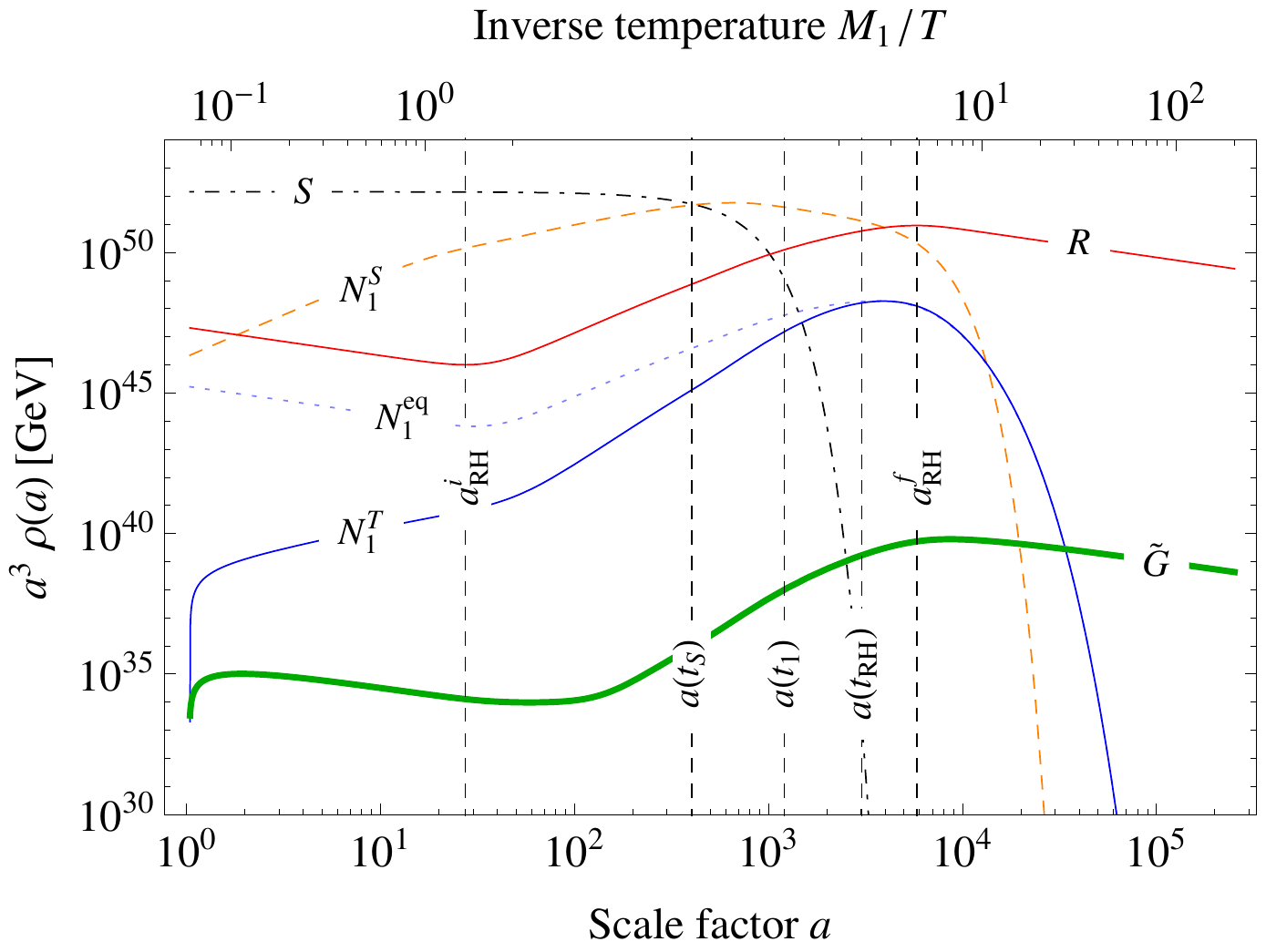}
\caption{Comoving number \textbf{(upper panel)} and energy \textbf{(lower panel)}
densities for $S$ bosons, $N_1$ neutrinos produced in $S$ decays ($N_1^S$),
thermally produced $N_1$ neutrinos ($N_1^T$),
$N_1$ neutrinos in thermal equilibrium ($N_1^{\eq}$, for comparison),
radiation ($R$), $B-L$ charge, and gravitinos ($\widetilde{G}$) as functions of
the scale factor $a$.
The corresponding values of the model parameters are
given in Tab.~\ref{tab:parapoints}.
\label{fig:engden}}
\end{center}
\end{figure}

\subsection{Comoving number and energy densities}

The evolution with the scale factor $a$ of the comoving number
densities and the various components of the energy density are
presented in Fig.~\ref{fig:engden}.
Both plots start at the time of $N_{2,3}$ decay, $a\left(t_2\right) \simeq 1.04$,
and end at a final scale factor of $a_f \simeq 2.53 \times 10^5$.
The values of the scale factor corresponding to
$S$ and $N_1$ decays as well as to reheating
(cf. Appendix~\ref{appendix:TR}) are also indicated
\begin{align}
a\left(t_S = t_2 + 1/\Gamma_S^0 \right) \simeq 400\,, \quad
a\left(t_1 = t_S + 1/\Gamma_{N_1}^0\right) \simeq 1200\,, \quad
a\left(t_{RH}\right) &\simeq 3000\,.
\end{align}

Tachyonic preheating results in an initial state
at $t = t_2$ that mainly consists of $S$ bosons and
to a smaller degree standard model radiation that stems from the decay of the
$N_{2,3}$ neutrinos, $\rho_R\left(t_2\right) / \rho_S \left(t_2\right) \sim 10^{-5}$,
and which inherits an initial $B-L$ asymmetry
equivalent to $\eta_B \simeq - 3.9 \times 10^{-5}$.
Around $t = t_S$ the $S$ bosons decay into relativistic
and nonthermal $N_1$ neutrinos.
Their subsequent decay into standard model
particles then washes out the initial (negative) asymmetry, builds up
a new (positive) asymmetry and leads to the production of the main
part of the radiation.
The energy transfer to the thermal bath, \textit{i.e.} the process of reheating,
takes place between $a_{RH}^i \simeq 27$ and $a_{RH}^f \simeq 5800$.
At these values of the scale factor the derivative of the comoving radiation energy
density $a^3 \rho_R$ vanishes.
Meanwhile, thermal neutrinos and gravitinos are continuously
produced in the thermal bath.
As both species inherit their momentum distributions from the bath,
they are always in approximate kinetic equilibrium
(cf. Appendix~\ref{subsec:PSDFN1T}). 
From $a \simeq 4100$ onwards the number density of the thermal $N_1$ neutrinos
exceeds the thermal equilibrium abundance.
At $a \sim 10^5$ the $B-L$ asymmetry and the gravitino abundance have
reached at their final values.

\subsection{Baryon asymmetry}

\label{subsec:Asymm}

The present value of the baryon asymmetry as well as its nonthermal
and thermal contributions are obtained from
\begin{align}
\eta_B = \frac{n_B^0}{n_{\gamma}^0} = \eta_B^S + \eta_B^T \,,\quad
\eta_B^{S,T} = c_{\mathrm{sph}} \frac{g_{\star,s}^0}{g_{\star,s}}
\left.\frac{N_{B-L}^{S,T}}{N_{\gamma}}\right|_{a_f}\,.
\label{eq:etaBdef}
\end{align}
In the supersymmetric standard model the sphaleron conversion factor is
$c_{\mathrm{sph}} = 8/23$, the effective number of degrees of freedom at
high and low temperatures is $g_{\star,s} = 915/4$ and $g_{\star,s}^0 = 43/11$,
respectively, and the number density of photons is
$N_{\gamma} = a^3 g_{\gamma}\zeta{(3)}/\pi^2 T^3$.
For our choice of parameters we obtain the asymmetries
\begin{align}\label{result}
\eta_B \simeq 1.9 \times 10^{-8}\,, \qquad \eta_B^S \simeq 1.9 \times
10^{-8}\,, \qquad \eta_B^T \simeq 2.8 \times 10^{-10}\,.
\end{align}
The calculated baryon asymmetry is larger than the observed one,
$\eta_B^{\mathrm{obs}} \simeq 6.2 \times 10^{-10}$ \cite{wmap10},
by about a factor 30. This is consistent since $\epsilon_1$ is an
estimate for the maximal CP asymmetry.
We find that $\eta_B$ is dominated by the nonthermal contribution
due to $N_1^S$ decays, $\eta_B^S \simeq \eta_B$.
The contribution from thermal neutrinos, even though it reaches
the right order of magnitude, slightly falls short of the observed
value.

Let us emphasize that given the choice of parameters in
Tab.~\ref{tab:parapoints} standard thermal
leptogenesis, with a given thermal bath, is able to produce the right amount
of baryon asymmetry.
Using a final efficiency factor of $\kappa_f (\widetilde{m}_1) \simeq 0.1$,
one obtains (cf. \cite{bdp04})
\begin{align}
\eta_B^{\mathrm{th}} =
\frac{3}{4} \frac{g_{*,s}^0}{g_{*,s}} \,c_{\mathrm{sph}} \,\epsilon_1
\,\kappa_f (\mt) \simeq 6\times 10^{-10}\,.
\end{align}
In the case under study $\eta_B^T$ turns out to be smaller than
$\eta_B^{\mathrm{th}}$ roughly by a factor of 2 because the entropy
production during $N_1^S$ decay enhances the washout rate
due to inverse $N_1$ decays.

\begin{figure}
\begin{center}
\includegraphics[width=12cm]{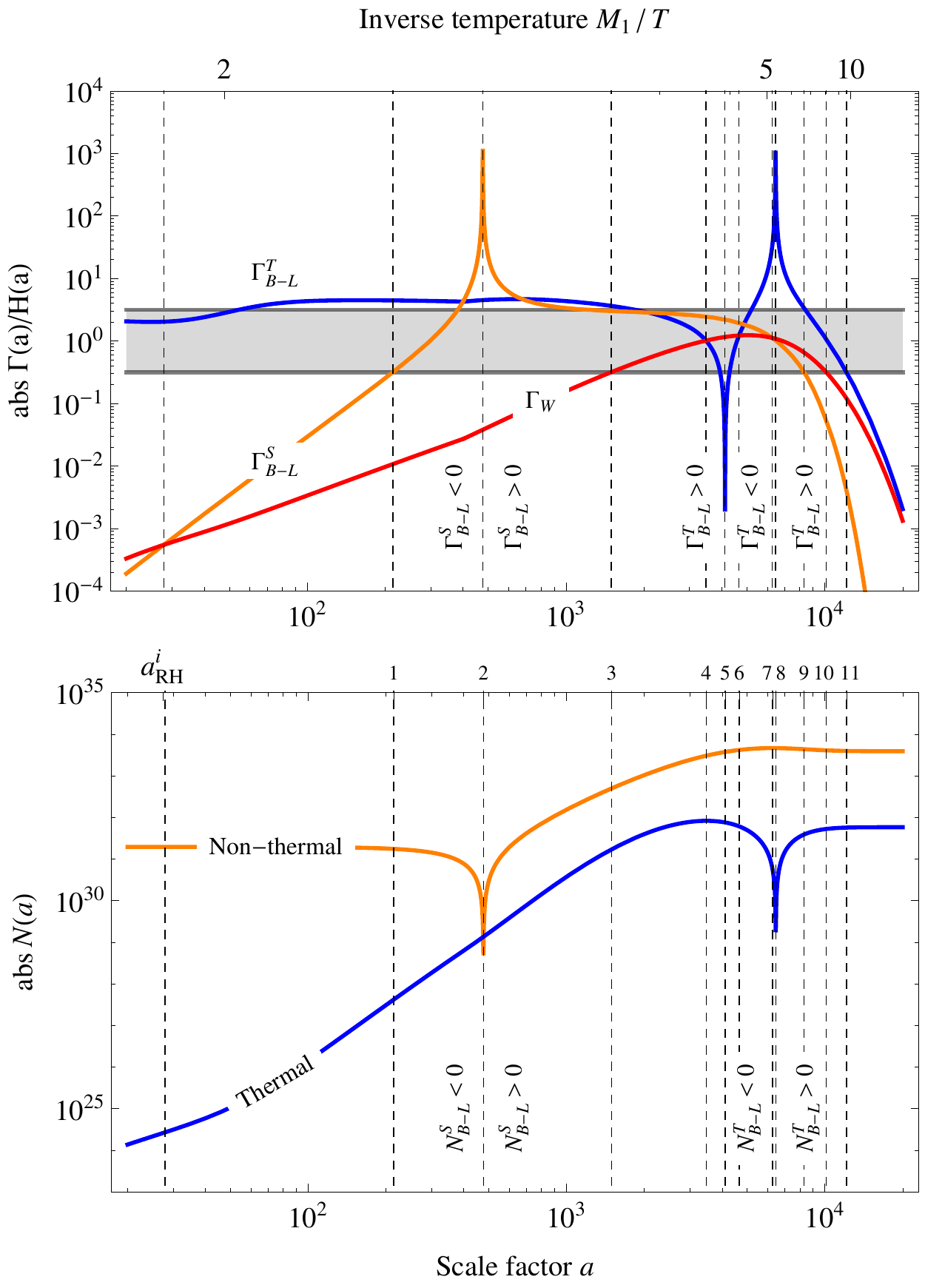}
\caption{Evolution with the scale factor $a$ of \textbf{(upper panel)}
  the  interaction
  rates $\Gamma_{B-L}^S$, $\Gamma_{B-L}^T$ and $\Gamma_W$ normalized to the Hubble rate $H$,
  and of \textbf{(lower panel)} the nonthermal and
  thermal parts $N_{B-L}^S$ and $N_{B-L}^T$  of the generated $B-L$ asymmetry.
The rates are defined in Eq.~\eqref{eq:BLrates}, and the asymmetries
were introduced in Eqs.~\eqref{eq:BEBLS} and \eqref{eq:BEBLT}.
The dashed lines and the integer numbers above the top frame edge in the lower panel
refer to the various values of the scale factor and their numbering as used in the
discussion of this figure in Section~\ref{subsec:Asymm}.
The gray band in the upper panel indicates where the interaction
rates are of the same order as the Hubble rate $H$.
\label{fig:BLrates}}
\end{center}
\end{figure}

The evolution of the nonthermal and thermal lepton asymmetries
is controlled by three different interaction rates which enter
the Boltzmann equations~\eqref{eq:BEBLS} and \eqref{eq:BEBLT}
\begin{align}
\Gamma_{B-L}^S = \epsilon_1 \frac{N_{N_1}^S}{N_{B-L}^S} \,
\Gamma_{N_1}^S\,, \qquad
\Gamma_{B-L}^T = \epsilon_1 \frac{N_{N_1}^T -
  N_{N_1}^\equi}{N_{B-L}^T}\, \Gamma_{N_1}^T\,, \qquad
\Gamma_W = \frac{N_{N_1}^\equi}{2 N_\ell^\equi} \,\Gamma_{N_1}^T\,.
\label{eq:BLrates}
\end{align}
They account for the decay of nonthermal neutrinos, the decay
of thermal neutrinos and the washout effects due to inverse neutrino
decays, respectively.
Their relative importance as well as their influence on the
generation of the asymmetries are illustrated in
Fig.~\ref{fig:BLrates}.
The respective interactions become efficient once
the corresponding rates are of the same order as the Hubble rate or 
larger.
This is why it takes until $a_1 \simeq 210$, when $\Gamma_{B-L}^S / H
\gtrsim \mathcal{O}(1)$ for the first time, for $N_{B-L}^S$ to begin to increase. 
At $a_2 \simeq 480$ the initial negative asymmetry has been
compensated by the generated positive one, and $N_{B-L}^S$ changes
sign.
Subsequently, for scale factors around $a(t_1)$, the ratio
$\Gamma_{B-L}^S / H$ remains approximately constant leading to the
generation of the main part of the asymmetry.
Meanwhile, due to the continuous entropy production
from nonthermal neutrino decays, the washout processes gain in importance.
At $a_3 \simeq 1500$ the rate $\Gamma_W$ becomes comparable to $H$,
which is reflected in a slight decrease of the slope of $N_{B-L}^S$.
From $a_7 \simeq 6300$ onwards, which is shortly after $\rho_R = \rho_{N_1}^S$,
the washout even dominates over the asymmetry production from $N_1^S$ decays.
Hence, the maximal nonthermal asymmetry reached at $a_7$ is slightly washed out
until it eventually freezes out when $\Gamma_W$ drops below $H$
at $a_{10} \simeq 10000$.
Notice that $\Gamma_{B-L}^S$ already becomes irrelevant at $a_9 \simeq 8300$.

The decays and inverse decays of thermal neutrinos lead
to a continuous production of a thermal asymmetry with a negative sign
until the rate $\Gamma_{B-L}^T$ is overcome by $\Gamma_W$ at $a_4
\simeq 3500$.
Following that moment, washout processes push $N_{B-L}^T$ back to
$N_{B-L}^T = 0$.
This development is reinforced by thermal neutrino decays once
$N_{N_1}^T$ has exceeded the equilibrium number density $N_{N_1}^\equi$
at $a_5 \simeq 4100$.
Until $a_8 \simeq 6500$ the thermal asymmetry is then converted
into a positive asymmetry.
After $a_6 \simeq 4700$ the rate $\Gamma_{B-L}^T$ permanently
dominates over $\Gamma_W$, and the thermal asymmetry does not decrease
anymore after $a_8$.
Instead it freezes out at its maximum value when $\Gamma_{B-L}^T / H
\sim \mathcal{O}(0.1)$ at $a_{11} \simeq 12000$.

Finally, parallel to the onset of entropy production at
$a_{RH}^i \simeq 27$, the rate $\Gamma_{B-L}^S$ exceeds
$\Gamma_W$, and $\Gamma_{B-L}^T$ slightly increases.

\subsection{Radiation temperature}

\begin{figure}
\begin{center}
\includegraphics[width=12cm]{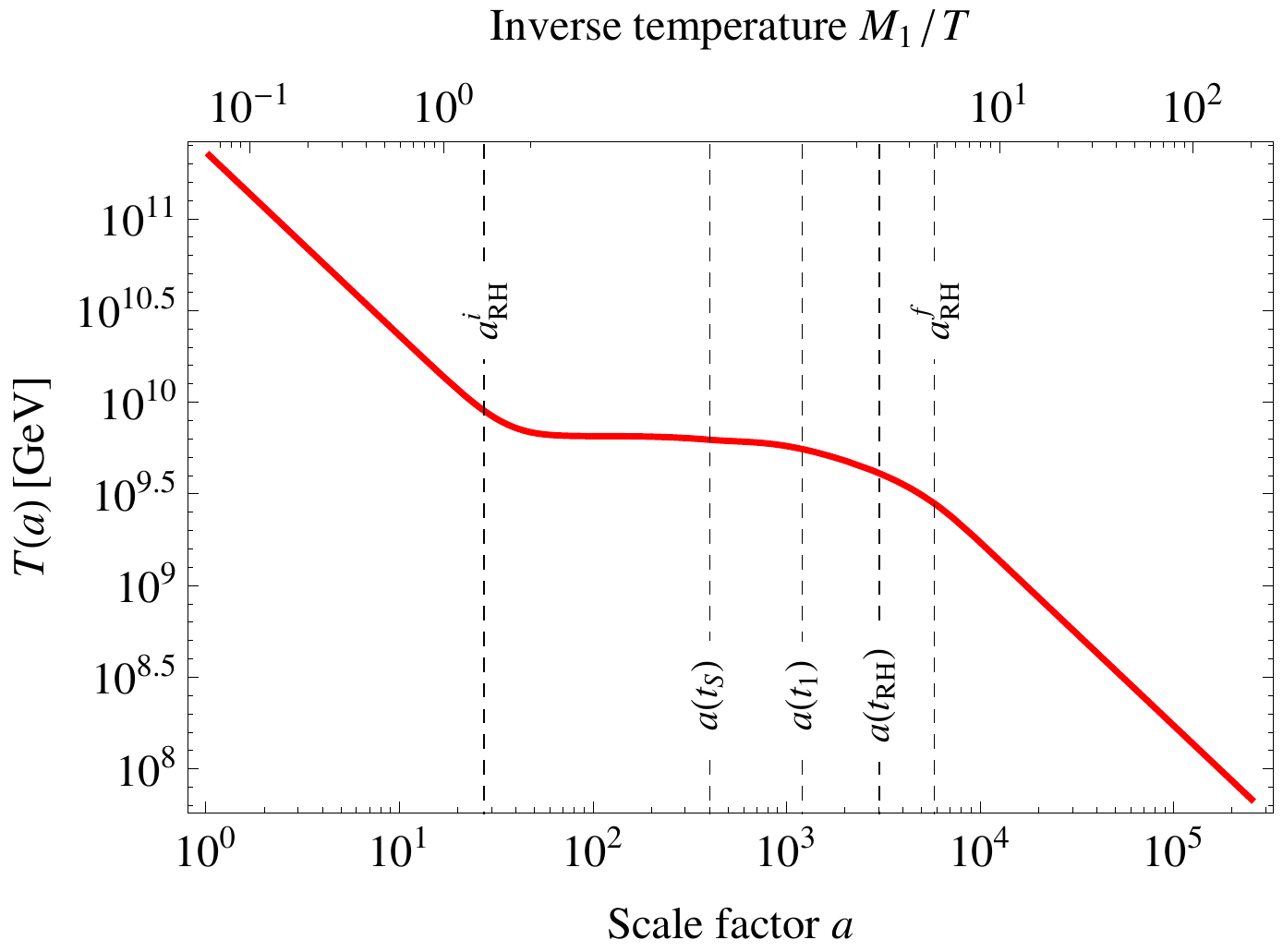}\vspace{10mm}
\includegraphics[width=12cm]{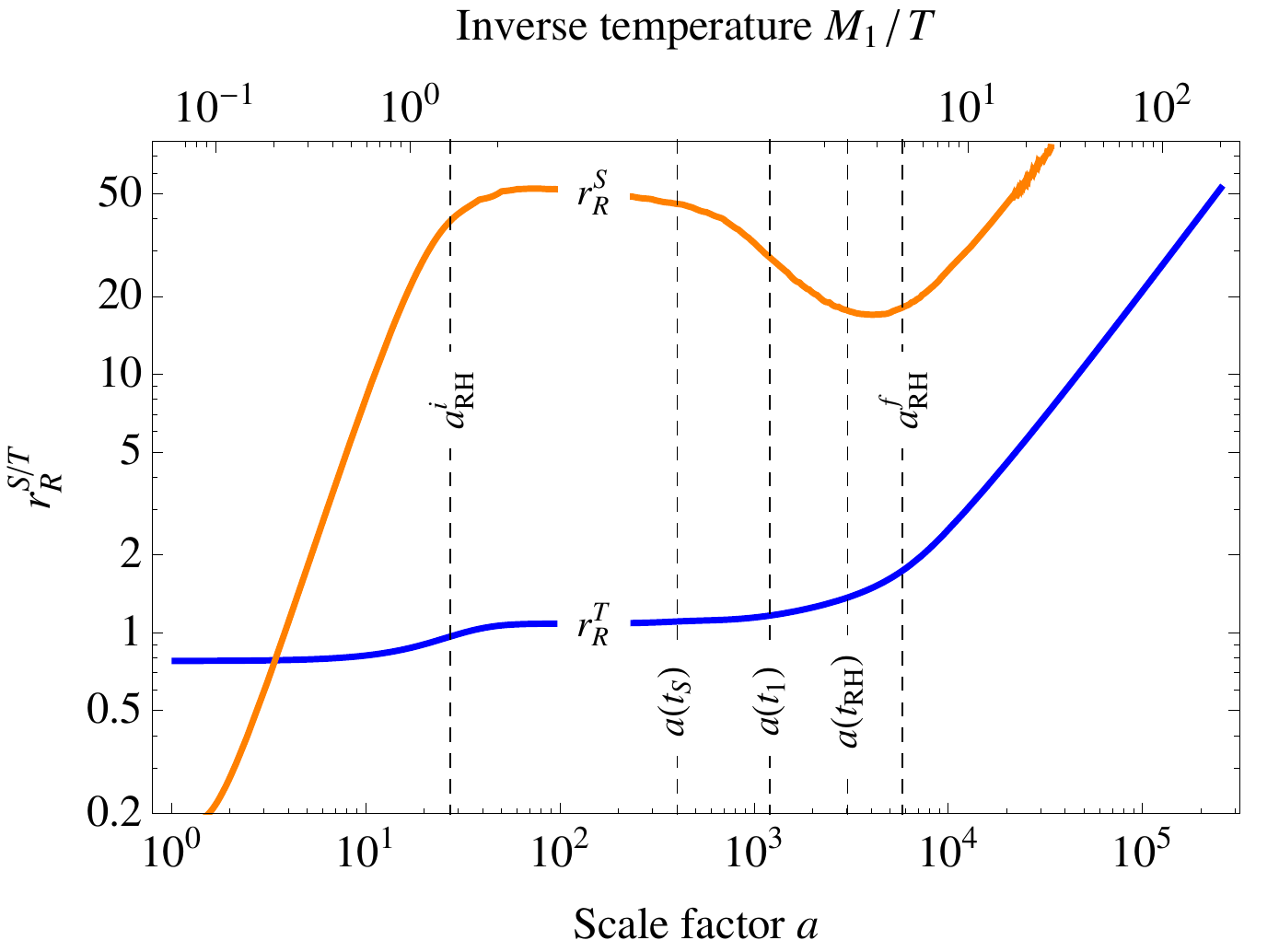}
\caption{\textbf{(Upper panel)} Temperature $T$ and \textbf{(Lower panel)}
the nonthermal and thermal correction factors $r_R^S$ and $r_R^T$ as
functions of the scale factor.
The temperature is calculated from Eq.~\eqref{eq:TNR}, the correction factors
were introduced in Eq.~\eqref{eq:rRSTDef} and are used in the Boltzmann
equation \eqref{eq:BENR} for radiation to ensure a correct counting of radiation
quanta.
\label{fig:genbth}}
\end{center}
\end{figure}

Having solved the Boltzmann equation \eqref{eq:BENR} for the number density
of radiation quanta, we obtain from Eq.~\eqref{eq:TNR}
the evolution of the plasma temperature $T$ which is plotted in Fig.~\ref{fig:genbth}.
We find that the reheating process between $a_{RH}^i \simeq 27$ and
$a_{RH}^f \simeq 5800$ is accompanied by an approximate temperature
plateau around $T \sim 6 \times 10^9\,\textrm{GeV}$.
Especially until $S$ boson decay around $a\left(t_S\right) \simeq 400$
the temperature is essentially constant.
This is due to the continuous production of nonthermal neutrinos which
do not efficiently decay before $a\left(t_1\right) \simeq 1200$.
With nonrelativistic $S$ bosons still representing the dominant
contribution to the energy density, the comoving $N_{N_1}^S$ number density
approximately scales like $N_{N_1}^S \propto \int_{t_2}^t dt^{\prime} \propto a^{3/2}$.
According to the Boltzmann equation \eqref{eq:BENR} for radiation, the
comoving number density $N_R$ then grows like the volume, implying a
constant temperature
\begin{align}
a H \frac{d}{da} N_R \propto N_{N_1}^S \propto a^{3/2} \,,\qquad
N_R \propto a^3 \,,\qquad
T = \textrm{const.}
\end{align}
Once the production of nonthermal neutrinos ceases, not as much
radiation is produced anymore and the temperature begins to drop.
During the phases of adiabatic expansion $T$ decreases like the inverse
of the scale factor, $T \propto 1/a$.

The actual reheating temperature $T_{RH}$ is reached once the Hubble rate
$H$ becomes as small as the effective decay rate $\Gamma_{N_1}^S$ of the
nonthermal neutrinos (cf. Appendix~\ref{appendix:TR})
\begin{align}
\Gamma_{N_1}^S (t_{RH}) = H (t_{RH})\,, \qquad
T_{RH} = T(t_{RH})\,. \end{align}
For the chosen set of parameters this happens at $a\left(t_{RH}\right)
\simeq 3000$,
with $H = \Gamma_{N_1}^S =52\,\textrm{GeV}$,
and the corresponding temperature turns out to be
\begin{align}
T_{RH} \simeq 4.1 \times 10^{9}\,\textrm{GeV}\,.
\label{eq:TRHres}
\end{align}
A detailed discussion of how this result for the reheating temperature
can be estimated on the basis of the input parameters is given
in Appendix~\ref{appendix:TR}.

The lower panel in Fig.~\ref{fig:genbth} presents the evolution
of $r_R^S$ and $r_R^T$ as functions of the scale factor $a$, the two
correction factors which effectively keep track of
the average energy per nonthermal / thermal
neutrino $\varepsilon_{N_1}^{S/T}$ in relation to the typical radiation
energy $\varepsilon_R$ as discussed in Sections~\ref{subsubsec:RelCOs} and
\ref{subsubsec:Rad}.
Until the onset of reheating, $r_R^S$ steeply rises.
This is the consequence of an adiabatically dropping temperature
and, on top of that, the progressively increasing effectiveness of
the $S$ boson decays which push the average $N_{N_1}^S$ energy more
and more towards $m_S / 2$.
Between $a_{RH}^i$ and $a(t_S)$ the temperature stays
rather constant and, as we have checked numerically,
$\varepsilon_{N_1}^S$ has saturated close to $m_S /3$.
Hence, $r_R^S$ only varies little around a value of
$r_R^S \simeq 50$ during that time.
After $a(t_S)$ the $S$ boson decays become less
frequent, the $N_{N_1}^S$ energies are redshifted and $r_R^S$
decreases.
This trend is stopped around $a(t_1)$ when the temperature
begins to fall again and the decay of the nonthermal neutrinos
themselves sets in.
These decays tend to remove rather long-lived and hence redshifted
neutrinos from the spectrum leading to an increase in $\varepsilon_{N_1}^S$.
Finally, after reheating the evolution of $r_R^S$ is again dominated
by the adiabatically decreasing temperature.

The initial value of the thermal correction factor, $r_R^T(t_2) \simeq 0.78$,
is close to $49/60$ and hence to what is expected for a relativistic fermion
coupled to the massless degrees of freedom of the MSSM.
The fact that it is even a bit larger is due to the negligible imprecision
of calculating $\varepsilon_{N_1}^T$ by means of classical statistics
(cf. Eq.~\eqref{eq:eRN1ST}). 
Once the temperature has dropped below $M_1$ around $a \simeq 17$, the
thermal neutrinos become nonrelativistic and $r_R^T$ increases.
This evolution is only delayed by the constant temperature
during reheating.
After reheating $r_R^T$ continuous to increase like $r_R^T \propto a$.

\subsection{Gravitino dark matter}

The present contribution from gravitinos to the total energy density
is given by
\begin{align}
\Omega_{\widetilde{G}} = \frac{m_{\widetilde{G}} n_{\gamma}^0}{\rho_c} \frac{g_{\star,s}^0}{g_{\star,s}}
\left.\frac{N_{\widetilde{G}}}{N_{\gamma}}\right|_{a_f} \,,
\label{eq:omegaG}
\end{align}
where $\rho_c = 1.052\times 10^{-5}\,h^2\,\mathrm{GeV}\,\mathrm{cm}^{-3}$
is the critical density and $n_\gamma^0 = 410 \,\textrm{cm}^{-3}$ is the
number density of CMB photons.
Recall that after fixing all other parameters we have chosen $M_1$
such the gravitino abundance equals the observed one for dark matter
\begin{align}
\Omega_{\widetilde{G}} h^2 \simeq   0.11\,.
\label{eq:OmegaGres}
\end{align}
In Appendix~\ref{appendix:RC} we demonstrate that this result can be
easily reproduced by means of certain semi-analytical estimations.

\begin{figure}[t!]
\begin{center}
\includegraphics[width=12cm]{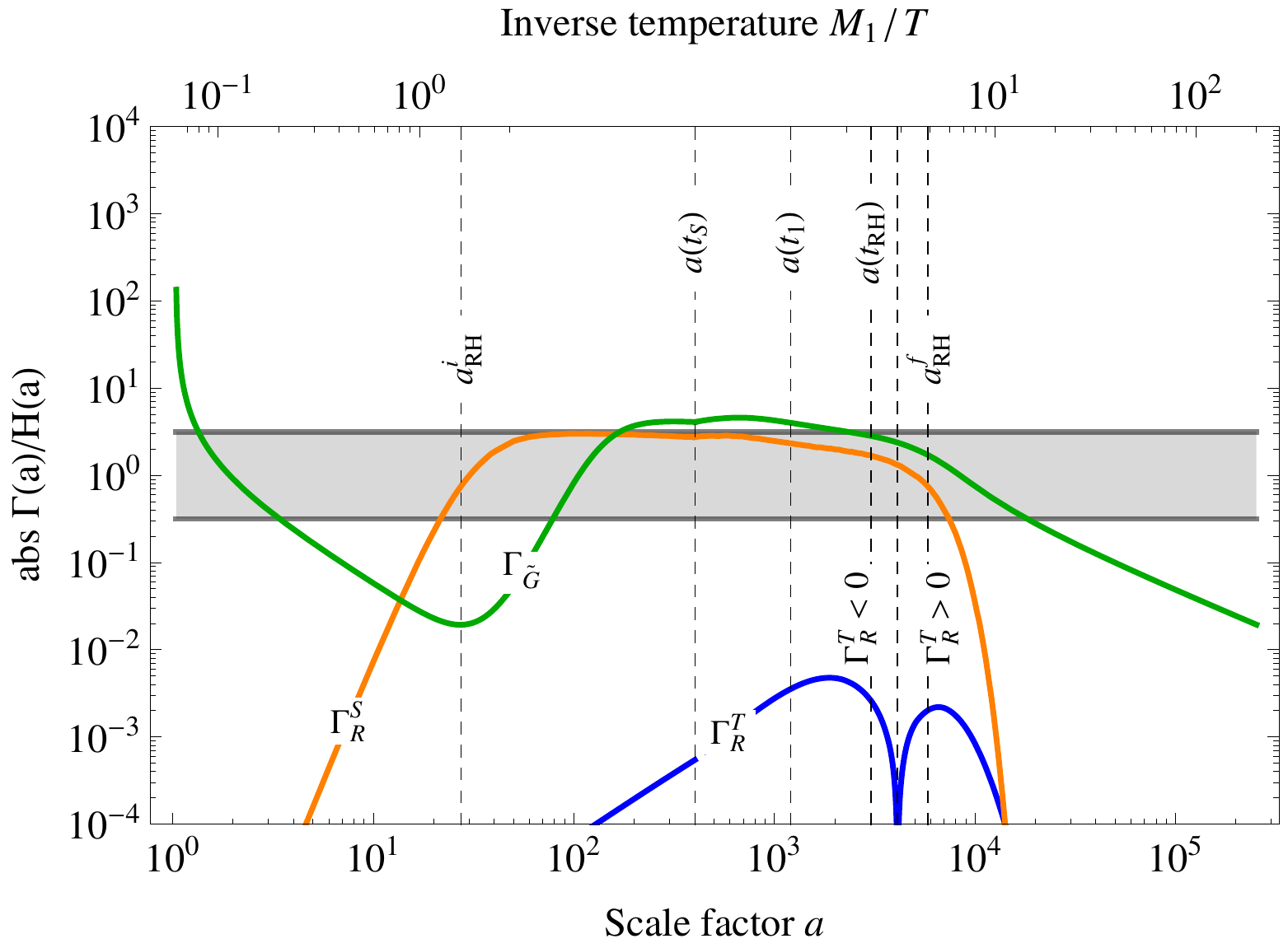}
\caption{Interaction rates $\Gamma_{\widetilde{G}}  = \gamma_{\widetilde{G}} /
  n_{\widetilde{G}}$, $\Gamma_R^S$ and $\Gamma_R^T$ relevant for the
  production of gravitinos and radiation, normalized to the Hubble
  rate $H$.
The rates $\Gamma_R^S$ and $\Gamma_R^T$ were introduced in
Eq.~\eqref{eq:GRST}.
The gray band indicates where the interaction rates are of the same
order as the Hubble rate $H$.
\label{fig:Gtrates}}
\end{center}
\end{figure}

The evolution of the gravitino abundance with time is controlled
by the production rate $\Gamma_{\widetilde{G}} =
\gamma_{\widetilde{G}} / n_{\widetilde{G}}$ (cf. Eqs.~\eqref{eq:BEGt}
and \eqref{eq:gammaGT}), which strongly depends on the temperature.
To demonstrate the close relation between the production of
gravitinos and that of radiation we plot $\Gamma_{\widetilde{G}}$
in Fig.~\ref{fig:Gtrates} together with the rates $\Gamma_R^S$ and
$\Gamma_R^T$ at which radiation is produced due to nonthermal
and thermal neutrino decay
\begin{align}
\Gamma_R^S = r_R^S \,\frac{N_{N_1}^S}{N_R} \, \Gamma_{N_1}^S\,, \qquad
\Gamma_R^T = r_R^T \,\frac{N_{N_1}^T - N_{N_1}^\equi}{N_R} \, \Gamma_{N_1}^T\,.
\label{eq:GRST}
\end{align}
As expected we find that only the decay of the nonthermal
neutrinos efficiently influences the radiation abundance.
Between $a \simeq 22 $ and $a \simeq 7400$, which is basically identical
to the time interval in which reheating takes place, $\Gamma_R^S$ is of
the same order as the Hubble rate.
$\Gamma_R^S / H$ reaches its maximal value before $S$ boson decay around
$a(t_S)$, subsequently decreases a bit and finally drops off
shortly after $a_{RH}^f$.
The ratio $\Gamma_R^T / H$ is at most of order $\mathcal{O}(10^{-2})$
which is the case towards the end of reheating when the thermal neutrinos
are close to thermal equilibrium.
The rate $\Gamma_{\widetilde{G}}$ traces the efficiency of the
nonthermal neutrino decays:
While the $N_1^S$ decays are not active yet, $\Gamma_{\widetilde{G}}/H$
decreases due to the falling temperature.
But as soon as $\Gamma_R^S$ becomes competitive with the Hubble rate around $a_{RH}^i$,
$\Gamma_{\widetilde{G}}/H$ bends over and eventually it reaches values
of order $\mathcal{O}(1)$ and even larger.
On the other hand, once the nonthermal neutrino decay has ended,
$\Gamma_{\widetilde{G}}$ returns to its ordinary behaviour that we
expect for adiabatic expansion.
In total gravitino production occurs between $a \simeq 79$ and $a \simeq 18000$.
The main part of the gravitino abundance is, hence, produced
towards the end or after reheating.
\section{Results and discussion}
\label{sec:results}
The parameter point selected in the previous section
was chosen such that we readily obtained the right baryon
asymmetry and gravitino abundance.
Now we extend our discussion to a quantitative analysis of
the entire parameter space and determine the bounds within which
consistency between successful leptogenesis and gravitino
dark matter can be reached.
According to the flavour model introduced in Section \ref{sec:flavor}
we are free to vary the neutrino mass parameters $M_1$,
$\widetilde{m}_1$ and $v_{B-L}$.
On the supergravity side the gravitino and gluino masses $m_{\widetilde{G}}$
and $m_{\tilde{g}}$ represent free parameters (cf. Section~\ref{subsubsec:Gravi}).
Moving in parameter space changes the interaction rates relevant to
our scenario, most notably the production and decay rates of the $N_1$
neutrino.
This has consequences for the reheating process (cf. Section \ref{subsec:TR}),
the generation of the baryon asymmetry (cf. Section \ref{subsec:BLasymmetry})
and the thermal production of gravitinos (cf. Section \ref{subsec:gravitino}).
By imposing the two conditions \cite{wmap10}
\begin{subequations}
\label{eq:conds}
\begin{align}
\eta_B \equiv \eta_B^S + \eta_B^T & \: \geq \eta_B^{\textrm{obs}}
\simeq 6.2 \times 10^{-10}\,,\label{eq:condeta}\\ 
\Omega_{\widetilde{G}} h^2 & = \: \Omega_{\textrm{DM}} h^2 \simeq 0.11\,,\label{eq:condomega}
\end{align}
\end{subequations}
we are able to identify the regions in parameter space in
which both, the present baryon-to-photon ratio and the dark matter density
are successfully generated.
In this manner, we obtain a link between neutrino and superparticle masses.
The parameter dependence of the reheating temperature and the interplay
of nonthermal and thermal leptogenesis follow along the way.

From the allowed range for the $B-L$ breaking scale (cf. Eq.\,\eqref{eq:vBLrange}),
we consider the boundary values and an intermediate scale.
All three values are associated with different ranges for
the heavy Majorana mass $M_1$ (cf. Eq.\,\eqref{eq:HN1mass}),
\begin{subequations}
\label{eq:vBLvalues}
\begin{align}
v_{B-L} &= 3.4\times 10^{12}\,\textrm{GeV}\,:  \qquad && \quad
1.3\times 10^5\,\textrm{GeV}\leq M_1 \leq 1.1\times
10^{10}\,\textrm{GeV}\,,\label{eq:vBLvalues1}\\
v_{B-L} &= 5.8\times 10^{13}\,\textrm{GeV}\,:  \qquad&& \quad
2.1\times 10^6\,\textrm{GeV}\leq M_1 \leq 1.9\times
10^{11}\,\textrm{GeV}\,,\label{eq:vBLvalues2}\\
v_{B-L} &= 1.0\times 10^{15}\,\textrm{GeV}\,: \qquad && \quad
3.7\times 10^7\,\textrm{GeV}\leq M_1 \leq 3.3\times 10^{12}\,\textrm{GeV}
\,.\label{eq:vBLvalues3}
\end{align}
\end{subequations}
Furthermore, in order to take into account the ${\cal O}(1)$
uncertainties in the Yukawa couplings $h^\nu$, we allow the effective
neutrino mass $\widetilde m_1$ to vary in the range
\begin{align}
10^{-5}\,\textrm{eV} \leq \widetilde m_1 \leq
10^{-1}\,\textrm{eV}\,.
\end{align}
The gravitino mass is taken from the interval
\begin{align}
30\,\textrm{MeV} \leq m_{\widetilde{G}} \leq 700\,\textrm{GeV}\,.
\end{align}
In view of the present bound on the gluino mass, $m_{\tilde g} \gtrsim 700$\,GeV,
imposed by collider searches \cite{Khachatryan:2011tk,Aad:2011hh},
we use a mass of 
$m_{\tilde g} = 800$\,GeV as a representative value in this section.
Different choices of $m_{\tilde g}$ would lead to similar qualitative
results, the only difference being that all values of $m_{\widetilde{G}}$
would have to be rescaled (cf. Section \ref{subsec:gravitino} and Appendix~\ref{appendix:RC}).

In all plots of the parameter space presented
in this section as well as in the Appendices~\ref{appendix:TR}
and \ref{appendix:RC} (Figs.~\ref{fig:treheat}, \ref{fig:BLasym},
\ref{fig:mGboundsM1}, \ref{fig:mGboundsT}, \ref{fig:gammant}, \ref{fig:mGrecon}
and \ref{fig:mGreconT})
we mark the positions of the two parameter points listed in
Tab.~\ref{tab:parapoints}: The point
the discussion in Section~\ref{sec:example} was based on as well as the
point that was investigated in Ref.~\cite{bsv10}.

\subsection{Reheating temperature}
\label{subsec:TR}

The concept of temperature is only applicable
as long as the interactions in the system under study
are in thermal or, at least, kinetic equilibrium.
Hence, regarding our scenario, it is not before the creation of an
initial thermal bath due to quick thermalization of the $N_{2,3}$ decay
products that we can meaningfully speak about a temperature.
Subsequently, the main part of the energy density continues to reside
in nonthermal particles.
At first most of the energy density is carried by the $S$ bosons, and
then, from $t \simeq t_S$ onwards, by the nonthermal $N_1$ neutrinos.
The energy transfer to the thermal bath, \textit{i.e.} the
reheating of the universe, becomes fully efficient
when the $N_1$ neutrinos decay into standard model radiation.
In first approximation, this happens once the Hubble rate $H$ has
dropped to the value of the effective decay rate $\Gamma_{N_1}^S$ of
the nonthermal $N_1$ neutrinos,
\begin{align}
\label{eq:TRHdef5}
\Gamma_{N_1}^S(t_{RH}) = H (t_{RH})\,, \qquad T_{RH} = T(t_{RH})\,,
\end{align}
where $\Gamma_{N_1}^S$ is the vacuum decay rate
$\Gamma_{N_1}^0$ weighted with the inverse time dilatation factor as
defined in Eq.\,\eqref{eq:InvDila}.
The reheating temperature $T_{RH}$ can then be obtained by applying
Eq.\,\eqref{eq:TRHdef5} to the solutions of the Boltzmann
equations.\footnote{As discussed in Section \ref{sec:boltz}, we use an
approximate solution of the Friedmann equation, $H=\dot a/a$, with
the scale factor $a(t)$ given in Eq.~\eqref{eq:scafac}.}

\begin{figure}
\begin{center}
\includegraphics[width=16cm]{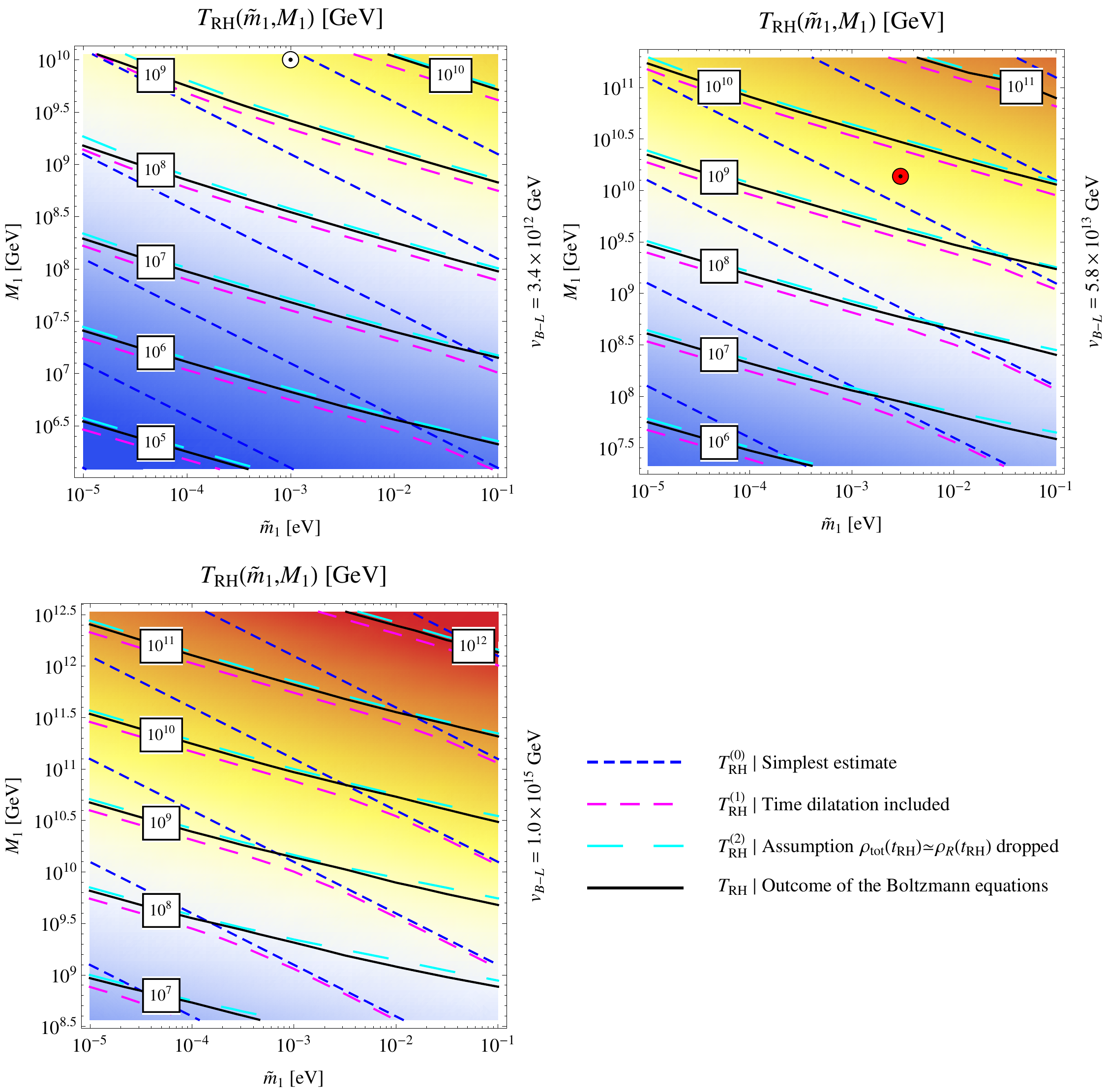}
\caption{Contour plots of the reheating temperature $T_{RH}$ as
a function of the parameters $\widetilde{m}_1$ and $M_1$ for the three
different choices of $v_{B-L}$ listed in Eq.~\eqref{eq:vBLvalues}.
The outcome of the Boltzmann equations $T_{RH}$, calculated
according to Eq.~\eqref{eq:TRHdef5}, is compared with three
different estimates $T_{RH}^{(0)}$, $T_{RH}^{(1)}$ and
$T_{RH}^{(2)}$ which are respectively defined in Eqs.~\eqref{eq:TRHdef2},
\eqref{eq:TRH1} and \eqref{eq:TRH2} .
The contour labels as well as the background colours
indicate the numerical values of $T_{RH}$.
Going to smaller values of $\widetilde{m}_1$ the
$T_{RH}^{(0)}$ and $T_{RH}^{(1)}$
contours approach the corresponding $T_{RH}$ contours
from below.
With respect to the $T_{RH}$ contours the $T_{RH}^{(2)}$
contours are shifted upwards by approximately
$\Delta \log_{10}T_{RH} \simeq 0.04$, cf. Eq.~\eqref{eq:TRH23}.
\label{fig:treheat}}
\end{center}
\end{figure}

For nonrelativistic Majorana neutrinos $N_1^S$, instantaneous
energy transfer and $H$ being determined from the Friedmann equation,
the reheating temperature is given by
\begin{align}
\label{eq:TRHdef2}
T_{RH}^{(0)} = \left( \frac{90}{8 \pi^3  g_{\star,\rho}} \right)^{1/4}
\sqrt{\Gamma_{N_1}^0 M_p}\,.
\end{align}
In our scenario we do not meet either of these
conditions which is why the simple estimate $T_{RH}^{(0)}$ has to be
augmented with several corrections in order to properly reproduce
the outcome of the Boltzmann equations:
\begin{itemize}
\item Being produced in $S$ decays, the
nonthermal neutrinos all carry initial energy
$m_S/2 \simeq 150 M_1$.
For most of the time they are, hence, highly
relativistic such that their decays occur at an effective rate
$\Gamma_{N_1}^S$ (cf. Eq.~\eqref{eq:TRHdef5}).
Replacing $\Gamma_{N_1}^0$ by $\Gamma_{N_1}^S$ results in an
approximation $T_{RH}^{(1)}$ for the reheating temperature;
\item At the time when $T_{RH}$ is evaluated
a large fraction of the energy density still resides in nonthermal neutrinos.
Taking into account that only a part of the total energy density
at $t = t_{RH}$ contributes to $T_{RH}$ yields an approximation
$T_{RH}^{(2)}$;
\item The fact that our approximation for the Hubble rate $H = \dot{a}/a$
does not fulfill the Friedmann equation exactly introduces a final imprecision
which effectuates the remaining small deviation of $T_{RH}^{(2)}$ from the actual reheating
temperature $T_{RH}$.
\end{itemize}
We refer the interested reader to Appendix \ref{appendix:TR} where the
reconstruction of the numerical result $T_{RH}$
starting from the simplest estimate $T_{RH}^{(0)}$ is discussed in greater detail.

\medskip
The reheating temperature $T_{RH}$ obtained from the Boltzmann equations
for the three values of $v_{B-L}$ in Eq.\,\eqref{eq:vBLvalues}
is presented in Fig.\,\ref{fig:treheat} together with the
different approximations $T_{RH}^{(i)}$ as a function of the neutrino mass
parameters $\widetilde m_1$ and $M_1$.
Notice that the behaviour of $T_{RH}^{(0)}$ is determined by the width
$\Gamma_{N_1}^0 \propto \widetilde m_1 \,M_1^2$ 
(cf. Eq.\,\eqref{eq:Ndecayrate}), which is independent of $v_{B-L}$.
The correction due to the time dilatation
factor mainly depends on the ratio of the Majorana
neutrino decay width Eq.\,\eqref{eq:Ndecayrate} and the $S$ boson decay width
Eq.\,\eqref{eq:GammaS0},
\begin{align}
\label{eq:gammacorrection}
\frac{\Gamma_{N_1}^0}{\Gamma_S^0} \propto \frac{\widetilde m_1\,
 v_{B-L}^2}{M_1 v_{EW}^2}\,.
\end{align}
For $\Gamma_{N_1}^0 \gg \Gamma_S^0$, the bulk of the nonthermal
neutrinos decaying at $t=t_{RH}$ is produced just shortly
before and is therefore relativistic.
On the other hand, for $\Gamma_{N_1}^0 \ll \Gamma_S^0$, most of the
nonthermal neutrinos decaying at the reheating time are
nonrelativistic.
For fixed $v_{B-L}$, this correction turns out to be marginal for
the smallest effective neutrino masses $\widetilde m_1$ and the
largest Majorana neutrino masses $M_1$.
The correction increases with the ratio in Eq.\,\eqref{eq:gammacorrection}
becoming larger.
Its maximum is given by the flavour model, $\sqrt{2M_1/m_S} \simeq
\sqrt{1/150}$.
The related correction corresponding to the overestimation of the
energy density of radiation at $t = t_{RH}$ has the same dependence on parameters.
Finally, the mismatch between the Hubble rate and the exact solution
of the Friedmann equation only slightly modifies the reheating temperature.
All in all, the global effect of these corrections is to increase
(decrease) the dependence of the reheating temperature on $M_1$
($\widetilde m_1$).

In each of the three panels of Fig.\,\ref{fig:treheat}, corresponding
to the three different values of $v_{B-L}$, the values of $\widetilde{m}_1$ and $M_1$
respectively span four orders of magnitude allowing for reheating temperatures ranging
over five orders of magnitude.
Reheating temperatures as small as $T_{RH} \simeq 10^5$\,GeV are
obtained for the lowest decay rates in association with the smallest
initial false vacuum energy density, \textit{i.e.} for the minimal values of
$v_{B-L}$, $\widetilde m_1$ and $M_1$.
Conversely, reheating temperatures as large as $T_{RH} \simeq
10^{12}$\,GeV are obtained for the maximal values of $v_{B-L}$,
$\widetilde m_1$ and $M_1$.

Lastly, we observe that the region where the reheating temperature
exceeds the Majorana neutrino mass significantly shrinks when going
from the simplest approximation $T_{RH}^{(0)}$ to the results of
the Boltzmann equations $T_{RH}$.
As for the former, $T_{RH}^{(0)} > M_1 $ for $\widetilde m_1 \gtrsim
2\times 10^{-3}$\,eV, independent of $M_1$, while in the latter case
$T_{RH} > M_1$ is only accomplished for the largest values of $\widetilde{m}_1$
and $M_1$.
The reasons for this relative decrease in the reheating temperature were
already mentioned following Eq.~\eqref{eq:TRHdef2}: the longer
neutrino lifetimes due to their relativistic nature and the overestimation
of radiation energy in deriving $T_{RH}^{(0)}$.
As the strength of the washout processes during the generation of
the baryon asymmetry crucially depends on the ratio of temperature $T$
and neutrino mass $M_1$ we expect the efficiency of leptogenesis to
severely drop in the region $T_{RH} > M_1$.

\subsection{Baryon asymmetry}
\label{subsec:BLasymmetry}
\begin{figure}
\begin{center}
\includegraphics[width=16cm]{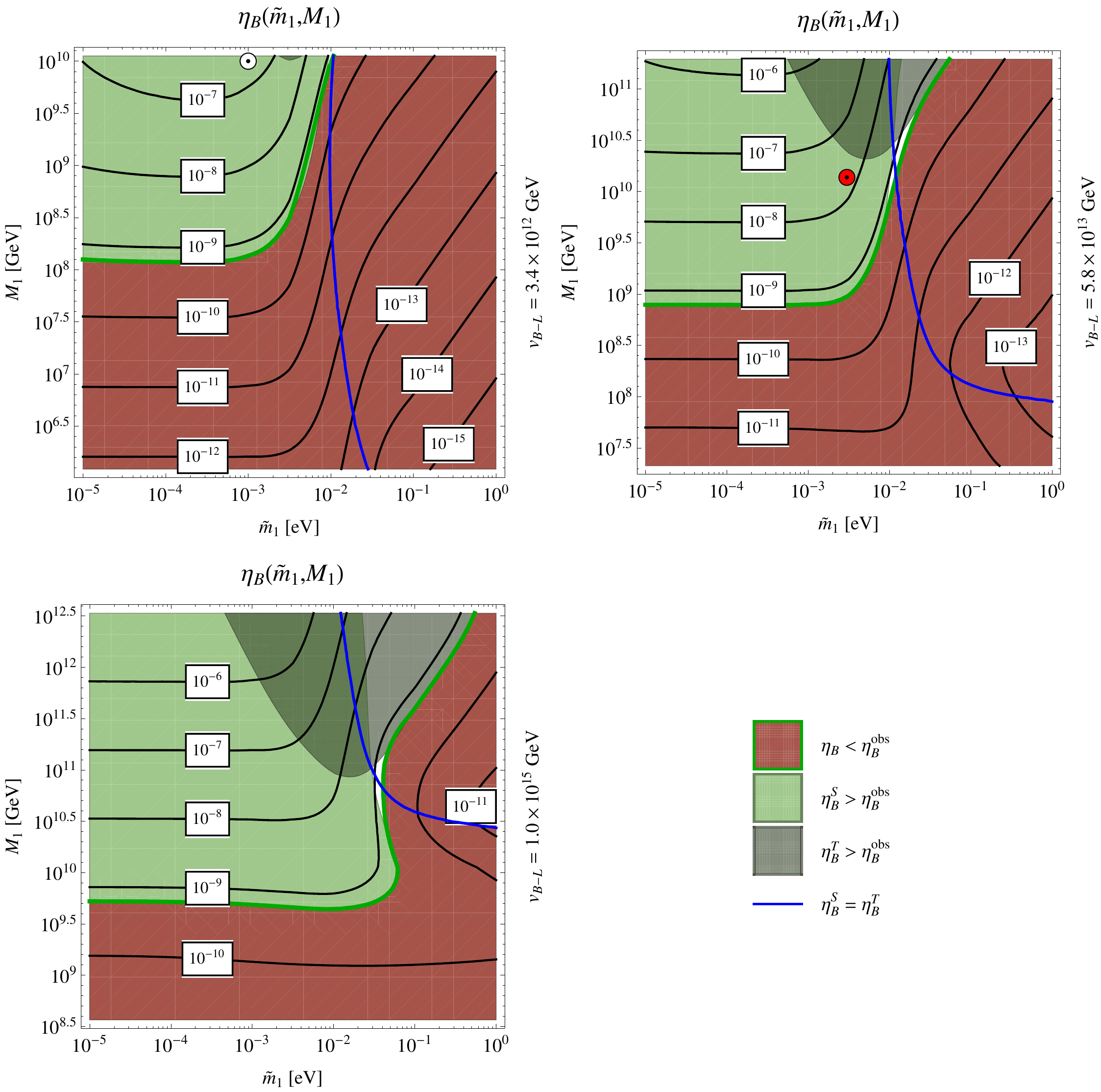}
\caption{Contour plots of the baryon-to-photon ratio $\eta_{B}$ as
defined in Eq.\,\eqref{eq:etaBdef} as a function of the parameters
$\widetilde m_1$ and $M_1$ for the three different choices of $v_{B-L}$
listed in Eq.~\eqref{eq:vBLvalues}.
In the light green (gray-green) region leptogenesis through the
decay of nonthermal (thermal) Majorana neutrinos successfully reproduces
the observed baryon asymmetry $\eta_{B}^{\textrm{obs}}$.
In the red region the total asymmetry is not able
to overcome the observational bound.
The blue line separates the regions in which each
leptogenesis variant is the dominant one.
\label{fig:BLasym}}
\end{center}
\end{figure}

The baryon asymmetry that is generated for a given choice of input
parameters follows from the 
respective solutions of the Boltzmann equations according
to Eq.~\eqref{eq:etaBdef}.
In this section we shall discuss in turn the contributions
it receives from the decays of the nonthermal and thermal neutrinos.
Our main results are displayed in Fig.\,\ref{fig:BLasym} which
presents the baryon asymmetry for the three values of the
$B-L$ breaking scale (cf. Eq.\,\eqref{eq:vBLvalues}) as
function of the neutrino mass parameters $\widetilde{m}_1$ and $M_1$.
In each panel of Fig.\,\ref{fig:BLasym}, we indicate the regions
in which leptogenesis from the decay of nonthermal (light green) and
thermal neutrinos (gray-green) successfully produces the observed
baryon asymmetry.
Notice that in some regions of parameter space (dark green)
both variants of leptogenesis manage to overcome the observational
bound individually while in others (white) $\eta_B^\textrm{obs}$ only is
exceeded after taking the sum of the two contributions.

The blue solid lines in Fig.\,\ref{fig:BLasym} separate the
parameter regions in which leptogenesis is respectively driven
either by the decay of the nonthermal or the thermal neutrinos.
In the viable regions of parameter space the nonthermal
contribution to the baryon asymmetry typically represents the
clearly dominant one.
As we have checked numerically, the generated asymmetry in that
case can be reconstructed to good approximation by assuming
that at $t = t_1$ the energy density of the nonthermal
neutrinos is almost instantaneously converted into radiation.
Nonthermal neutrinos of average energy $\varepsilon_{N_1}^S$
(cf. Eq.\,\eqref{eq:eRN1ST}), that rapidly decay around $t = t_1$,
lead to an asymmetry (cf. \cite{kt90})
\begin{align}
\label{eq:etarapid}
\eta_B^{\mathrm{rapid}}
\simeq 7 \,\frac{3}{4} \,c_{\mathrm{sph}}
\,\epsilon_1 \left.\frac{T}{\varepsilon_{N_1}^S}\right|_{t=t_1}\,.
\end{align}
We emphasize that neglecting the relativistic motion of the
nonthermal neutrinos, \textit{i.e.} employing simply the mass $M_1$
instead of the full energy per particle $\varepsilon_{N_1}^S$
in Eq.~\eqref{eq:etarapid}, would entail an asymmetry proportional
to the temperature, $\eta_B^{\mathrm{rapid}} \propto T(t_1)$.
Such an estimate fails to reproduce our results
except for some accidental points in parameter space.\footnote{This
actually happens in Ref.~\cite{bsv10} in which $\eta_B^{\mathrm{rapid}}$
is calculated for the corresponding set of parameter values given in
Tab.~\ref{tab:parapoints}.}

For fixed $v_{B-L}$ and $\widetilde{m}_1 \lesssim \mathcal{O}(10^{-3})$\,eV
the nonthermal baryon asymmetry does not depend on
$\widetilde{m}_1$ anymore.
This observation can be easily understood in terms of the
Boltzmann equation~\eqref{eq:BEBLS} for $N_{B-L}^S$.
For very small effective neutrino masses, the washout processes
become inefficient, leaving us only with the production term.
The size of the final asymmetry then only depends on the maximal
$N_1^S$ abundance that can be reached in the course of $S$ boson decay,
which is reminiscent of standard thermal leptogenesis in the
weak washout regime.
Since the collision operator accounting for the production of nonthermal neutrinos
through the decays of $S$ bosons as well as the CP parameter
$\epsilon_1$ are solely controlled by the Majorana neutrino mass,
the resulting baryon asymmetry ends up being exclusively determined by $M_1$.

Increasing $v_{B-L}$ for fixed neutrino masses
$\widetilde{m}_1$ and $M_1$ reduces the produced baryon asymmetry.
This is due to several effects whose influence is apparent
in Eq.~\eqref{eq:etarapid}:
On the one hand a higher $B-L$ breaking scale implies a
larger relativistic correction resulting in a smaller effective
decay rate $\Gamma_{N_1}^S$, on the other hand it leads to a
faster Hubble expansion.
The former increases $\varepsilon_{N_1}^S \sim \left<M_1/E_{N_1}\right>_S^{-1} M_1$
and delays the neutrino decays such that Eq.~\eqref{eq:etarapid}
needs to be evaluated at a later time $t_1$ corresponding to a smaller
temperature $T$.
The faster Hubble rate $H$ reinforces the drop-off in the temperature.
We may reformulate this argument in terms of the rates
$\Gamma_{N_1}^S$ and $H$ by saying that a smaller ratio $\Gamma_{N_1}^S / H$
reflects a lower efficiency of the nonthermal neutrino decays.
From this point of view, the generation of the asymmetry struggles to keep pace with the
expansion of the universe resulting in a more diluted asymmetry.

At values of $\widetilde{m}_1$ larger than $\mathcal{O}(10^{-3})$\,eV
resonant $\ell H \leftrightarrow \bar{\ell}\bar{H}$ scatterings
that wash out the generated asymmetry at a rate $\Gamma_W$ 
(cf. Eq.~\eqref{eq:BLrates}) dramatically
decrease the efficiency of nonthermal leptogenesis:
For small $\widetilde{m}_1$ we have $T_{RH} \ll M_1$
(cf. Fig. \ref{fig:treheat}) and the production of
on-shell $N_1$ neutrinos out of the thermal bath is
Boltzmann suppressed,
\begin{align}
T \ll M_1 \,:\quad \Gamma_W = \frac{N_{N_1}^\equi}{2 N_{\ell}^\equi} \Gamma_{N_1}^T
\propto \left(\frac{M_1}{T}\right)^{3/2} e^{-M_1/T} \,\Gamma_{N_1}^0 \ll \Gamma_{N_1}^0\,.
\label{eq:GWBoltz}
\end{align}
But as $\widetilde{m}_1$ becomes larger, the reheating temperature
approaches $M_1$ and the final asymmetry is depreciated due to washout.
For given $\widetilde{m}_1$ and $M_1$, increasing
$v_{B-L}$ results in a decrease of the reheating
temperature (cf. Section \ref{subsec:TR}).
In particular, this reduces the region in parameter space
where $T_{RH}>M_1$, consequently extending the region in which
nonthermal leptogenesis can successfully proceed without being
much affected by washout.

\medskip
The decay of the nonthermal neutrinos is not the only mechanism
by means of which the baryon asymmetry is generated in our scenario.
It also receives a contribution $\eta_B^T$ from the decays of the thermally
produced neutrinos $N_1^T$.
If $M_1$ is sufficiently large, $\eta_B^T$ can
exceed the observed baryon asymmetry on its own.
Its behaviour in parameter space is similar to the
one of standard thermal leptogenesis with vanishing initial neutrino abundance.
However, it is important to note that standard thermal leptogenesis
differs from our thermal mechanism in the sense that in the former the thermal
bath out of which the Majorana neutrinos are produced is assumed to have an
independent origin (\textit{e.g.}, inflaton decay) and the initial
temperature usually is taken to be arbitrarily high.
By contrast, the generation of our thermal asymmetry $\eta_B^T$
is tightly coupled to the dynamics of reheating in the course of
the nonthermal neutrino decays.
In standard thermal leptogenesis the CP asymmetry
$\epsilon_1$ (cf. Eq.\,\eqref{eq:epsiloni}) as well as the evolution
of the $N_1$ and $B-L$ abundances are controlled by the neutrino
mass parameters $\widetilde{m}_1$ and $M_1$.
To guarantee successful leptogenesis, $M_1$ is constrained
to be at least ${\cal O}(10^9)$\,GeV if $\widetilde{m}_1$ is fixed at
$\widetilde{m}_1 \simeq 10^{-3}\,\textrm{GeV}$.
Effective neutrino masses $\widetilde{m}_1$ different from that
result in larger bounds on $M_1$.

\begin{table}
\begin{center}
\begin{tabular}{cccccc}
Panel & $v_{B-L}$ [GeV] & $\widetilde{m}_1$ [eV] & $M_1$ [GeV] &
$T_{RH}$ [GeV] & $T_{RH} / M_1$ \\ 
\hline\hline
1 & $3.4 \times 10^{12}$ & $3.2 \times 10^{-3}$ & $1.0 \times 10^{10}$
& $6.4 \times 10^{9}$  & $0.63$ \\ 
2 & $5.8 \times 10^{13}$ & $5.2 \times 10^{-3}$ & $2.1 \times 10^{10}$
& $7.9 \times 10^{9}$  & $0.38$ \\ 
3 & $1.0 \times 10^{15}$ & $1.6 \times 10^{-2}$ & $8.5 \times 10^{10}$
& $2.0 \times 10^{10}$ & $0.23$ \\ 
\hline\hline
\end{tabular}
\caption{Parameter points in the three panels of Fig.~\ref{fig:BLasym}
corresponding to the lowest possible values of $M_1$ for which the decay
of the thermal neutrinos suffices to reproduce the observed baryon asymmetry.
The values for the reheating temperature follow from Fig.~\ref{fig:treheat}.}
\label{tab:etaBT}
\end{center}
\end{table}

In our scenario, the values of $M_1$ above which thermal
leptogenesis is efficient are comparatively one to two orders
of magnitude larger.
Tab.~\ref{tab:etaBT} summarizes the respective bounds on $M_1$
for the three different $B-L$ breaking scales together with the
corresponding values of $\widetilde{m}_1$ and $T_{RH}$.
The fact that now $M_1$ has to be much larger than
$\mathcal{O}(10^9)$\,GeV finds its origin
in the interplay between the specific reheating process at work
and the temperature dependence of thermal leptogenesis:
First of all, in the discussion of the decay of the nonthermal neutrinos
we saw that the temperature is bounded from above to prevent
complete washout of the asymmetry.
The same holds for thermal leptogenesis; but in this case
the temperature also must not be too low in order to ensure
an efficient neutrino production from the thermal bath.
Consequently, as a compromise between very small $\left(T \ll
  M_1\right)$ and very large $\left(T \gg M_1\right)$ temperatures,
thermal leptogenesis is most efficient at $T \sim M_1$
(cf. Tab.~\ref{tab:etaBT}).\footnote{Note that our scenario also differs
  from standard thermal leptogenesis because we only consider decays
  and inverse decays. We have 
  checked that including $\Delta L = 1$ and $\Delta L = 2$ scatterings
would enforce the production of Majorana neutrinos for $\widetilde
m_1 \lesssim 3\times10^{-3}$\,eV as it is the case in standard thermal
leptogenesis \cite{bdp04}, resulting in a slight expansion of the
allowed region in the weak washout regime.}
Second, as for standard thermal leptogenesis, our thermal mechanism
prefers an intermediate value of $\widetilde{m}_1$.
Taking $\widetilde{m}_1$ to large values increases the strength
of the washout processes.
Small $\widetilde{m}_1$ results in a low temperature and a
small neutrino decay rate $\Gamma_{N_1}^0$ such that the
$N_1^T$ production becomes suppressed.
When asking for the lower bounds on $M_1$ we thus have to
look for the smallest values of $M_1$ for which the condition
$T \sim M_1$ holds and $\widetilde{m}_1$ is neither too large nor too small.
In contrast to standard thermal leptogenesis, in our scenario
the accessible temperatures now also depend on the mass parameter $M_1$.
As can be seen from Fig.\,\ref{fig:treheat}, the considered reheating
process simply does not manage to satisfy the condition $T \sim M_1$
for $M_1 \sim 10^9 \,\textrm{GeV}$ without entering the strong washout regime.
Instead, $M_1$ has at least to be as large as indicated
in Tab.~\ref{tab:etaBT} to avoid too large values of $\widetilde{m}_1$
while still fulfilling $T \sim M_1$.

Comparing the three points in Tab.~\ref{tab:etaBT} we note that
the ratio of the reheating temperature $T_{RH}$ to $M_1$ decreases
as $v_{B-L}$ becomes larger.
The production of thermal neutrinos is, consequently,
less efficient for high $v_{B-L}$.
This is, however, compensated
for by the increase in the CP asymmetry parameter $\epsilon_1$
for heavier $N_1$ neutrinos (cf. Eq.\,\eqref{eq:epsilon123}).
Likewise, the corresponding effective neutrino masses $\widetilde{m}_1$
increase when going to larger $B-L$ breaking scales.
This effect is based on the fact that for fixed $\widetilde{m}_1$ and $M_1$
an increase in $v_{B-L}$ entails a drop in the temperature.
The factor representing the Boltzmann suppression in the washout
rate $\Gamma_W$ (cf. Eq.~\eqref{eq:GWBoltz}) then becomes smaller
which enables one to raise the neutrino decay width $\Gamma_{N_1}^0$
by increasing $\widetilde{m}_1$.

Standard thermal leptogenesis predicts a final baryon
asymmetry of
\begin{align}
\eta_B^{\mathrm{th}} =
\frac{3}{4} \frac{g_\star^0}{g_\star} c_{\mathrm{sph}} \epsilon_1
\kappa_f (\mt)\,.
\label{eq:etathermal}
\end{align}
where the final efficiency factor $\kappa_f$ only depends on $\widetilde{m}_1$.
For $\widetilde m_1 \gtrsim 10^{-3}$\,eV it may be parametrized as
\cite{bdp04}
\begin{align}
\kappa_f (\mt) = 2\times 10^{-2}\left(
 \frac{10^{-2}\,\textrm{eV}}{\widetilde m_1}\right)^{1.1}\,.
\label{eq:kappaf}
\end{align}
Combining Eqs.~\eqref{eq:etathermal} and \eqref{eq:kappaf}
with Eq.~\eqref{eq:epsilon123}, one finds that $\eta_B^{\mathrm{th}}$
evolves as $\eta_B^{\textrm{th}} \propto \widetilde m_1^{-1}M_1 $.
This is exactly the behaviour of the total baryon asymmetry
one observes in the regions where the thermal contribution
dominates over the nonthermal one, \textit{i.e.} the
regions on the right-hand side of the blue lines in Fig.\,\ref{fig:BLasym}.
As the number density of nonthermal neutrinos usually
exceeds the number density of thermal neutrinos at the time
the asymmetry is created, the relative size of the two asymmetries
$\eta_B^S$ and $\eta_B^T$ is controlled by the efficiency of the
nonthermal mechanism.
Only when the nonthermal asymmetry is suppressed due to
efficient washout, the baryon
asymmetry due to the decay of the thermal neutrinos
has a chance to dominate.

In conclusion, it is remarkable that leptogenesis through
the decay of the nonthermal Majorana neutrinos is able to widely extend
the region in parameter space in which the observed baryon asymmetry
can successfully be reproduced.
For the lowest $B-L$ breaking scale $v_{B-L} = 3.4 \times
10^{12}$\,GeV, Majorana neutrinos as light as $M_1 \simeq 10^8$\,GeV
are sufficient to generate the observed baryon asymmetry.

\subsection{Gravitino dark matter}
\label{subsec:gravitino}
Having discussed leptogenesis on its own in the
last section, we now ask for the regions in parameter
space where both conditions of Eq.~\eqref{eq:conds} are satisfied,
\textit{i.e.} in which we obtain gravitino dark matter along with a sufficient
baryon asymmetry.
As outlined in Section \ref{subsubsec:Gravi} the thermal production
of gravitinos is controlled by three parameters: the gravitino and
gluino masses $m_{\widetilde G}$ and $m_{\tilde g}$ as well as the
temperature $T$.
The latter is determined by the reheating process,
$T_{RH} = T_{RH}(v_{B-L},\widetilde{m}_1,M_1)$, such that 
$\Omega_{\widetilde G} h^2 $, the present contribution
from gravitinos to the energy density of the universe, depends
on all free parameters of our scenario.
For each point in parameter space the respective solutions
of the Boltzmann equations allow us to calculate
$\Omega_{\widetilde G} h^2 $ according to Eq.~\eqref{eq:omegaG}.
By imposing the condition that gravitinos be the constituents
of dark matter we can then derive relations between neutrino
and superparticle masses.
For instance, if we fix the gluino mass at $800\,\textrm{GeV}$,
\begin{align}
\left.\Omega_{\widetilde{G}} h^2 (v_{B-L},M_1,\widetilde{m}_1,m_{\widetilde{G}},m_{\tilde{g}})
\right|_{m_{\tilde{g}} = 800\,\textrm{GeV}} =  \Omega_{\textrm{DM}} h^2\,,
\label{eq:condomega800}
\end{align}
we can solve for $M_1$ as a function of $v_{B-L}$, $\widetilde{m}_1$
and $m_{\widetilde{G}}$,
\begin{align}
M_1 = M_1 (v_{B-L},\widetilde{m}_1,m_{\widetilde{G}})\,.
\label{eq:M1func}
\end{align}
We consider those choices of the parameters $v_{B-L}$, $\widetilde{m}_1$ and
$m_{\widetilde{G}}$ as viable, which actualize gravitino dark matter for
$M_1$ values that are accessible in the context of the flavour model
(cf. Eq.~\eqref{eq:HN1mass}), 
\begin{align}
M_1 (v_{B-L},\widetilde{m}_1,m_{\widetilde{G}}) \leq \eta^2 v_{B-L}\,.
\label{eq:condM1}
\end{align}
Furthermore, applying Eq.~\eqref{eq:M1func} to the results
of Section~\ref{subsec:TR} allows one to trade the $M_1$
dependence of the reheating temperature $T_{RH}$ for a
dependence on $m_{\widetilde{G}}$,
\begin{align}
T_{RH} = T_{RH} (v_{B-L},M_1,\widetilde{m}_1) \:\rightarrow\:
T_{RH} (v_{B-L},\widetilde{m}_1,m_{\widetilde{G}})\,.
\label{eq:TRHfunc}
\end{align}
Similarly, Eq.~\eqref{eq:M1func} can be used to translate the bounds on $\widetilde{m}_1$ and $M_1$
shown in Fig.~\ref{fig:BLasym} that were obtained by requiring successful
leptogenesis into bounds on $\widetilde{m}_1$ and $m_{\widetilde{G}}$,
\begin{align}
\eta_B(v_{B-L},M_1,\widetilde{m}_1) \:\rightarrow\:
\eta_B(v_{B-L},\widetilde{m}_1,m_{\widetilde{G}}) \geq \eta_B^{\textrm{obs}}\,.
\label{eq:condeta2}
\end{align}

\begin{figure}
\begin{center}
\includegraphics[width=16cm]{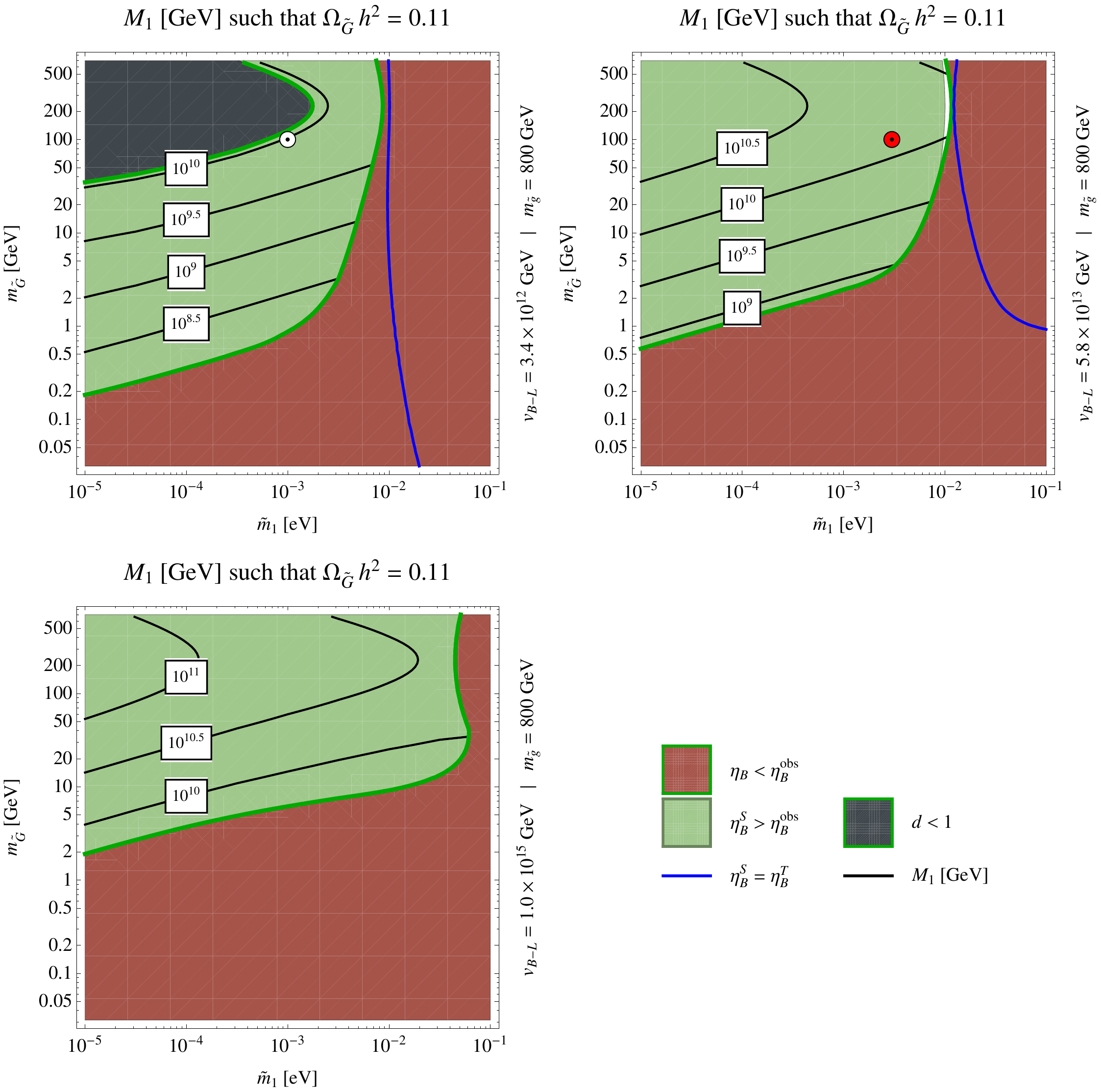}
\caption{
Gravitino mass range consistent with 
gravitino dark matter (cf. Eq.\,\eqref{eq:condomega800}) and successful leptogenesis
(cf. Eq.\,\eqref{eq:condeta2}) depending on
the effective light neutrino
mass. 
The contour lines refer to the neutrino mass $M_1$
  (cf. Eq.\,\eqref{eq:condM1}) as a function of 
 $\widetilde m_1$ and $m_{\widetilde G}$ such that the gravitino
 abundance is $\Omega_{\widetilde G} h^2 = 0.11$. 
 In addition to the
 colour code introduced in Fig.\,\ref{fig:BLasym}, the black region in
 the upper-left panel
 represents the $M_1$ values that are not allowed by the flavour
 model.
\label{fig:mGboundsM1}}
\end{center}
\end{figure}

\begin{figure}
\begin{center}
\includegraphics[width=16cm]{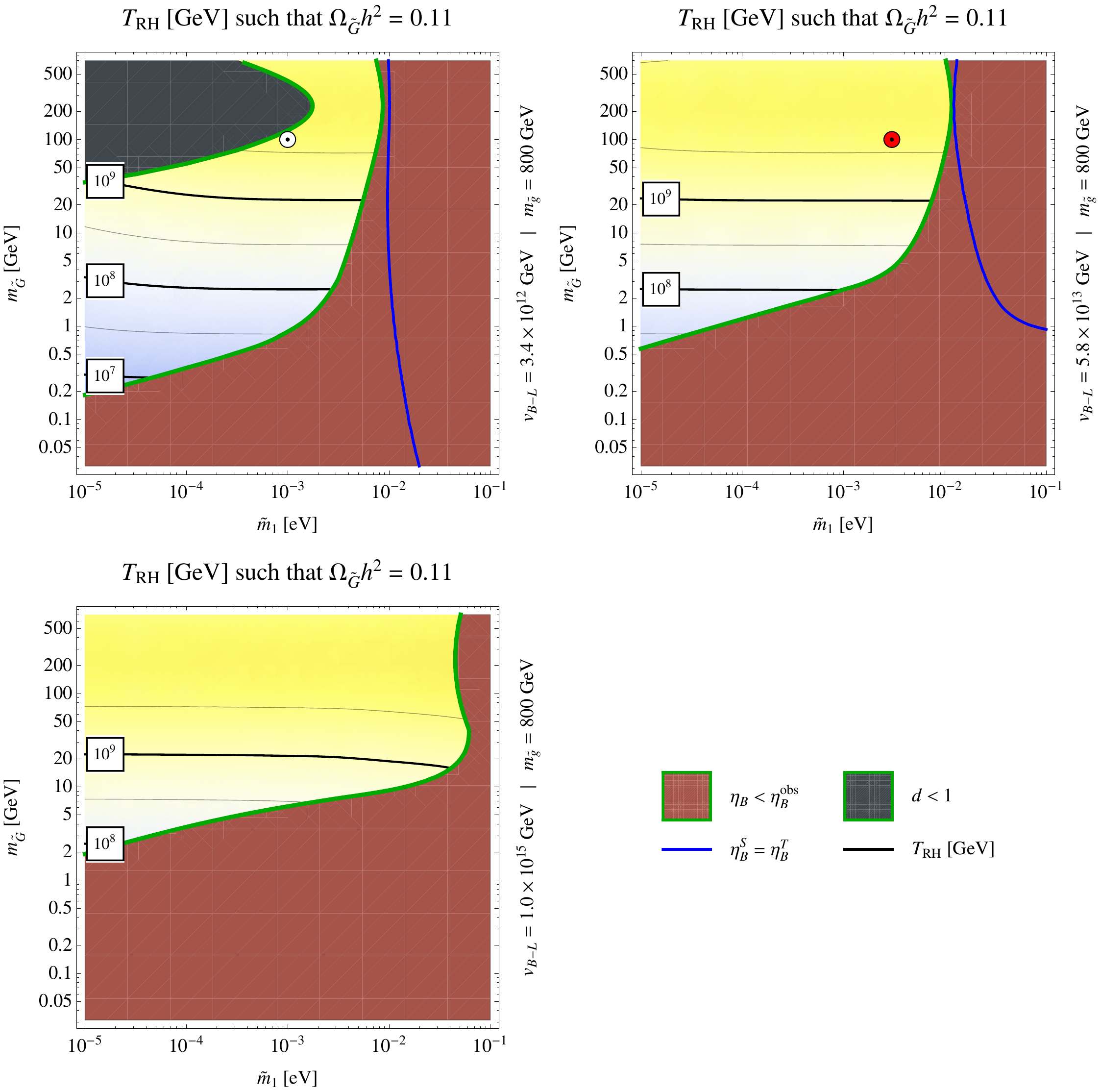}
\caption{Like Fig.\,\ref{fig:mGboundsM1} but with contours for the reheating
 temperature $T_{RH}$ (cf. Eq.\eqref{eq:TRHfunc}) instead of the
 neutrino mass $M_1$.
\label{fig:mGboundsT}}
\end{center}
\end{figure}

The parameter points we are after, \textit{i.e.} the points at which
the baryon asymmetry is accounted for by leptogenesis and gravitinos constitute
the dark matter, now correspond to those values of $v_{B-L}$, $\widetilde{m}_1$
and $m_{\widetilde{G}}$ that satisfy the two conditions in Eqs.~\eqref{eq:condM1}
and \eqref{eq:condeta2} simultaneously.
On the basis of our numerical study of the Boltzmann equations
we are able to identify the regions of interest in parameter space:
Fig.\,\ref{fig:mGboundsM1} presents our results in combination with
the associated values of $M_1$ (cf. Eq.~\eqref{eq:M1func}),
Fig.~\ref{fig:mGboundsT} features the related reheating temperatures
(cf. Eq.~\eqref{eq:TRHfunc}) instead.
Again, both figures consists of three panels each that respectively
take care of the the three different $B-L$ breaking scales specified
in Eq.~\eqref{eq:vBLvalues}.
We refer the interested reader to Appendix~\ref{appendix:RC} which
gives a detailed account of how Figs.~\ref{fig:mGboundsM1}
and \ref{fig:mGboundsT} can be reconstructed by means of simple analytic
expressions and with the aid of our numerical findings for $T_{RH}$
and $\eta_B$.

Notice that we also consider gravitino masses almost as
large as the gluino mass, $m_{\widetilde{G}} \leq 700\,\textrm{GeV}$ while
$m_{\tilde{g}} = 800\,\textrm{GeV}$.
Imposing gaugino mass unification at the GUT scale would, however, forbid such
a nearly degenerate superparticle spectrum.
The running of the renormalization group equations would then
imply a gaugino mass relation $M_3/M_1 \simeq 5.9$ at low energies.
Given that it is the lightest supersymmetric particle, the gravitino
would have to be lighter than the bino resulting in an upper mass bound of
$m_{\widetilde G} \lesssim 140$\,GeV.

If we were to select a gluino mass other than $m_{\tilde{g}} = 800\,\textrm{GeV}$
all values of $m_{\widetilde{G}}$ in Figs.~\ref{fig:mGboundsM1}
and \ref{fig:mGboundsT} would have to be rescaled while
$v_{B-L}$, $M_1$ and $\widetilde{m}_1$ could remain unchanged.
This follows from the fact that the gravitino abundance in Eq.~\eqref{eq:condomega800}
can be kept constant by compensating a change $m_{\tilde g} \rightarrow a\, m_{\tilde g}$
in the gluino mass by a change $m_{\widetilde G} \rightarrow b\,m_{\widetilde G}$
in the gravitino mass without altering the reheating temperature $T_{RH}$.
As long as $m_{\widetilde{G}}$ is much smaller than $m_{\tilde{g}}$,
the factor $b$ simply corresponds to $a^2$.
The general relation between $a$ and $b$ is discussed in Appendix~\ref{appendix:RC}.
To sum up, thanks to this relation the results presented in Figs.\,\ref{fig:mGboundsM1} and
\ref{fig:mGboundsT} can be generalized to different gluino masses by
correspondingly relabeling the gravitino axis.

As a general trend in Figs.\,\ref{fig:mGboundsM1} and
\ref{fig:mGboundsT} we observe that for fixed $\widetilde{m}_1$
and $m_{\widetilde{G}} \lesssim 230 \,\textrm{GeV}$
the neutrino mass $M_1$ and the reheating temperature $T_{RH}$
continuously become larger when increasing the gravitino mass.
For $m_{\widetilde{G}} \gtrsim 230 \,\textrm{GeV}$ this behaviour
is reversed:
The $M_1$ contours in Fig.\,\ref{fig:mGboundsM1} bend over
as soon as $m_{\widetilde{G}} \simeq 230 \,\textrm{GeV}$ is exceeded.
In Fig.~\ref{fig:mGboundsT} the temperature remains rather
constant at $T_{RH} \sim 5 \times 10^{9}\,\textrm{GeV}$ for
$150\,\textrm{GeV} \lesssim m_{\widetilde{G}} \lesssim 400\,\textrm{GeV}$.
Beyond $m_{\widetilde{G}} \gtrsim  400\,\textrm{GeV}$
it begins to decrease again.\footnote{Cf. the contour
corresponding to $T_{RH} = 10^{9.5}\,\textrm{GeV}$ reentering the
second panel of Fig.~\ref{fig:mGboundsT} at $\widetilde{m}_1 \simeq 10^{-5}\,\textrm{eV}$
and $m_{\widetilde{G}} \simeq 700\,\textrm{GeV}$.}
The physical origin of these two regimes can be traced back
to the rate $\Gamma_{\widetilde{G}} =
\gamma_{\widetilde{G}} / n_{\widetilde{G}}$ (cf. Eqs.~\eqref{eq:BEGt}
and \eqref{eq:gammaGT}) at which gravitinos are created from the
thermal bath (cf. Eq.~\eqref{eq:OmegaGRes}) \cite{bbp98,bbb00},
\begin{align}
\Gamma_{\widetilde{G}} = \Gamma_{\widetilde{G}} \left(T,m_{\widetilde{G}},m_{\tilde{g}}\right)
\propto \left(1 + \frac{m_{\tilde{g}}^2(T)}{3 m_{\widetilde{G}}^2}\right)\,.
\label{eq:GammaGprop}
\end{align}
In the regime $m_{\widetilde{G}} \ll m_{\tilde{g}}(T)$ the second term in Eq.~\eqref{eq:GammaGprop}
is the dominant one and it is mainly the goldstino part of the gravitino, \textit{i.e.} its
components with helicity $\pm \frac{1}{2}$, that is produced.
A larger gravitino mass then implies a smaller rate $\Gamma_{\widetilde{G}}$
necessitating a stronger reheating in order to still generate the right abundance.
Correspondingly, the neutrino mass $M_1$ also has to increase to
bring about the higher temperature.
Evolving a gluino mass of $800\,\textrm{GeV}$ from the electroweak scale
to a temperature $T \sim 5 \times 10^{9}\,\textrm{GeV}$ results in a
high-scale mass of $m_{\tilde{g}}(T) \sim 400\,\textrm{GeV}$.
Because of that, $\Gamma_{\widetilde{G}}$ is dominated by the first term in
Eq.~\eqref{eq:GammaGprop} from $m_{\widetilde{G}} \simeq 400/\sqrt{3}\,\textrm{GeV}
\simeq 230\,\textrm{GeV}$ onwards.
This means that, for such large values of $m_{\widetilde{G}}$, primarily
the transverse degrees of freedom of the gravitino, \textit{i.e.} its
components with helicity $\pm \frac{3}{2}$, are excited.
The production rate $\Gamma_{\widetilde{G}}$ then becomes independent of
$m_{\widetilde{G}}$ turning into a function of the temperature $T$ only.
In such a case the final gravitino abundance $\Omega_{\widetilde{G}}h^2$ simply
scales linearly with $m_{\widetilde{G}}$ (cf. Eq.~\eqref{eq:OmegaG}).
Hence, larger gravitino masses have to be balanced by smaller reheating
temperatures to keep $\Omega_{\widetilde{G}}h^2$ fixed.
This explains the decrease in $T_{RH}$ and $M_1$ at very large gravitino masses.

On the other hand, varying $\widetilde{m}_1$ at constant $m_{\widetilde{G}}$
has almost no effect on the reheating temperature, which is expected
since $\Gamma_{\widetilde{G}}$ inherently is
a function of $T$, $m_{\widetilde{G}}$ and $m_{\tilde{g}}$.
As each gravitino mass is associated with an appropriate
rate $\Gamma_{\widetilde{G}}$, the choice of $m_{\widetilde{G}}$
already implies a unique reheating temperature
$T_{RH} \approx T_{RH}(m_{\widetilde{G}})$,
independent of the underlying neutrino parameters
(cf. Fig.~\ref{fig:TRHrecon} in Appendix~\ref{appendix:RC}).
Meanwhile, the neutrino mass $M_1$ becomes smaller
when increasing $\widetilde{m}_1$ in order to ensure that $T_{RH}$
remains approximately constant for fixed $m_{\widetilde{G}}$
(cf. Fig.~\ref{fig:treheat}).

In Figs.~\ref{fig:mGboundsM1} and \ref{fig:mGboundsT}
we also indicate the regions in parameter space that are not compatible
with our scenario because either of the two conditions in
Eqs.~\eqref{eq:condM1} and \eqref{eq:condeta2} is not satisfied.
Bounds coming from the flavour model (cf. Eq.~\eqref{eq:condM1})
only show up for $v_{B-L} = 3.4 \times 10^{12}\,\textrm{GeV}$:
The requirement that $M_1$ be smaller than $1.1\times 10^{10}\,\textrm{GeV}$
(cf. Eq.~\eqref{eq:vBLvalues}) excludes gravitino masses larger than
$35\,\textrm{GeV}$ for $\widetilde{m}_1 = 10^{-5}\,\textrm{eV}$.
At $\widetilde{m}_1 = 10^{-3}\,\textrm{eV}$ it rules out
masses in the range between $120\,\textrm{GeV}$ and $430\,\textrm{GeV}$ and for
$\widetilde{m}_1 \gtrsim 1.8 \times 10^{-3}\,\textrm{eV}$ it does not constrain
$m_{\widetilde{G}}$ any longer at all.
In the case of the two other choices for $v_{B-L}$
the respective flavour bounds on $M_1$ are never reached because the
corresponding reheating temperatures are too high.
Demanding a sufficient baryon asymmetry (cf. Eq.~\eqref{eq:condeta2})
yields lower bounds on $m_{\widetilde{G}}$ in the weak washout regime
and limits the maximal value of $\widetilde{m}_1$.
Notice that these bounds are in one-to-one correspondence with the
constraints on $M_1$ and $\widetilde{m}_1$ in Fig.\,\ref{fig:BLasym}.
For instance, at small $\widetilde{m}_1$ the gravitino mass can only decrease as long as $M_1$ is
large enough so that the observed baryon asymmetry is reproduced.
Similarly, at large $\widetilde{m}_1$ the sharp drop-off in the efficiency of leptogenesis due to
stronger washout limits the viable range of $\widetilde{m}_1$.
In Tab.~\ref{tab:mGbounds} we present the smallest gravitino masses
that are accessible for certain representative values of $\widetilde{m}_1$.
As in Fig.~\ref{fig:mGboundsM1} the contour lines of constant $M_1$ slightly
fall off with decreasing
$\widetilde{m}_1$, we find the lowest bounds on $m_{\widetilde{G}}$
at $\widetilde{m}_1 = 10^{-5}\,\textrm{eV}$.
For weak washout higher $B-L$ breaking scales lead to
tighter bounds on $m_{\widetilde{G}}$, just as it is the case for
the neutrino mass $M_1$ (cf. Fig.\,\ref{fig:BLasym}).
In the strong washout regime we encounter the opposite behavior.
Here, the contour line corresponding to $\eta_B = \eta_B^{\textrm{obs}}$, which separates
the allowed and excluded regions in parameter space, steeply rises.
In Section~\ref{subsec:TR} we argued that the larger the value of $v_{B-L}$
the later this rise sets in when increasing $\widetilde{m}_1$
(cf. Eq.~\eqref{eq:GWBoltz}).
Therefore, the tightest bounds on $m_{\widetilde{G}}$ are now obtained for low $B-L$ breaking scales.

\begin{table}
\begin{center}
\begin{tabular}{ccccccc}
Panel & $v_{B-L}$ [GeV] / $\widetilde{m}_1$ [eV] & $10^{-5}$ &
$10^{-4}$ & $10^{-3}$ & $10^{-2}$ & $10^{-1}$\ \\ 
\hline\hline
1 & $3.4 \times 10^{12}$ & $180\,\textrm{MeV}$ & $360\,\textrm{MeV}$ &
$870\,\textrm{MeV}$ & --- & --- \\ 
2 & $5.8 \times 10^{13}$ & $570\,\textrm{MeV}$ & $1.2\,\textrm{GeV}$ & $2.5\,\textrm{GeV}$ & $70\,\textrm{GeV}$ & --- \\
3 & $1.0 \times 10^{15}$ & $1.9\,\textrm{GeV}$ & $3.7\,\textrm{GeV}$ & $6.2\,\textrm{GeV}$ & $9.2\,\textrm{GeV}$ & --- \\
\hline\hline
\end{tabular}
\caption{Lower bounds on the gravitino mass according to Figs.~\ref{fig:mGboundsM1}
and \ref{fig:mGboundsT} for the three different choices of $v_{B-L}$
listed in Eq.~\eqref{eq:vBLvalues} and five different values of $\widetilde{m}_1$.
A dash (---) indicates that leptogenesis is not efficient enough to produce the observed
baryon asymmetry as long as the requirement of gravitino dark matter is kept.}
\label{tab:mGbounds}
\end{center}
\end{table}

We also note that for $\widetilde{m}_1 = 0.1\,\textrm{eV}$ it is not
possible to produce the observed baryon asymmetry while sticking to the assumption
of gravitino dark matter, independent of the value chosen for $v_{B-L}$.
As $\widetilde{m}_1$ is bounded from below by $m_1$, the smallest eigenvalue
of the standard model neutrino mass matrix, this observation opens up the possibility
of falsifying our proposed scenario in future neutrino experiments.
The measurement of a light neutrino mass of $0.1\,\textrm{eV}$,
combined with the known differences of the light neutrino masses squared, would imply
that $\widetilde{m}_1 \gtrsim 0.1\,\textrm{eV}$, thereby ruling out
our mechanism of entropy production.
Likewise, any lower limit on the absolute neutrino mass scale coming from,
\textit{e.g.} cosmological observations would restrict the allowed
range for the gravitino mass.
A determination of the gravitino mass on the basis of cosmic gamma-ray
observations or decays of the next-to-lightest-superparticle (NLSP) in collider
experiments could, in turn, constrain the neutrino mass spectrum.

In standard thermal leptogenesis the reheating temperature has to be
at least $T_{RH} \gtrsim 10^9$\,GeV, independent of the
initial conditions, to guarantee a successful generation of
the baryon asymmetry \cite{bdp04}.
Together with the lower bound on the gluino mass imposed by collider
searches, this constrains the gravitino mass to lie in the range
$m_{\widetilde G} \simeq 10 \div 100$\,GeV in order to be compatible with
the observed dark matter abundance.
By contrast, the present scenario allows for a much broader range of
gravitino masses since the reheating temperature can be significantly lower
than in the case of thermal leptogenesis.
As apparent in Fig.~\ref{fig:mGboundsT}, $T_{RH}$ can decrease down
to values of $\mathcal{O}(10^7)\,\textrm{GeV}$ if $\widetilde{m}_1$
and $M_1$ are chosen such that the nonthermal neutrinos decay extremely slowly.
We note that for such reheating temperature 
production of gravitinos from inflaton decay is usually negligible \cite{nty10}.
This paves the way for gravitino masses as small as $200\,\textrm{MeV}$
(cf. Tab.~\ref{tab:mGbounds}).
On the other hand, as our scenario can as well accomodate
neutrino masses $M_1$ of order
$10^{10} \div 10^{11}$\,GeV,
the gravitino can also be almost as heavy as the gluino,
$m_{\widetilde{G}} \simeq \textrm{few} \times 100\,\textrm{GeV}$.
Note, however, that requirements such as gaugino mass
unification will further constrain the superparticle spectrum.

\medskip
Finally, we would like to point out that low gravitino masses
have interesting consequences for the decay of the
next-to-lightest superparticle (NLSP).
If $R$-parity is conserved, the lower bound on the mass of
stable gravitinos from standard thermal leptogenesis, $m_{\widetilde
  G} \gtrsim 10$\,GeV, implies a long NLSP lifetime which could
jeopardize the success of primordial nucleosynthesis (BBN).

A study of general neutralino NLSPs with gravitino LSP has been
performed in \cite{Covi:2009bk}.
In this work, lower mass bounds for different NLSP types
have been extracted from the bounds on the decay of heavy neutral particles
during BBN for given gravitino masses.
Assuming a gravitino of $10\,\textrm{GeV}$, it turns out that
the predictions of primordial nucleosynthesis are not affected
for pure bino, pure wino and mixed gaugino-Higgsino neutralino NLSPs
that are heavier than 3, 0.8 and 1\,TeV, respectively.
These bounds are now significantly softened for the light
gravitino masses which can occur in our scenario:
For $m_{\widetilde G} = 200$\,MeV, pure bino,
pure wino and mixed gaugino-Higgsino neutralino NLSPs as light as
$800, 150$ and $200$\,GeV, respectively, are in agreement with the BBN bounds.
These lower masses are interesting for two reasons: First,
they allow a smaller scale of supersymmetry breaking and
second, they can be probed more easily at the LHC.
For a charged NLSP like a scalar $\tau$-lepton, its lifetime has to be
sufficiently short, $\tau _{\textrm{NSLP}} \lesssim 10^3-10^4$\,s
\cite{Pospelov:2006sc,Cyburt:2006uv}, which typically requires
$m_{\widetilde G} < 1$\,GeV for $m_{\textrm{NLSP}} = {\cal O}(100)$\,GeV.

Remarkably, the small gravitino masses required from such constraints
on NLSP decays can be accommodated in our scenario.
We thus emphasize that reheating through the decays of
heavy neutrinos provides a way to alleviate the existing
tension between the generation of the baryon asymmetry, stable gravitino dark
matter and BBN constraints.
\section{Summary and outlook}

We have studied the production of the entropy of the hot early universe
through the decays of heavy Majorana neutrinos. As an example, we considered
the decay of a false vacuum with unbroken $B-L$ symmetry.
Tachyonic preheating leads to a state whose energy density is dominated by
nonrelativistic $S$ bosons, the Higgs boson associated with spontaneous
$B-L$ breaking, with a subdominant admixture of radiation. Subsequent
production of the lightest heavy Majorana neutrinos $N_1$ from $S$ decays and 
from the thermal bath leads to a phase whose energy density is dominated
by $N_1$ neutrinos. Their decay then produces all entropy of the hot thermal
universe, together with the baryon asymmetry via a mixture of nonthermal and
thermal leptogenesis. Simultaneously, thermal production of gravitinos accounts
for the observed dark matter.

We have studied the time evolution of this system by means of a set of 
Boltzmann equations for distribution functions which take into account the 
differences between thermally and nonthermally produced $N_1$ neutrinos.
Details of the initial state are rather unimportant for the final result.
What matters is the existence of a phase dominated by $N_1$ neutrinos.
Their lifetime determines the reheating temperature that is crucial for the
final gravitino abundance and hence for the produced amount of dark matter.

Our analysis has been based on a flavour model that allows to vary the 
key parameters for leptogenesis, the neutrino masses $M_1$ and 
$\widetilde{m}_1$, over a wide range, consistent with the masses and flavour
mixings of charged leptons and neutrinos. The final baryon asymmetry and the 
dark matter abundance
have been calculated in terms of several parameters of the Lagrangian, 
independent of initial conditions: the scale $v_{B-L}$ of $B-L$ breaking,
the heavy Majorana neutrino mass $M_1$, the effective light neutrino mass
$\widetilde{m}_1$, the gravitino mass $m_{\tilde{G}}$ and the gluino mass 
$m_{\tilde{g}}$. For generalisations of the flavour model the produced
matter-antimatter asymmetry and dark matter can depend on further parameters.

Particularly interesting is the resulting relation between the lightest 
neutrino mass and the gravitino mass. For instance, for a typical gluino
mass of $800~\mathrm{GeV}$ and a light neutrino mass of $10^{-5}~\mathrm{eV}$ 
the gravitino mass can be as small as $200~\mathrm{MeV}$,
whereas a lower neutrino mass bound of $0.01~\mathrm{eV}$ implies a lower 
bound of $9~\mathrm{GeV}$ on the gravitino mass. The measurement of a light 
neutrino mass of $0.1~\mathrm{eV}$ would falsify the proposed mechanism of 
entropy production. These results provide an important connection between 
collider searches for superparticles and neutrino mass determinations in 
laboratory experiments and by cosmological observations. Measurements of
the absolute neutrino mass scale and superparticle masses consistent with
our predictions would provide important indirect evidence for the origin
of entropy, matter and dark matter and for the maximal temperature of the
hot thermal universe. 

We are planning to extend our analysis in several directions: Boltzmann
equations for the superpartners of $S$ bosons and heavy Majorana neutrinos have
to be incorporated in a fully consistent calculation of baryon asymmetry and
gravitino dark matter. Further important questions concern the connection
with inflation and the possible relation between $B-L$ breaking and
supersymmetry breaking. \\

\vspace{5mm}
\noindent
{\bf\large Acknowledements}\\
The authors thank L.~Covi, V.~Domcke and J.~Hasenkamp 
for helpful discussions.
This work has been supported by the German Science Foundation (DFG) within 
the Collaborative Research Center 676 ``Particles, Strings and the Early
Universe''.

\appendix
\numberwithin{equation}{section}

\section{Conventions for the Boltzmann equations}
\label{appendix:BE}

The Boltzmann equation for a particle species $\psi$ describes the
time evolution of the distribution function
$f_\psi\left(t,x^\mu,p^\mu\right)$ in the one-particle phase space 
$\Phi_\psi$ \cite{kt90}, where $f_\psi$ is defined such that $f_\psi d\Phi_\psi$
gives the average number of $\psi$ particles in the phase space volume
$d\Phi_\psi$ at time $t$.
Imposing homogeneity and isotropy of spacetime, $f_\psi$ only
depends on the absolute value $p$ of the 3-momentum
$\vec{p}$ as well as on time $t$.
The Boltzmann equation for $\psi$ particles in the
Friedmann-Lema\^itre framework then reads
\begin{align}
\hat{\mathcal{L}} f_\psi(t,p) =
\left(\frac{\partial}{\partial t} - H p \frac{\partial}{\partial p}\right)
f_\psi(t,p) =  \mathcal{C}_\psi\,,
\label{eq:BEpsi}
\end{align}
where with $\hat{\mathcal{L}}$ we denote the Liouville operator, and
with $\mathcal{C}_\psi$ the collision operator.
The latter keeps track of changes in $f_\psi$ due to inelastic
interactions, and may be decomposed into contributions from decays and
scatterings,
\begin{align}
\mathcal{C}_\psi = \sum_{ij...}
C_\psi(\psi \leftrightarrow ij...) +
\sum_{a}\sum_{ij...}
C_\psi(\psi a \leftrightarrow ij...) + ... \,.
\end{align}

The operators $C_\psi$ are obtained from quantum mechanical
transition probabilities integrated over the multi-particle phase
space
\begin{align}
C_\psi(\psi ab... \leftrightarrow ij...) = & \:
\frac{1}{2 g_\psi E_\psi} \int d\Pi\left(\psi|a,b,...;i,j,...\right) \left(2\pi\right)^4
\delta^{(4)}\left(\textstyle \sum p_{\textrm{out}} - \textstyle \sum p_{\textrm{in}}\right) \label{eq:Cpsi}\\
&\times  \: \big[f_i f_j...\left(1\pm f_\psi\right)\left(1\pm f_a\right)\left(1\pm f_b\right) ...
\left|\mathcal{M}\left(i j ... \rightarrow \psi a b ...\right)\right|^2 \nonumber\\
& \: - f_\psi f_a f_b ...\left(1\pm f_i\right)\left(1\pm f_j\right) ...
\left|\mathcal{M}\left(\psi a b ... \rightarrow i j ...\right)\right|^2\big]\,, \nonumber
\end{align}
where $g_\psi$ is the number of internal degrees of freedom of $\psi$ and
$d\Pi$ subsumes all Lorentz invariant momentum space elements
$d\tilde{p} = \left(2\pi\right)^{-3} d^3 p/2E$ along with a statistical factor $S$
that prevents double counting in the case of identical particles
\begin{align}
d\Pi(\psi|a,b,...;i,j,...) = S(\psi,a,b,...;i,j,...)
d\tilde{p}_a d\tilde{p}_b ...  d\tilde{p}_i  d\tilde{p}_j ...\,.
\end{align}
The amplitudes squared $\left|\mathcal{M}\right|^2$ are
understood to be summed over all internal degrees of freedom. 
Since they are expected to yield only small corrections
\cite{HahnWoernle:2009qn}, the Bose enhancement $(1+f)$ and Pauli
blocking $(1-f)$ quantum statistical factors related to boson and
fermion production respectively, are neglected in this work.
Their influence may partly be canceled by other quantum corrections,
like off-shell effects \cite{Anisimov:2010dk}.
Additionally, the $C_\psi$ operators can be split into direct and
inverse processes
\begin{align}
C_\psi(\psi a b... \leftrightarrow ij...) =
C_\psi(ij... \rightarrow \psi a b...) - C_\psi(\psi a b... \rightarrow i j ...)\,.
\end{align}

\medskip
If the $\psi$ particles are in kinetic equilibrium, the integration of
Eq.\,\eqref{eq:BEpsi} over the $\psi$ phase space leads to a Boltzmann
equation for the $\psi$ number density $n_\psi$
\begin{align}
\dot{n}_\psi + 3 H n_\psi = \sum_{ij...}
\gamma(\psi \leftrightarrow ij...) +
\sum_{a}\sum_{ij...}
\gamma(\psi a \leftrightarrow ij...) + ... \,,
\label{eq:BEnpsi}
\end{align}
where $n_\psi$ and the interaction densities $\gamma$
are defined as
\begin{align}
n_\psi(t) = & \: \frac{g_\psi}{\left(2\pi\right)^3} \int
d^3 p \; f_\psi(t,p) \,,
\label{eq:numdendef}\\
\gamma(\psi ab... \leftrightarrow ij...) = & \:
\frac{g_\psi}{(2\pi)^3} \int d^3 p \; C_\psi(\psi
  ab... \leftrightarrow ij...)\,.
\label{eq:gammadef}
\end{align}
The Boltzmann equation Eq.\,\eqref{eq:BEnpsi} can alternatively be
written as an equation for the comoving number
density $N_\psi = a^3 n_\psi$ as a function of the scale factor $a$
\begin{align}
a H \frac{d}{da} N_\psi =  a^3 \left[\sum_{ij...}
\gamma(\psi \leftrightarrow ij...) +
\sum_{a}\sum_{ij...}
\gamma(\psi a \leftrightarrow ij...) + ...\right]\,.
\end{align}

\section{Phase space distribution of thermal neutrinos}
\label{subsec:PSDFN1T}
\begin{figure}
\begin{center}
\includegraphics[width=12cm]{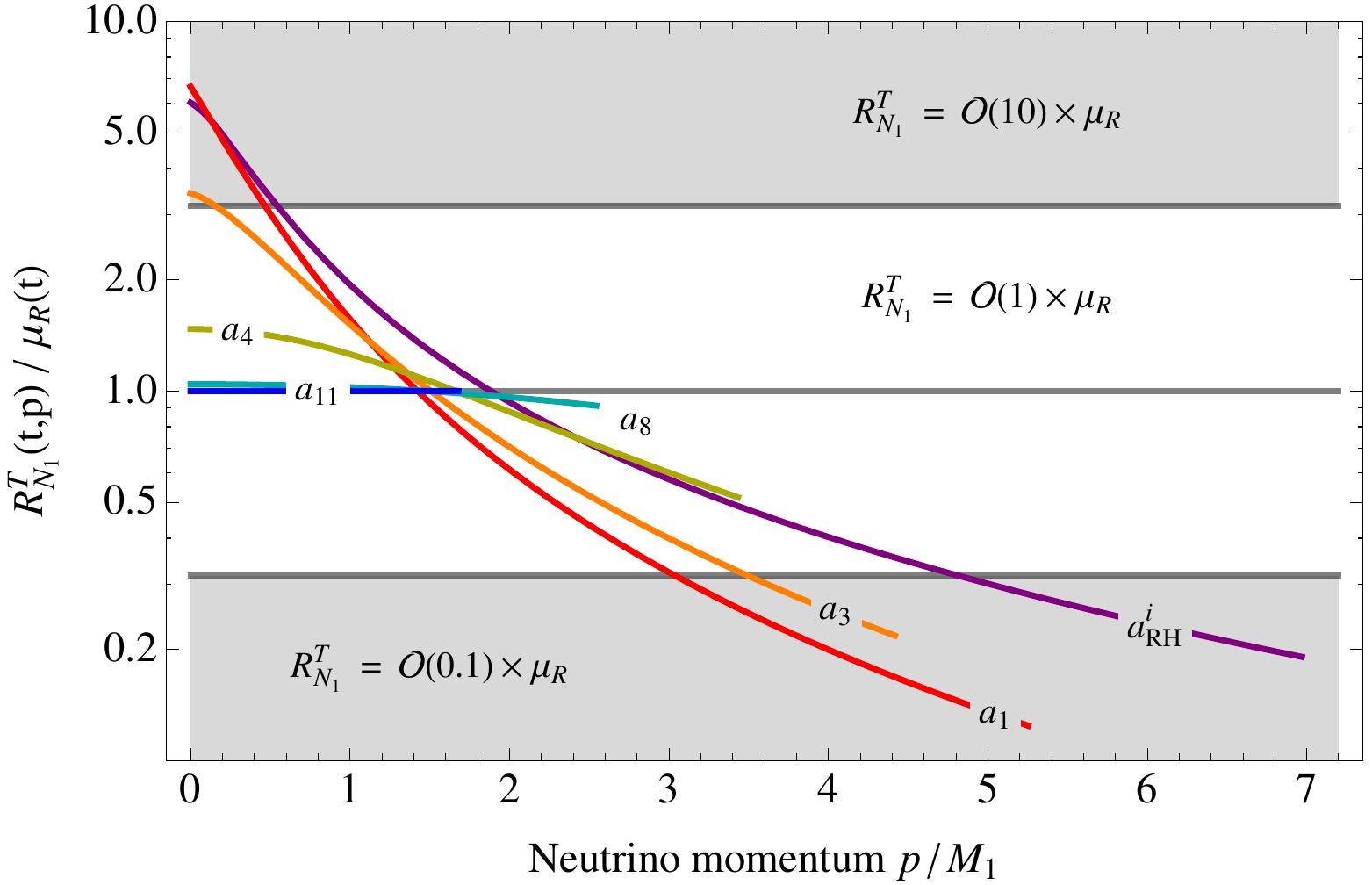}
\caption{Ratio $R_{N_1}^T = f_{N_1}^T/f_{N_1}^\equi$ normalized to its
mean value $\mu_R$ as a function of the neutrino momentum $p$ at
different times, cf. Eqs.~\eqref{eq:RN1TDef} and \eqref{eq:musigmaR}
as well as Tab.~\ref{tab:kinequ}.
The input parameters $v_{B-L}$, $\widetilde{m}_1$ and $M_1$ are chosen
as in Section~\ref{sec:example} (cf. Tab.~\ref{tab:parapoints}).
The respective values of the scale factor are discussed in
Section~\ref{subsec:Asymm}.
The white and gray bands separate the different orders of magnitude.
\label{fig:kinequ}}
\end{center}
\end{figure}
When deriving the Boltzmann equation \eqref{eq:BEN1T} for thermal neutrinos
in Section~\ref{subsubsec:N1neutr}, we asserted that these are
approximately in kinetic equilibrium (cf. Eq.~\eqref{eq:fN1Tapprox}).
Given a numerical solution for $T(a)$ for a specific choice
of input parameters $v_{B-L}$, $\widetilde{m}_1$ and $M_1$,
we can check the self-consistency of this approach
by comparing our approximate distribution function proportional
to $f_{N_1}^\equi$ with the exact expression in Eq.~\eqref{eq:fN1Tex}.
In this appendix we perform such a comparison for
the exemplary parameter point 
discussed in Section~\ref{sec:example} (cf. Tab.~\ref{tab:parapoints}).
To begin with, we introduce the following ratio
\begin{align}
R_{N_1}^T\left(t,p\right) =  &\: \frac{f_{N_1}^T(t,p)}{f_{N_1}^\equi(t,p)}\,, \nonumber\\=  &\:
\int\limits_{t_2}^t dt'
\exp\left(-M_1 \Gamma_{N_1}^0 \int\limits_{t'}^t dt'' E_{N_1}^{-1}(t'')\right)
\frac{M_1}{E_{N_1}(t')}\, \Gamma_{N_1}^0 \frac{f_{N_1}^\equi(t',p)}{f_{N_1}^\equi(t,p)} \,.
\label{eq:RN1T2}
\end{align}
with $f_{N_1}^T$ taken from Eq.~\eqref{eq:fN1Tex}, and determine its
momentum dependence at different times.
A momentum independence of $R_{N_1}^T$ would then reflect an
exact kinetic equilibrium.
In this case the equation $f_{N_1}^T = R_{N_1}^T f_{N_1}^\equi$ can
easily be integrated over phase space yielding
\begin{align}
R_{N_1}^T = R_{N_1}^T(t) \quad\Leftrightarrow\quad
N_{N_1}^T(t) = R_{N_1}^T(t) N_{N_1}^\equi(t) \quad\Leftrightarrow\quad
f_{N_1}^T(t,p) = \frac{N_{N_1}^T(t)}{N_{N_1}^\equi(t)} f_{N_1}^\equi(t,p) \,.
\label{eq:RN1TDef}
\end{align}
A convenient measure for the deviation from kinetic equilibrium at a given time
is the coefficient of variation $c_R = \sigma_R/\mu_R$, \textit{i.e.} the
standard deviation $\sigma_R$ of $R_{N_1}^T$ in relation to its
mean value $\mu_R$ with respect to an appropriate momentum interval $\Delta p$,
\begin{align}
\mu_R(t) = \left<R_{N_1}^T\right>_p\,,\quad
\sigma_R(t) = \left(\left<\left(R_{N_1}^T\right)^2\right>_p - \left<R_{N_1}^T\right>_p^2\right)^{1/2}
\label{eq:musigmaR}
\end{align}
where $\left<\cdot\right>_p$ is defined as
\begin{align}
\left<f\right>_p (t)= \frac{1}{\Delta p} \int_0^{\Delta p} dp \,f(t,p)\,,
\end{align}
and $\Delta p$, by convention, is always chosen as
\begin{align}
f_{N_1}^\equi\left(t,\Delta p\right) = 10^{-4} \times f_{N_1}^\equi\left(t,0\right)\,,
\label{eq:DeltapDef}
\end{align}
such that the relevant range of momenta is covered.

\medskip
We compute $R_{N_1}^T$, $\mu_R$, $\sigma_R$, $c_R$ and $\Delta p$ for
six representative values of the scale factor, and summarize the
corresponding results in Fig.~\ref{fig:kinequ} and Table~\ref{tab:kinequ}.
At early times small momenta are much more frequent than in
kinetic equilibrium and states with large momenta are underpopulated.
For extreme momenta, $R_{N_1}^T$ can become ten times as large or small as its
mean value $\mu_R$.
As time goes on, this tilt in $R_{N_1}^T$, however, disappears
and for $a\simeq 12000$, kinetic equilibrium is eventually reached.
On average $R_{N_1}^T$ deviates from $\mu_R$ not more than one
order of magnitude and, from this perspective, the approximation
of kinetic equilibrium may be regarded as justified.
The steady convergence to kinetic equilibrium is also reflected in the behaviour
of the coefficient of variation $c_R$
which starts out at a value of $c_R \sim \mathcal{O}(1)$ and decreases to
$c_R \sim \mathcal{O}(10^{-4})$.
For other choices of the model parameters we expect the $N_1^T$
phase space distribution to behave similarly.

Scatterings of the thermal neutrinos involving standard model
quark pairs such as  $N_1 \ell \leftrightarrow q \bar u$,
$N_1 \bar u \leftrightarrow \ell \bar q$ and $N_1 q\leftrightarrow \ell u$,
speed up the equilibration of the neutrino distribution function \cite{HahnWoernle:2009qn}.
This results in a larger abundance of thermal neutrinos at
high temperatures.
On the other hand, scatterings also tend to increase the efficiency of washout processes
such that, after all, their impact on the generated thermal asymmetry is negligible for our purposes.

There are two main reasons why $R_{N_1}^T$ is not flat from the
beginning:
The first is directly related to the momentum dependence
of the production and decay terms in the Boltzmann equation
\eqref{eq:BEN1T} for thermal neutrinos.
In both terms the effective rate $\Gamma_{N_1}^T$ comes weighted
with the inverse time dilatation factor $\langle M_1 / E_{N_1} \rangle$.
It is, thus, larger at smaller momenta which is why initially,
when $f_{N_1}^T \ll f_{N_1}^\equi$, neutrinos with smaller momenta are
overproduced.
Once $f_{N_1}^T$ has overshot $f_{N_1}^\equi$ the decay term dominates,
again preferably changing the abundance of low-momentum states.
This interplay between production and decay is balanced such
that $R_{N_1}^T$ is eventually flattened out.
A numerical investigation of the different factors in the integrand
of Eq.~\eqref{eq:RN1T2} confirms this simple argument.
The second reason is the intermediate stage of reheating
between the phases of adiabatic expansion.
Assuming an equilibrium distribution $f_{N_1}^\equi$
misconceives the evolution of the temperature
in the sense that higher temperatures and thus more high-momentum
neutrinos are expected at earlier times.
By contrast, the actual distribution $f_{N_1}^T$ takes
the temperature plateau into account and is, hence, aware of
the correspondingly less efficient production at high momenta.

\begin{table}
\begin{center}
\begin{tabular}{cccccc}
\# & $a$ & $\mu_R$ & $\sigma_R$ & $c_R$ & $\Delta p \left[M_1\right]$ \\
\hline\hline
1 &    $27$ & $8.4 \times 10^{-4}$ & $1.0 \times 10^{-3}$ &                $1.2$ &  $7.0$ \\
2 &   $210$ & $1.0 \times 10^{-2}$ & $1.4 \times 10^{-2}$ &                $1.3$ &  $5.3$ \\
3 &  $1500$ &               $0.29$ &               $0.25$ &               $0.87$ &  $4.4$ \\
4 &  $3500$ &               $0.75$ &               $0.24$ &               $0.32$ &  $3.4$ \\
5 &  $6500$ &               $1.0$  & $4.0 \times 10^{-2}$ & $3.9 \times 10^{-2}$ &  $2.5$ \\
6 & $12000$ &               $1.0$  & $4.0 \times 10^{-5}$ & $3.9 \times 10^{-5}$ &  $1.7$ \\
\hline\hline
\end{tabular}
\caption{Indicators for the deviation of the thermal neutrinos from
kinetic equilibrium at different times:
$\mu_R$, $\sigma_R$ and $\Delta p$ are introduced in Eqs.~\eqref{eq:musigmaR}
and \eqref{eq:DeltapDef}, $c_R$ is defined as $c_R = \sigma_R/\mu_R$.
The values of the scale factor correspond to $a_{RH}^i$, $a_1$, $a_3$,
$a_4$, $a_8$ and $a_{11}$, cf. Section~\ref{subsec:Asymm} and Fig.~\ref{fig:kinequ}.}
\label{tab:kinequ}
\end{center}
\end{table}
\section{Reheating temperature}
\label{appendix:TR}
The energy transfer to the thermal bath, \textit{i.e.} the
reheating of the universe, becomes fully efficient
when the nonthermally produced $N_1$ neutrinos decay into
standard model radiation.
This happens once the Hubble rate $H$ has dropped to the value
of the effective decay rate $\Gamma_{N_1}^S$ of the nonthermal $N_1$ neutrinos.
The temperature at this time, $t = t_{RH}$, defines the reheating temperature
$T_{RH}$
\begin{align}
\Gamma_{N_1}^S (t_{RH}) =
H (t_{RH})\,,\qquad
T_{RH} = T(t_{RH})\,.
\label{eq:TRHdef}
\end{align}
Notice that this coincides with the common definition
of $T_{RH}$ in scenarios in which the universe
is reheated through the decay of some species
with effective decay rate $\Gamma$.

The decay rate relevant to our scenario, $\Gamma_{N_1}^S$,
corresponds to the $T = 0$ neutrino decay width $\Gamma_{N_1}^0$
weighted with the average inverse time dilatation factor
for nonthermal neutrinos (cf. Section~\ref{subsubsec:N1neutr})
\begin{align}
\Gamma_{N_1}^S = \gamma_t^{-1} \Gamma_{N_1}^0 \,,\qquad
\gamma_t = \left<\frac{M_1}{E_{N_1}}\right>_S^{-1}\,, \qquad
\gamma = \gamma_t (t_{RH})\,.
\label{eq:GN1SGN10}
\end{align}
In the course of our numerical analysis Eqs.~\eqref{eq:TRHdef} and
\eqref{eq:GN1SGN10} are used to determine
the reheating temperature as a function of the model parameters,
$T_{RH} = T_{RH}(v_{B-L},M_1,\tilde{m}_1)$.
The result of this computation is presented in Fig.~\ref{fig:treheat}
in Section~\ref{sec:results}.
In this appendix we shall illustrate how one can understand the exact
outcome of the Boltzmann equations in terms of increasingly
accurate approximations (also shown in Fig.~\ref{fig:treheat}).

We begin with Eq.~\eqref{eq:TRHdef} and try to solve it
analytically for $T_{RH}$.
First, the Friedmann equation allows us to express
$H(t_{RH})$ through the total
energy density at time $t_{RH}$
\begin{align}
H^2(t_{RH})
= \left(\frac{\dot{a}}{a}\right)_{t_{RH}}^2
= \beta^2 \frac{8\pi}{3M_p^2} \rho_{\textrm{tot}}
(t_{RH})\,.
\label{eq:IIFriedEq}
\end{align}
Here $\beta$ is a correction factor that accounts for
the fact that we do not determine the scale factor $a$
dynamically but simply approximate it by means of
constant effective coefficients $\omega$ in the equation
of state.
This imprecision in $a$ is then transmitted to $H$ such that
it does not fulfill the Friedmann equation exactly.
Second, let us denote the fraction of the total energy
density that is stored in radiaton at time $t_{RH}$ by $\alpha^{-1}$.
With the aid of Eq.~\eqref{eq:rnRT} we may then write
\begin{align}
H^2(t_{RH})
= \alpha\beta^2 \frac{8\pi}{3M_p^2}
\frac{\pi^2}{30}\,g_{\star,\rho} \,T_{RH}^4\,.
\label{eq:HtRH}
\end{align}
Combining Eqs.~\eqref{eq:TRHdef}, \eqref{eq:GN1SGN10}
and \eqref{eq:HtRH} one finds
\begin{align}
T_{RH} = \alpha^{-1/4} \beta^{-1/2} \gamma^{-1/2} 
\left(\frac{90}{8\pi^3 g_{\star,\rho}}\right)^{1/4} \sqrt{\Gamma_{N_1}^0 M_p}\,.
\label{eq:TRH3}
\end{align}
By construction this formula yields the same results as
Eq.~\eqref{eq:TRHdef}.
Its main advantage over Eq.~\eqref{eq:TRHdef}, however,
is that it allows us to estimate $T_{RH}$ with varying precision depending
on how accurately the correction factors
$\alpha$, $\beta$ and $\gamma$ are taken into account.

\begin{figure}
\begin{center}
\includegraphics[width=16cm]{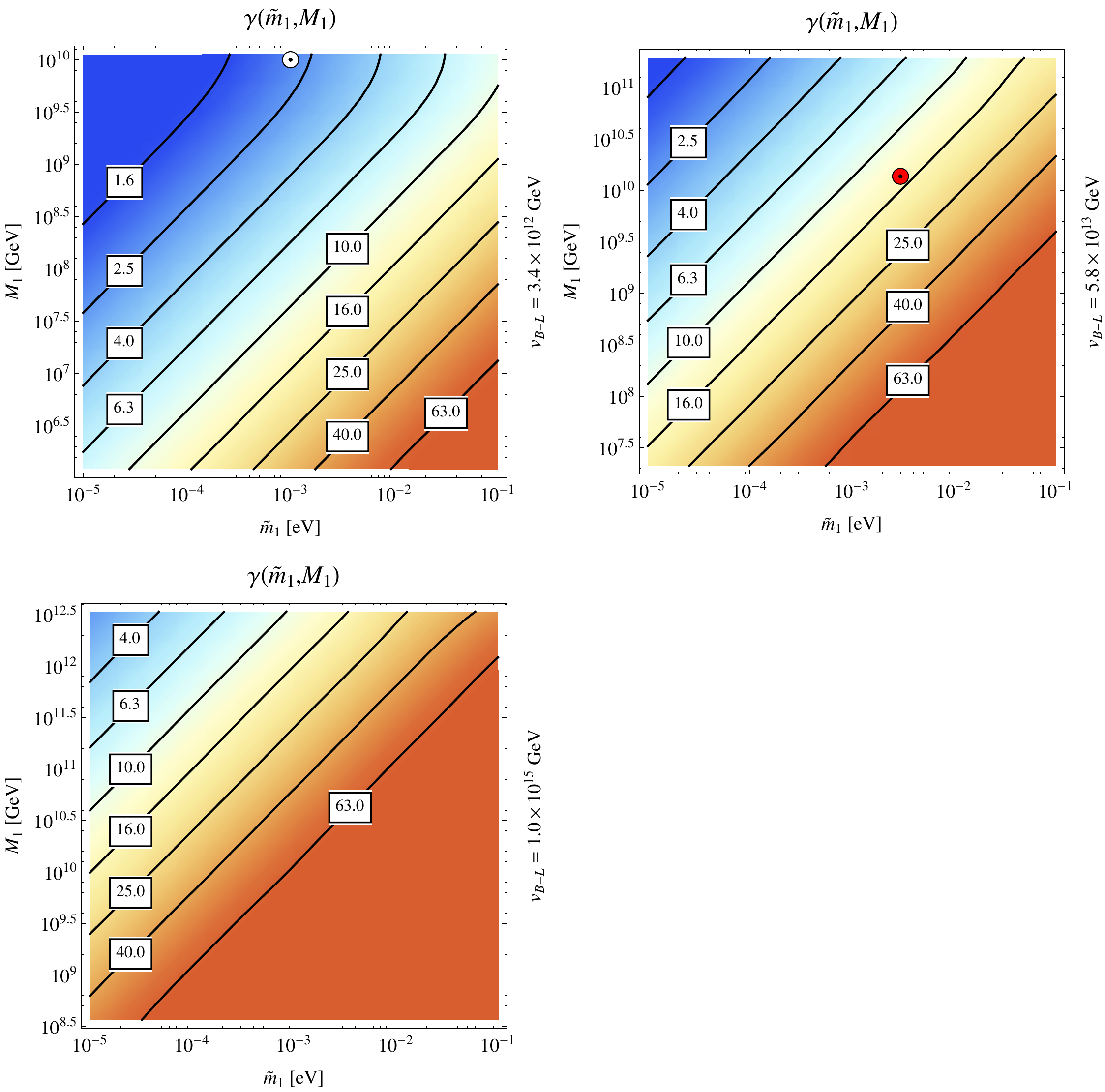}
\caption{Contour plots of the relativistic correction
factor $\gamma$ as a function of the parameters $\widetilde{m}_1$ and $M_1$
for the three different choices of $v_{B-L}$ (cf. Eq.\,\eqref{eq:vBLvalues}).
$\gamma^{-1}$ is defined as the average inverse time dilatation
factor for nonthermal $N_1$ neutrinos at $t = t_{RH}$, cf. Eq.~\eqref{eq:GN1SGN10}.
The background colours reflect the parameter dependence of
$\gamma$ as also indicated by the contour lines and labels.
In the reddish regions (large $\gamma$) the nonthermal neutrinos are relativistic
at $t = t_{RH}$, in the bluish regions (small $\gamma$) rather nonrelativistic.
In principal, $\gamma$ can take on values between 1 and
$\frac{1}{2}m_S / M_1 \simeq \frac{1}{2}\eta^{-2} \simeq 150$.
Its behavior in parameter space is mainly controlled by
$\Gamma_S^0 / \Gamma_{N_1}^0 \propto
\left(v_{B-L}^2\widetilde{m}_1\right)/\left(v_{EW}^2 M_1\right)$.
\label{fig:gammant}}
\end{center}
\end{figure}

When solving the Boltzmann equations numerically we have also
determined these factors along the way.
The dependence of the time dilatation factor $\gamma$
on the neutrino parameters is shown in Fig.~\ref{fig:gammant}.
As all nonthermal neutrinos are produced with initial energy
$\frac{1}{2}m_S \simeq \frac{1}{2}\eta^{-2}M_1 \simeq 150 M_1$
it is clear that $\gamma$ is bounded from above:
$1 \leq \gamma \lesssim 150$.
In practice, we find that $\gamma$ takes on values roughly between
1.1 and 88 entailing $\gamma^{-1/2}$ factors in Eq.~\eqref{eq:TRH3}
approximately between 0.95 and 0.11.
The general behaviour of $\gamma$ as a function
of the model parameters is mainly controlled by the ratio
of the $S$ and $N_1$ decay widths
\begin{align}
\gamma = \gamma(\Gamma_{N_1}^0 / \Gamma_S^0)\,, \qquad
\frac{\Gamma_{N_1}^0}{\Gamma_S^0} \propto
\frac{v_{B-L}^2\widetilde{m}_1}{v_{EW}^2 M_1}\,.
\end{align}
The larger $\Gamma_{N_1}^0$ compared to $\Gamma_S^0$ the less
contribute very long-lived nonthermal $N_1$ neutrinos to $\gamma$.
Most nonthermal neutrinos present at $t = t_{RH}$ were then
produced just shortly before and are hence relativistic.
On the other hand, if the $S$ bosons decay very fast,
$\Gamma_S^0 \gg \Gamma_{N_1}^0$, the nonthermal neutrinos
are mainly produced at the early stages of reheating and
$\gamma$ is rather dominated by elder, nonrelativistic
neutrinos.

The correction factor $\alpha$, the total-to-radiation energy
density ratio at $t=t_{RH}$, increases when going to larger
$v_{B-L}$ or $\widetilde{m}_1$ and decreases for smaller $M_1$.
It hence essentially shows the same trends in parameter space as
$\gamma$.
The physical reason for this is that large $\gamma$
implies a rather long-lasting stage of $N_1$ production
through $S$ decay which persists until shortly before
the $N_1$ neutrinos decay themselves.
Thereby, the $N_1$ abundance at $t = t_{RH}$ ends up being still
quite large.
On top of that, in the case of highly relativistic neutrinos,
the effective decay rate $\Gamma_{N_1}^S$ changes faster with
time than for nonrelativistic neutrinos.
Or, put into mathematical terms:
$\gamma_t$ (cf. Eq.~\eqref{eq:GN1SGN10})
is a monotically decreasing function of time which means that large
values of $\gamma_t$ entail large values of $\frac{d}{dt} \gamma_t^{-1}$.
Thus, if at $t = t_{RH}$ the decay rate $\Gamma_{N_1}^S$ equals the
Hubble rate $H$ and it is fast changing it was much smaller before.
Then not as many nonthermal neutrinos decay at times $t < t_{RH}$
and $\rho_{N_1}^S (t_{RH})$ may contribute much
more to $\rho_{\textrm{tot}}(t_{RH})$ than
$\rho_R(t_{RH})$.
Numerically, we find that $\alpha$ lies in the range between
3 and 4 for almost two thirds of the investigated parameter space.
In approximately 85 \% it is of $\mathcal{O}(10)$, in 10 \% of
$\mathcal{O}(100)$ and in 5 \% even larger up to $\alpha \simeq 7.8 \times 10^4$.
In Eq.~\eqref{eq:TRH3} the factor $\alpha^{-1/4}$ typically
has a size between $0.75$ and $0.49$.
But in extreme cases it can become as small as $\alpha \simeq 6.0 \times 10^{-2}$.

The correction factor in the Friedmann equation $\beta$
turns out to be quite constant in parameter space.
We find that it varies between 0.53 and 0.92.
Its standard deviation with respect to its mean value
is rather small: $\beta = 0.82 \pm 0.06$.
A $\beta$ factor smaller than one is the expected consequence
of our approach to the calculation of the scale factor:
After $t = t_S$ we assume pure radiaton domination,
$\omega = \rho/p = 1/3$, although for times $t \gtrsim t_S$
surely still some nonrelativistic $S$ bosons contribute
to the total energy density.
This leads to an overestimation of the speed at which
the Hubble rate decreases and in Eq.~\eqref{eq:IIFriedEq}
to a too small Hubble rate compared to the right-hand
side of the equation.
A correction factor of $\beta^{-1/2} \simeq 1.1$ in Eq.~\eqref{eq:TRH3}
is, however, almost insignificant.

In conclusion, we can say that $\gamma$ constitutes
the largest correction, followed by $\alpha$, the factor $\beta$
can almost be neglected.
This observation leads us to three increasingly accurate
approximations for the reheating temperature $T_{RH}^{(0)}$,
$T_{RH}^{(1)}$ and $T_{RH}^{(2)}$, all of which are also shown
in Fig.~\ref{fig:treheat}.
First, we set $\alpha = \beta = \gamma = 1$.
With $g_{\star,\rho} = 915/4$ and Eq.~\eqref{eq:Ndecayrate}, we then have
\begin{subequations}
\begin{align}
T_{RH}^{(0)} & \: = \left(\frac{90}{8\pi^3
    g_{\star,\rho}}\right)^{1/4} \sqrt{\Gamma_{N_1}^0 M_p}\\
& \: \simeq 0.2 \,\sqrt{\Gamma_{N_1}^0 M_p} \label{eq:TRH0}\\
& \: \simeq 8\times 10^9\,\textrm{GeV}
\left(\frac{\widetilde{m}_1}{10^{-3} \,\textrm{eV}}\right)^{1/2} 
\left(\frac{M_1}{10^{10}\,\textrm{GeV}}\right)\,.
\end{align}
\end{subequations}
This estimate can be improved by including the time
dilatation factor $\gamma$,
\begin{align}
T_{RH}^{(1)} = \gamma^{-1/2} \,T_{RH}^{(0)} 
\simeq 0.2 \,\sqrt{\Gamma_{N_1}^S M_p}\,.
\label{eq:TRH1}
\end{align}
Finally, we drop the assumption that at $t = t_{RH}$ the
entire energy resides in radiation,
\begin{align}
T_{RH}^{(2)} = \alpha^{-1/4}\gamma^{-1/2} T_{RH}^{(0)} 
\simeq \alpha^{-1/4} \, 0.2 \,\sqrt{\Gamma_{N_1}^S M_p}\,.
\label{eq:TRH2}
\end{align}
The remaining difference between the outcome of the Boltzmann
equations and $T_{RH}^{(2)}$ is then quantified by $\beta$
\begin{align}
T_{RH} = & \: \beta^{-1/2} \,T_{RH}^{(2)}\,, \\
\quad\Rightarrow\quad
\log_{10} T_{RH} = & \: \log_{10} T_{RH}^{(2)} + \Delta \log_{10} T_{RH}\,,
\qquad \Delta \log_{10} T_{RH} \simeq 0.04\,.
\label{eq:TRH23}
\end{align}

To conclude, let us apply the above formul\ae\, to the specific parameter
example which we discussed in Section~\ref{sec:example}.
For the parameter values listed in Tab.~\ref{tab:parapoints}, solving the
Boltzmann equations leads to a reheating temperature of $T_{RH} = 4.1\times 10^9\,\textrm{GeV}$
(cf. Eq.~\eqref{eq:TRHres}).
This result can be compared to the three estimates
introduced in this appendix.
For the three correction factors $\alpha$, $\beta$ and $\gamma$ we obtain
\begin{align}
\alpha \simeq 3.2\,, \qquad \beta \simeq 0.84\,, \qquad \gamma \simeq 14\,.
\label{eq:abgRes}
\end{align}
At $t = t_{RH}$ there is, hence, still roughly twice as much energy in
nonthermal neutrinos as in radiation.
Equality in terms of the energy content is not reached before $a \simeq 4000$
in our parameter example.
Moreover, with a typical energy $E_{N_1} \sim\mathcal{O}(10) M_1$
the nonthermal neutrinos are clearly still relativistic during reheating.
According to Eq.~\eqref{eq:abgRes} the estimates $T_{RH}^{(0)}$, $T_{RH}^{(1)}$
and $T_{RH}^{(2)}$ yield reheating temperatures of
\begin{align}
T_{RH}^{(0)} \simeq 1.9 \times 10^{10} \,\textrm{GeV}\,, \qquad
T_{RH}^{(1)} \simeq 5.0 \times 10^{9} \,\textrm{GeV}\,, \qquad
T_{RH}^{(2)} \simeq 3.7 \times 10^{9} \,\textrm{GeV}\,.
\end{align}
Notice that $T_{RH}^{(2)}$ multiplied by $\beta^{-1/2}$ reproduces again the
numerical result for the reheating temperature $T_{RH}$ in Eq.~\eqref{eq:TRHres}.
We conclude that the most naive estimate $T_{RH}^{(0)}$ is off the actual
value by roughly an order of magnitude.
$T_{RH}^{(1)}$ and $T_{RH}^{(2)}$ respectively deviate from $T_{RH}$ by $10 \%$
and $20\%$.
\section{Semi-analytic reconstructions}

\label{appendix:RC}

Our study of the parameter space in Section~\ref{sec:results}
allowed us to determine the $N_1$ neutrino mass $M_1$
and the reheating temperature $T_{RH}$ as functions of
$\widetilde{m}_1$ and $m_{\widetilde{G}}$ such that the
gravitino abundance always has the right size to account
for dark matter (cf. Figs.~\ref{fig:mGboundsM1} and
\ref{fig:mGboundsT}).
In this appendix we now attempt to reconstruct these
results by means of simple analytic expressions
and with the aid of our numerical findings for $\eta_B$
and $T_{RH}$.

\begin{figure}
\begin{center}
\includegraphics[width=12cm]{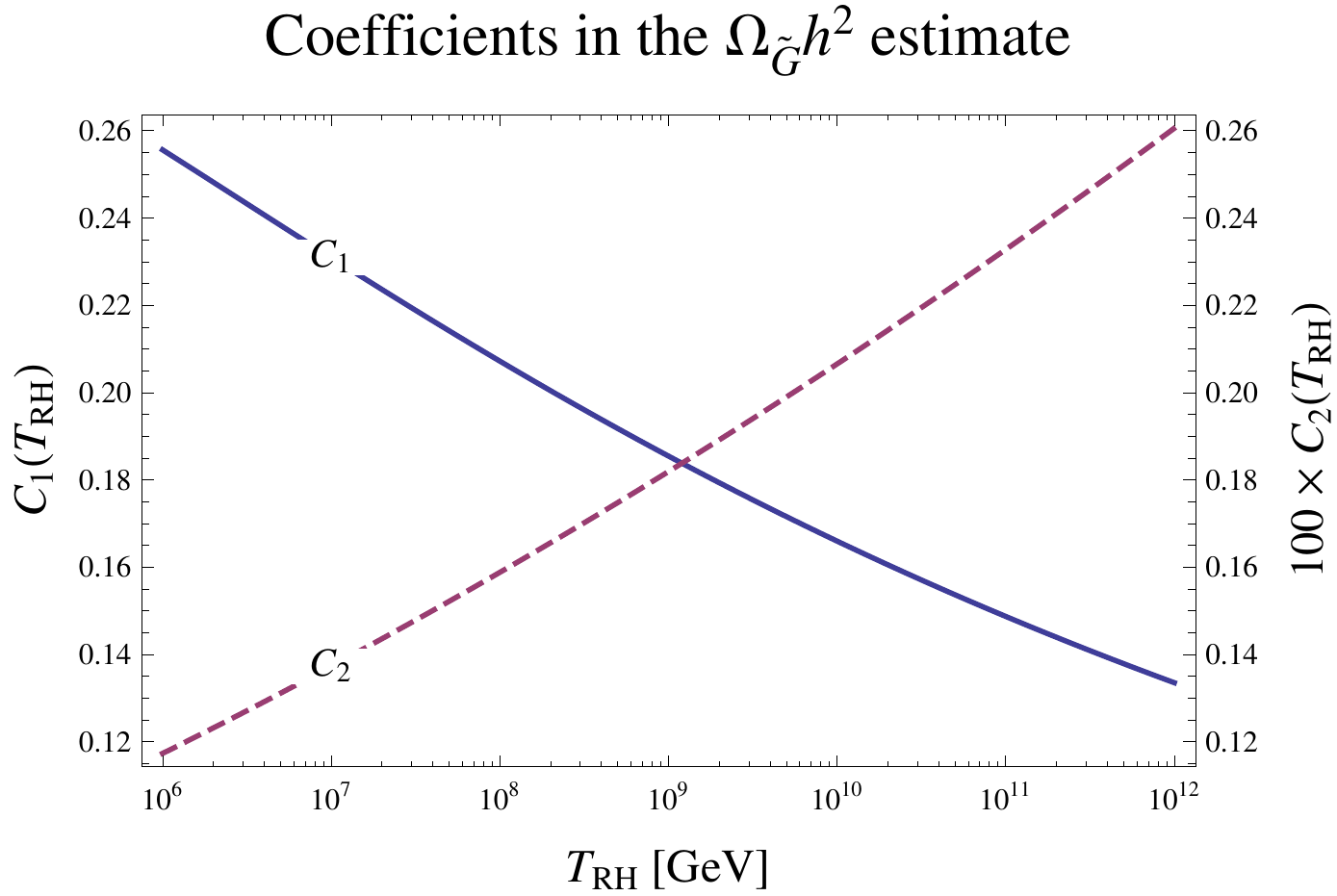}
\caption{Dependence of the coefficients $C_1$ and
$C_2$ in Eq.~\eqref{eq:OmegaGRes2} on the reheating
temperature $T_{RH}$.
\label{fig:C1C2coefficients}}
\vspace{5mm}
\includegraphics[width=12cm]{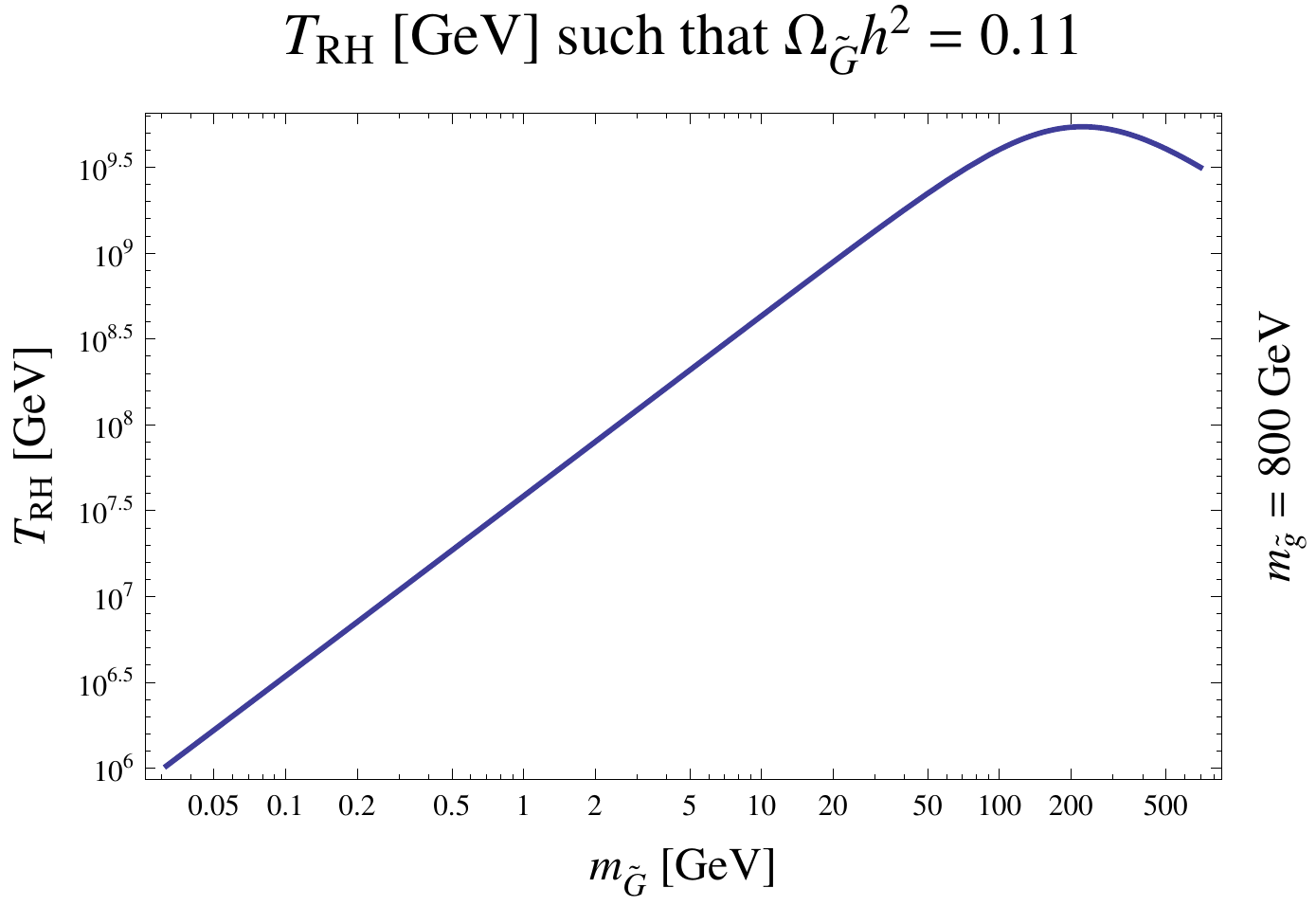}
\caption{Analytic estimate for the reheating temperature $T_{RH}$
as implicitly defined by Eq.~\eqref{eq:OmegaGRes2} with $\varepsilon = 1.32$
for varying gravitino mass $m_{\widetilde{G}}$ and fixed gluino mass,
$m_{\tilde{g}} = 800\,\textrm{GeV}$.
\label{fig:TRHrecon}}
\end{center}
\end{figure}

As gravitinos are nonrelativistic, their present
contribution to the energy density of the universe is given as
\begin{align}
\Omega_{\widetilde{G}} h^2 =
\Omega_{\widetilde{G}} h^2
(v_{B-L},M_1,\widetilde{m}_1,m_{\widetilde{G}},m_{\tilde{g}})
= m_{\widetilde{G}} \,\eta_{\widetilde{G}} \,n_\gamma^0 \,h^2 / \rho_c\,,
\qquad 
\eta_{\widetilde{G}} = n_{\widetilde{G}}^0 / n_\gamma^0\,.
\label{eq:OmegaG}
\end{align}
In order to relate the gravitino-to-photon ratio $\eta_{\widetilde{G}}$
to the corresponding number densities during reheating
we make two simplifying assumptions.
First, we say that after $t = t_{RH}$ the entropy of
the thermal bath is not increased much further which
leads us to
\begin{align}
\quad n_{\gamma}^0 = \delta_1  \left(\frac{a\left(t_{RH}\right)}{a_0}\right)^3
\frac{g_{\star,s}}{g_{\star,s}^0}\, n_\gamma(t_{RH})\,.
\label{eq:ngamma0}
\end{align}
Second, we assume that at $t = t_{RH}$ the gravitino production
becomes inefficient such that at later times not many
further gravitinos are produced,
\begin{align}
n_{\widetilde{G}}^0 = \delta_2  \left(\frac{a\left(t_{RH}\right)}{a_0}\right)^3
n_{\widetilde{G}}(t_{RH})\,.
\label{eq:ngravi0}
\end{align}
Meanwhile, this second assumption also implies that at $t = t_{RH}$
the gravitino production rate $\gamma_{\widetilde{G}}$ is of the
same order as the Hubble rate $H$
\begin{align}
\frac{\gamma_{\widetilde{G}}(t_{RH})}{n_{\widetilde{G}}(t_{RH})}
= \delta_3^{-1}   H(t_{RH}) \quad\Leftrightarrow\quad
n_{\widetilde{G}}(t_{RH}) = \delta_3 \,
\frac{\gamma_{\widetilde{G}}(t_{RH})}{H(t_{RH})}\,.
\label{eq:ngraviRH}
\end{align}
The three correction factors $\delta_1 \gtrsim 1$,
$\delta_2 \gtrsim 1$ and $\delta_3 \sim \mathcal{O}(1)$,
introduced in Eqs.~\eqref{eq:ngamma0}, \eqref{eq:ngravi0}
and \eqref{eq:ngraviRH}, respectively, quantify the deviations
of the actual values of $n_\gamma^0$, $n_{\widetilde{G}}^0$ and
$n_{\widetilde{G}}(t_{RH})$ from our approximations.
Combining them in one factor $\delta = \delta_2 \delta_3 / \delta_1$
we may write for $\eta_{\widetilde{G}}$
\begin{align}
\eta_{\widetilde{G}} = \delta \, \frac{g_{\star,s}^0}{g_{\star,s}}
\frac{\gamma_{\widetilde{G}}(t_{RH})}
{n_\gamma(t_{RH})H(t_{RH})}\,,
\label{eq:etaG}
\end{align}
where $n_\gamma(t_{RH})$, $\gamma_{\widetilde{G}}(t_{RH})$
and $H(t_{RH})$ directly follow from
Eqs.~\eqref{eq:rnRT}, \eqref{eq:gammaGT} and \eqref{eq:HtRH}.
Inserting Eq.~\eqref{eq:etaG} back into Eq.~\eqref{eq:OmegaG}
we find for $\Omega_{\widetilde{G}}h^2$
\begin{align}
\Omega_{\widetilde{G}}h^2 = \varepsilon 
f_{\widetilde{G}}(T_{RH}) \left(m_{\widetilde{G}} +
\frac{m_{\tilde{g}}^2 (T_{RH})}{3 m_{\widetilde{G}}}\right) T_{RH}\,,
\qquad \varepsilon = \alpha^{-1/2} \beta^{-1} \delta\,,
\label{eq:OmegaGRes}
\end{align}
where $f_{\widetilde{G}}(T_{RH})$ stands for
\begin{align}
f_{\widetilde{G}}(T_{RH}) = \frac{n_\gamma^0 h^2}{\rho_c}
\frac{g_{\star,s}^0}{g_{\star,s}} \left(\frac{90}{8 \pi^3 g_{\star,\rho}}\right)^{1/2}
\frac{54 \,g_s^2(T_{RH})}{g_\gamma M_p} \left[\ln\left(\frac{T_{RH}^2}
{m_g^2(T_{RH})}\right) + 0.8846 \right]\,.
\end{align}
Eq.~\eqref{eq:OmegaGRes} may conveniently be rewritten as
\begin{align}
\Omega_{\widetilde{G}}h^2 = \varepsilon 
C_1(T_{RH}) \left(\frac{T_{RH}}{10^9 \,\textrm{GeV}}\right)
\left[C_2(T_{RH})\bigg(\frac{m_{\widetilde{G}}}{10\,\textrm{GeV}}\bigg)
+ \left(\frac{10\,\textrm{GeV}}{m_{\widetilde{G}}}\right)
\left(\frac{m_{\tilde{g}}(\mu_0)}{800\,\textrm{GeV}_{}}\right)^2\right]
\label{eq:OmegaGRes2}
\end{align}
with $C_1$ and $C_2$ being defined as
\begin{align}
C_1(T_{RH}) = & \: \frac{\left(800\,\textrm{GeV}\right)^2}{10\,\textrm{GeV}}
\times 10^9\,\textrm{GeV} \times \frac{g_s^4(T_{RH})}{3 \,g_s^4(\mu_0)}
\times f_{\tilde{G}}(T_{RH})\,, \\
C_2(T_{RH}) = & \: \frac{10\,\textrm{GeV}}{\left(800\,\textrm{GeV}\right)^2}
\times 10\,\textrm{GeV} \times \frac{3\, g_s^4(\mu_0)}{g_s^4(T_{RH})}\,. \label{eq:C2def}
\end{align}
The dependence of $C_1$ and $C_2$ on the reheating
temperature is presented in Fig.~\ref{fig:C1C2coefficients}.
We find that $C_1 / C_2 \sim \mathcal{O}(100)$ which means
that for $m_{\tilde{g}} \gg m_{\widetilde{G}}$ the term linear
in $m_{\widetilde{G}}$ in Eq.~\eqref{eq:OmegaGRes2} can usually
be neglected.
Notice that doing so and setting $\varepsilon = 1$
turns Eq.~\eqref{eq:OmegaGRes2} into Eq.~\eqref{eq:GDMestimate} in the
introduction.

Confronting Eq.~\eqref{eq:OmegaGRes2} with our numerical
data shows that $\varepsilon$ usually differs from 1 and is
slightly parameter-dependent, preventing us from determining it
\textit{a priori}.
However, $\varepsilon$ can be determined \textit{a posteriori}.
In the region of parameter space in which $\Omega_{\widetilde{G}}h^2 =
\Omega_{DM}h^2$ we find that $\varepsilon = 1.20 \pm 0.24$.
Restricting ourselves further to parameter values for
which gravitino dark matter also is in accordance with successful
leptogenesis we obtain $\varepsilon = 1.32 \pm 0.15$.
In Eq.~\eqref{eq:GDMestimate} such a correction factor would be
reflected in a change of the numerical coefficient from 0.26 to 0.34.

\begin{figure}
\begin{center}
\includegraphics[width=16cm]{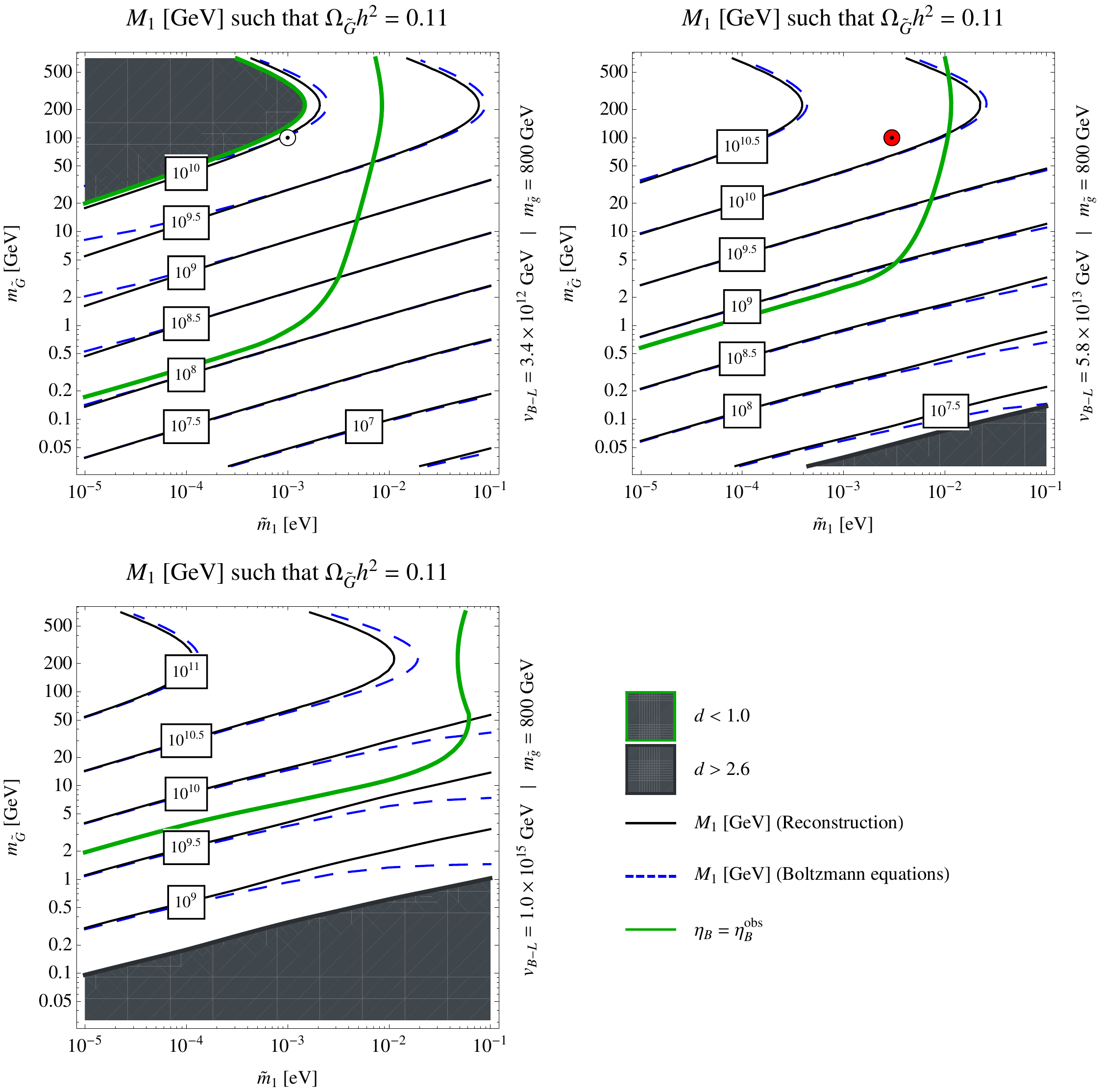}
\caption{Semi-analytical reconstruction of Fig.~\ref{fig:mGboundsM1}
 (solid lines)
on the basis of Eq.~\eqref{eq:OmegaGRes2} with $\varepsilon = 1.32$
and the numerical results for $T_{RH}$ and $\eta_B$.
For comparison also the $M_1$ contours deduced from the
Boltzmann equations (dashed lines) are shown.
They deviate from the reconstructed results as the correction
factor $\varepsilon = \alpha^{-1/2}\beta^{-1}\delta$ actually is
parameter-dependent.
\label{fig:mGrecon}}
\end{center}
\end{figure}

\begin{figure}
\begin{center}
\includegraphics[width=16cm]{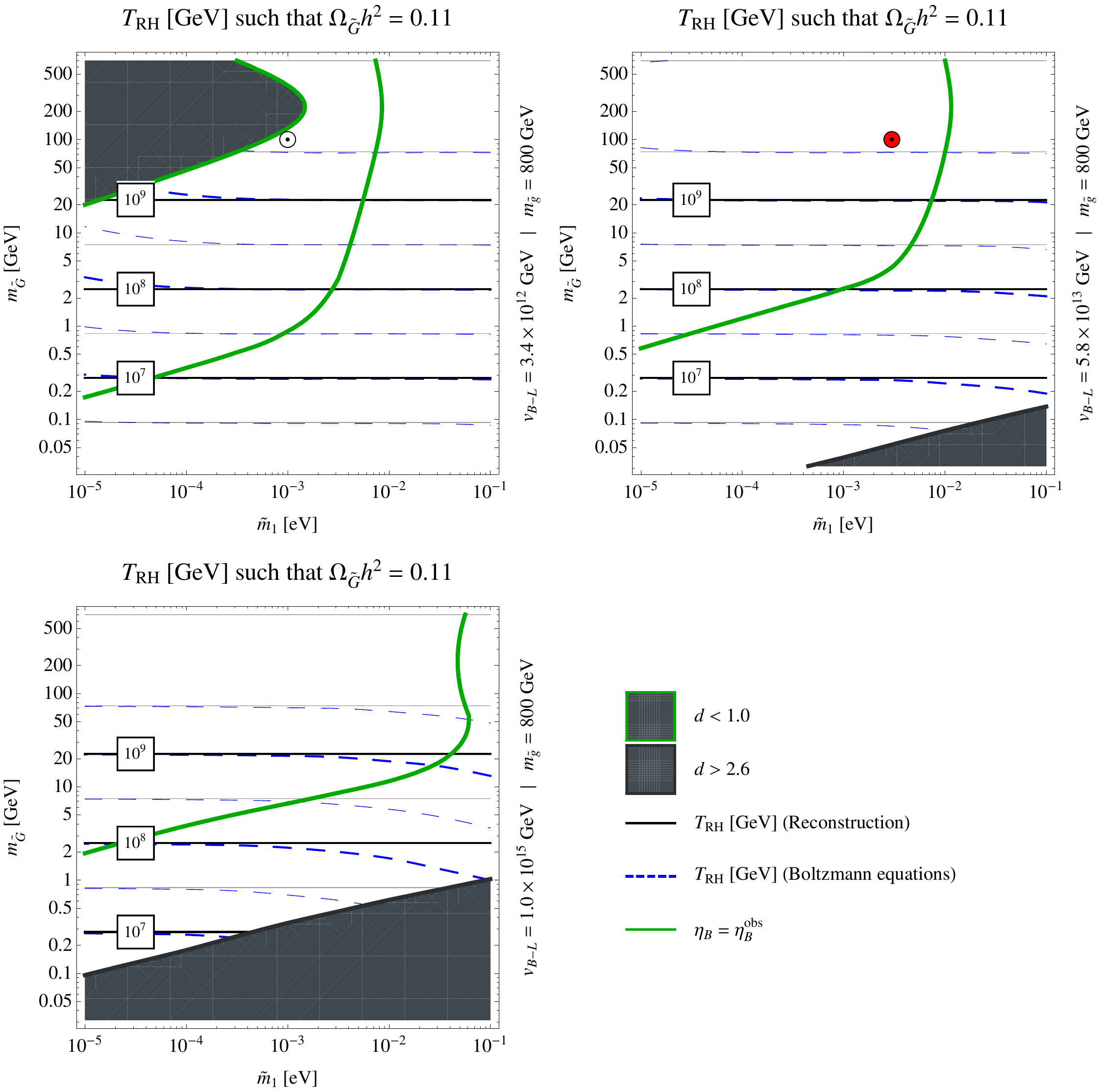}
\caption{Semi-analytical reconstruction of Fig.~\ref{fig:mGboundsT} (solid lines)
on the basis of Eq.~\eqref{eq:OmegaGRes2} with $\varepsilon = 1.32$
and the numerical results for $T_{RH}$ and $\eta_B$.
For comparison also the $T_{RH}$ contours deduced from the
Boltzmann equations (dashed lines) are shown.
They deviate from the reconstructed results as the correction
factor $\varepsilon = \alpha^{-1/2}\beta^{-1}\delta$ actually is
parameter-dependent.
\label{fig:mGreconT}}
\end{center}
\end{figure}

Eq.~\eqref{eq:OmegaGRes2} then implicitly determines the reheating
temperature as a function of $m_{\widetilde{G}}$ and
$m_{\tilde{g}}(\mu_0)$.
Fixing the gluino mass at $800\,\textrm{GeV}$ and solving
Eq.~\eqref{eq:OmegaGRes2} for $T_{RH}$ provides us with an estimate
for the reheating temperature solely dependent on $m_{\widetilde{G}}$ 
(cf. Fig.~\ref{fig:TRHrecon}).
This is all we need to be able to reconstruct Figs.~\ref{fig:treheat}
and \ref{fig:BLasym}:
With $T_{RH} = T_{RH}(m_{\widetilde{G}})$ at hand
we can compute the reheating temperature for all values of
the parameter triple
$\left(v_{B-L},\widetilde{m}_1,m_{\widetilde{G}}\right)$.
From our numerical results for $T_{RH}$ as a function of
$v_{B-L}$, $M_1$ and $\widetilde{m}_1$, shown in Fig.~\ref{fig:treheat},
we then infer the corresponding values of $M_1$.
Our results for $\eta_B$ in Fig.~\ref{fig:BLasym} finally point us
to the respective baryon asymmetries
\begin{align}
\Omega_{\widetilde{G}}h^2 \overset{!}{=} 0.11
\quad & \: \Rightarrow \quad T_{RH} = T_{RH} (m_{\widetilde{G}})\,,
\quad && \textrm{(Eq.~\eqref{eq:OmegaGRes2})}\nonumber\\
\quad & \: \Rightarrow \quad M_1 = M_1 \left(v_{B-L}, T_{RH},
 \widetilde{m}_1\right)\,, 
\quad && \textrm{(Fig.~\ref{fig:treheat})}\nonumber\\
\quad & \: \Rightarrow \quad \eta_B = \eta_B \left(v_{B-L}, M_1,
 \widetilde{m}_1\right)\,.\nonumber 
\quad && \textrm{(Fig.~\ref{fig:BLasym})}
\end{align}
The outcome of this procedure is presented in Figs.~\ref{fig:mGrecon}
and \ref{fig:mGreconT}. 
As it turns out our reconstructed results fit the exact data
from the Boltzmann equations amazingly well.
We thus conclude that our numerical results for
the reheating temperature $T_{RH}$ and the baryon asymmetry
$\eta_B$ when combined with Eq.~\eqref{eq:OmegaGRes2}
essentially suffice to reproduce Figs.~\ref{fig:mGboundsM1} and
\ref{fig:mGboundsT}.

\medskip
Let us check how well the result for $\Omega_{\widetilde{G}}h^2$
(cf. Eq.~\eqref{eq:OmegaGRes2}) that we obtained for the parameter
example discussed 
in Section~\ref{sec:example} can be reproduced with the formul\ae\, developed in this appendix.
The three correction factors $\delta_1$, $\delta_2$ and $\delta_3$
turn out to be
\begin{align}
\delta_1 = \frac{N_\gamma(t_f)}{N_\gamma(t_{RH})} \simeq 2.7\,, \quad
\delta_2 = \frac{N_{\widetilde{G}}(t_f)}{N_{\widetilde{G}}(t_{RH})} \simeq 15\,, \quad
\delta_3 = \frac{H(t_{RH})\,n_{\widetilde{G}}(t_{RH})}{\gamma_{\widetilde{G}}(t_{RH})}\simeq 0.35\,,
\label{eq:d123Res}
\end{align}
which implies that entropy production has almost completed at $t = t_{RH}$.
By contrast the gravitino production rate is still roughly three times as large as the
Hubble rate at this time such that the bulk part of the gravitinos is, in fact,
produced at the last stages of reheating and later.
In combination with Eq.~\eqref{eq:abgRes} the three factors yield
values of $\delta$ and $\varepsilon$ of
\begin{align}
\delta = \delta_2\delta_3 / \delta_1 \simeq 2.0\,, \qquad
\varepsilon = \alpha^{-1/2}\beta^{-1}\delta \simeq 1.3\,.
\end{align}
This result for $\varepsilon$ coincides with the fit value
used for the reconstruction of Figs.~\ref{fig:mGrecon}
and \ref{fig:mGreconT}. 
Notice also that the effects of the various approximations parametrized
by $\alpha$ and $\delta$ tend to cancel such that overall factor $\varepsilon$
represents a correction of only $30\%$ in the end.
Given the reheating temperature in Eq.~\eqref{eq:TRHres} the coefficients
$C_1$ and $C_2$ in Eq.~\eqref{eq:OmegaGRes2} take on the following values
\begin{align}
C_1 (T_{RH}) \simeq 0.17\,, \qquad
C_2 (T_{RH}) \simeq 2.0 \times 10^{-3}\,.
\end{align}
Based on Eq.~\eqref{eq:OmegaGRes2} we can then estimate the
gravitino abundance
\begin{subequations}
\begin{align}
\Omega_{\widetilde{G}}h^2 = & \:  1.3 \times
0.17 \left(\frac{T_{RH}}{10^9 \,\textrm{GeV}}\right)
\left[0.002\left(\frac{m_{\widetilde{G}}}{10\,\textrm{GeV}_{}}\right)
+ \left(\frac{10\,\textrm{GeV}}{m_{\widetilde{G}}}\right)
\left(\frac{m_{\tilde{g}}(\mu_0)}{800\,\textrm{GeV}_{}}\right)^2\right] \\
\simeq & \: 1.3 \times 0.17 \times 4.1 \times \left(0.02 + 0.1\right) \simeq 0.11\,,
\end{align}
\end{subequations}
which is exactly the value we obtained solving the Boltzmann equations.
Without the correction factor $\varepsilon$, we would end up
with a too small value, $\Omega_{\widetilde{G}}h^2 \simeq 8.5 \times 10^{-2}$.
If we were to neglect the term linear in $m_{\widetilde{G}}$ in addition,
our estimate would come out even smaller, $\Omega_{\widetilde{G}}h^2
\simeq 7.1 \times 10^{-2}$.
In concluding, we also mention that our result for the reheating temperature
in Eq.~\eqref{eq:TRHres} coincides, by construction, with the value required
for gravitino dark matter:
In Fig.~\ref{fig:TRHrecon} we read off that to $m_{\widetilde{G}} = 100\,\textrm{GeV}$
corresponds a temperature of $T_{RH} \simeq 4.0 \times
10^9\,\textrm{GeV}$.

\medskip
Finally, our results may be easily generalized to gluino masses other than
$800\,\textrm{GeV}$.
In fact, for given values of $v_{B-L}$, $\widetilde{m}_1$, $M_1$
and $m_{\widetilde{G}}$ it is possible to keep
$\eta_B$ and $\Omega_{\widetilde{G}}h^2$ constant, when changing $m_{\tilde{g}}$,
by simply rescaling the gravitino mass,
\begin{align}
m_{\widetilde{G}}^0 \rightarrow m_{\widetilde{G}} =
m_{\widetilde{G}}\left(m_{\tilde{g}},m_{\widetilde{G}}^0\right)\,, \qquad
m_{\widetilde{G}}\left(800\,\textrm{GeV},m_{\widetilde{G}}^0\right) = m_{\widetilde{G}}^0\,.
\end{align}
As for the baryon asymmetry, this is a trivial consequence
of the fact that $\eta_B$ is a function of $v_{B-L}$, $\widetilde{m}_1$
and $M_1$ only.
In the case of the gravitino abundance we observe that for
fixed reheating temperature, $T_{RH} = T_{RH}\left(v_{B-L},\widetilde{m}_1,M_1\right)$,
$\Omega_{\widetilde{G}}h^2$ remains constant
as long as $m_{\widetilde{G}}\big(m_{\tilde{g}},m_{\widetilde{G}}^0\big)$
is chosen such that the term in square brackets in
Eq.~\eqref{eq:OmegaGRes2} does not change,
\begin{align}
\left[C_2\left(\frac{m_{\widetilde{G}}^0}{10\,\textrm{GeV}}\right)
+ \left(\frac{10\,\textrm{GeV}}{m_{\widetilde{G}}^0}\right)\right] =
\left[C_2\bigg(\frac{m_{\widetilde{G}}}{10\,\textrm{GeV}}\bigg)
+ \left(\frac{10\,\textrm{GeV}}{m_{\widetilde{G}}}\right)
\left(\frac{m_{\tilde{g}}}{800\,\textrm{GeV}_{}}\right)^2\right]\,,
\label{eq:mgscaled}
\end{align}
where $C_2 = C_2(T_{RH})$ is defined in Eq.~\eqref{eq:C2def}.
From Eq.~\eqref{eq:mgscaled} we can determine
the rescaled gravitino mass $m_{\widetilde{G}}$
as a function of the rescaled gluino mass $m_{\tilde{g}}$ and
the original gravitino mass $m_{\widetilde{G}}^0$.
As Eq.~\eqref{eq:mgscaled} is a quadratic equation in $m_{\widetilde{G}}$,
it generically has two solutions $m_{\widetilde{G}}^\pm$,
one of which is typically closer to the original gravitino mass than the other.
$m_{\widetilde{G}}^0$ lies right in between $m_{\widetilde{G}}^-$
and $m_{\widetilde{G}}^+$ once the two terms in square
brackets in Eq.~\eqref{eq:OmegaGRes2} are of equal size, \textit{i.e.} when
gravitinos in helicity $\pm\frac{1}{2}$ states contribute exactly as much to
the total abundance as gravitinos in helicity $\pm\frac{3}{2}$ states.
One easily sees that this is the case when $m_{\widetilde{G}}^0 \simeq 230\,\textrm{GeV}$
(cf. Eq.~\eqref{eq:GammaGprop}).
When going to larger $m_{\tilde{g}}$, we have
$m_{\widetilde{G}}^0 \gtrsim m_{\widetilde{G}}^+ \gg m_{\widetilde{G}}^-$
above $230\,\textrm{GeV}$ and $m_{\widetilde{G}}^0 \lesssim m_{\widetilde{G}}^- \ll m_{\widetilde{G}}^+$
below $230\,\textrm{GeV}$.
At $m_{\tilde{g}}$ smaller than $800\,\textrm{GeV}$, we always
find $m_{\widetilde{G}}^- < m_{\widetilde{G}}^0 < m_{\widetilde{G}}^+$.

\begin{figure}[t!]
\begin{center}
\includegraphics[width=8cm]{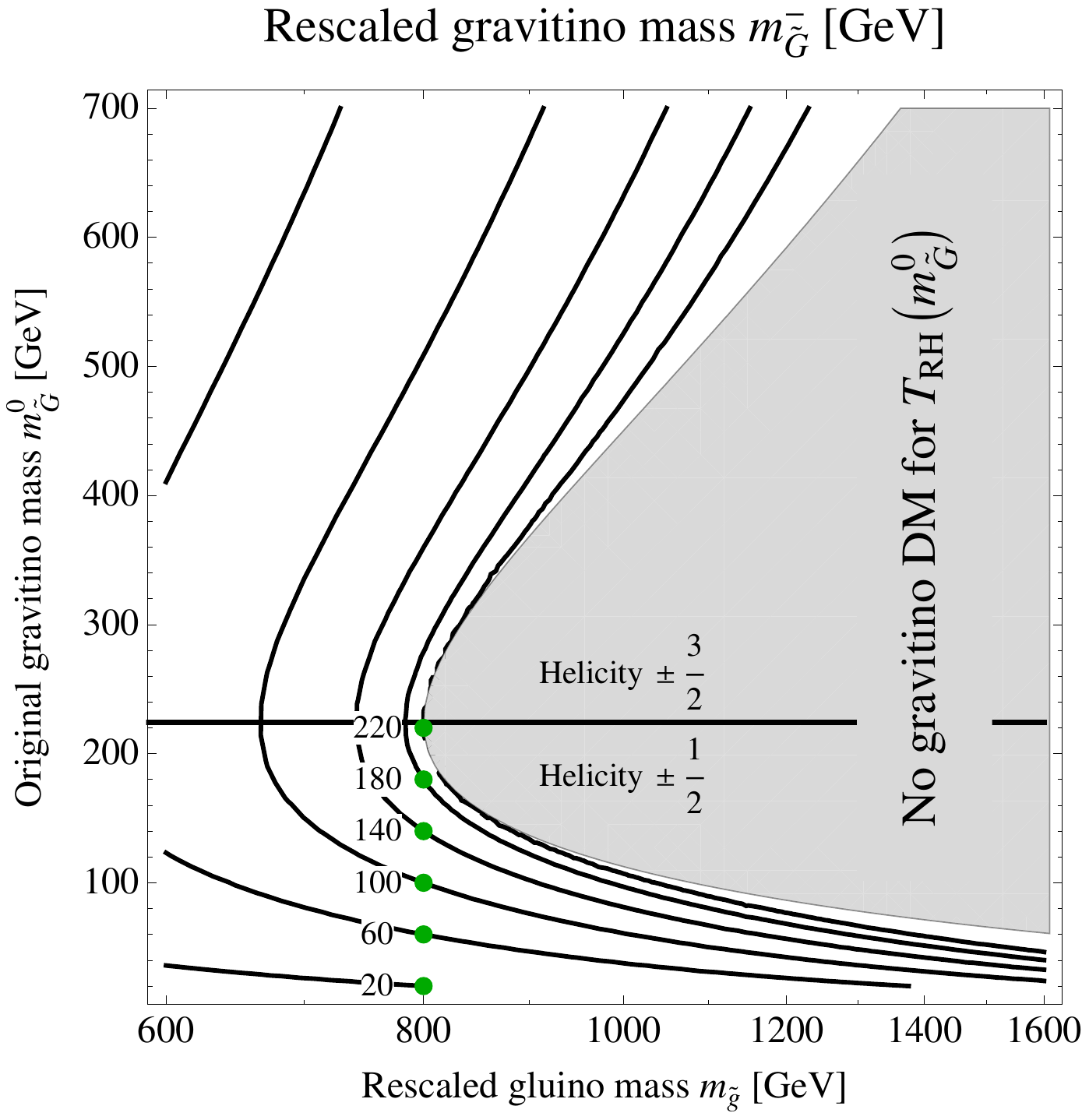}
\includegraphics[width=8cm]{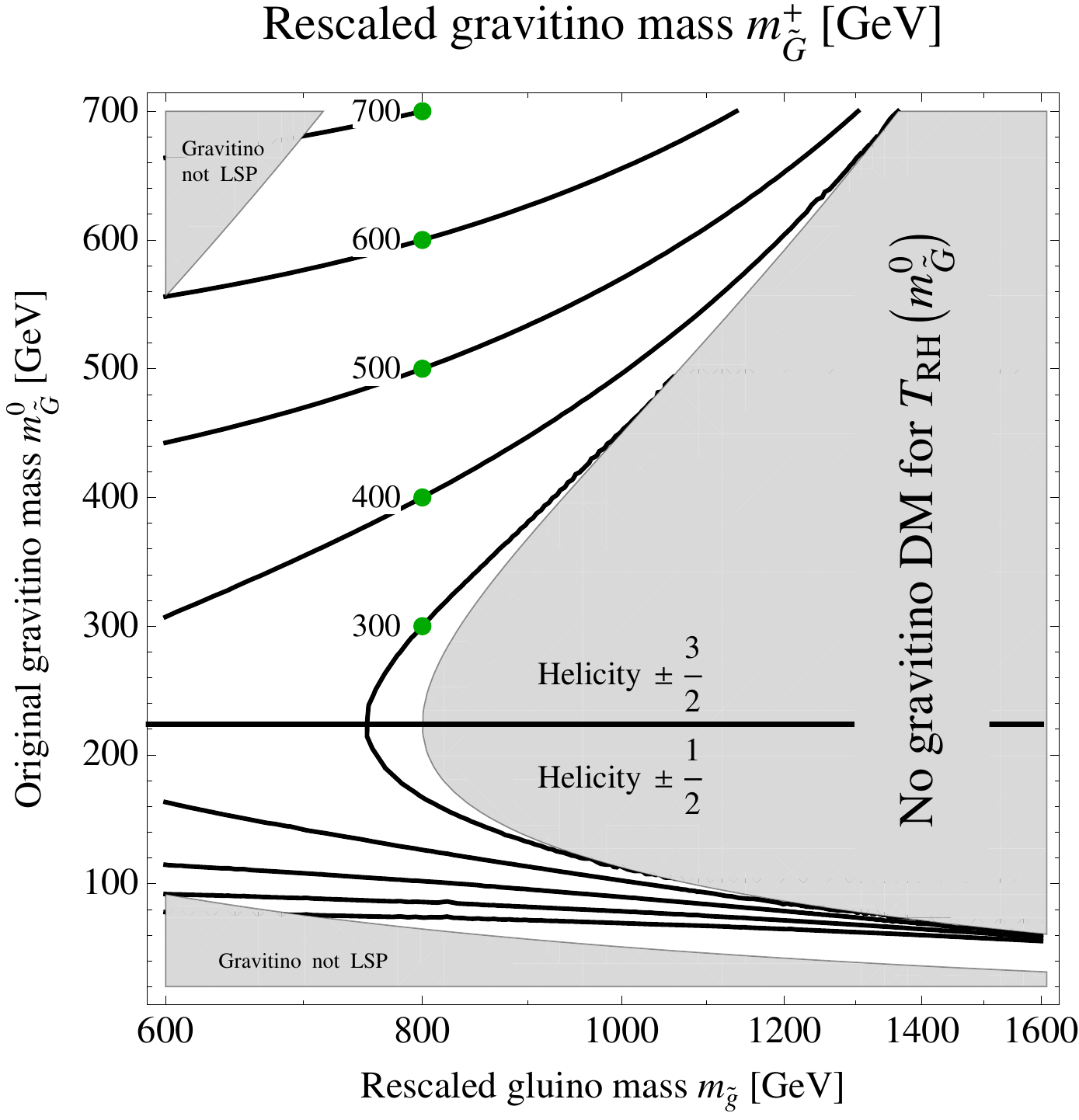}
\caption{Contour plots of the two solutions $\big(m_{\widetilde{G}}^\pm\big)$
of Eq.~\eqref{eq:mgscaled}
for the rescaled gravitino mass $m_{\widetilde{G}}$ as a function of the
rescaled gluino mass $m_{\tilde{g}}$ and the 
original gravitino mass $m_{\widetilde{G}}^0$.
The black solid contours correspond to constant values of $m_{\widetilde{G}}$
(given next to the green dots).
They serve as level curves that allow a determination
of $m_{\widetilde{G}}$ for arbitrary points in the $\big(m_{\tilde{g}},m_{\widetilde{G}}^0\big)$-plane.
They can also be regarded as function graphs of $m_{\widetilde{G}}^0$
as a function of $m_{\tilde{g}}$ for constant $m_{\widetilde{G}}$.
We restrict ourselves to the interval
$20\,\textrm{GeV} \leq m_{\widetilde{G}}^0 \leq 700\,\textrm{GeV}$
in this figure.
Below $20\,\textrm{GeV}$, Eq.~\eqref{eq:naivesc} provides
an excellent approximation.
\label{fig:mGscaled}}
\end{center}
\end{figure}

If the gravitino mass is much smaller than the gluino mass,
almost only the goldstino part of the gravitino is produced
and the term linear in $m_{\widetilde{G}}$ in Eq.~\eqref{eq:OmegaGRes2}
can be neglected.
The scaling behaviour of the gravitino mass then becomes trivial
\begin{align}
m_{\widetilde{G}}^0 \ll m_{\tilde{g}}\,: \quad
m_{\widetilde{G}} =
m_{\widetilde{G}}^0 \left(\frac{m_{\tilde{g}}}{800\,\textrm{GeV}_{}}\right)^2\,.
\label{eq:naivesc}
\end{align}

Actually, the rescaled gravitino mass $m_{\widetilde{G}}$ also is a function
of $T_{RH}$ as it depends on the coefficient $C_2(T_{RH})$.
But as discussed in this appendix, there is an almost unique correspondence
between the gravitino mass and the reheating temperature.
In order to solve Eq.~\eqref{eq:mgscaled} we thus simply read off
$T_{RH}$ from Fig.~\ref{fig:TRHrecon} as a function of the input gravitino
mass, $T_{RH} = T_{RH}\big(m_{\widetilde{G}}^0\big)$.
As can be seen from Fig.~\ref{fig:mGreconT}, this simplified reheating temperature
is in good agreement with the exact outcome of the Boltzmann equations.
Our solutions $m_{\widetilde{G}}^\pm$ for the rescaled gravitino mass are
presented in the two panels of Fig.~\ref{fig:mGscaled}, respectively.
In the gray shaded regions there are either no real solutions of
Eq.~\eqref{eq:mgscaled} or the rescaled gravitino mass is larger
than the corresponding gluino mass, $m_{\widetilde{G}} > m_{\tilde{g}}$.
The former case implies that it is impossible to keep the gravitino
abundance constant when going to larger $m_{\tilde{g}}$
while sticking to the reheating temperature $T_{RH}\big(m_{\widetilde{G}}^0\big)$.
In the latter case, the gravitino would not be the LSP any longer.

\end{document}